\documentclass[onefignum,onetabnum]{siamart190516}

\usepackage{amssymb,amsmath}
\usepackage{bm}
\usepackage{lipsum}
\usepackage{amsfonts}
\usepackage{graphicx,subfigure}
\usepackage{algorithmic}
\usepackage{color}
\usepackage{tikz}

\headers{Continuum Model and Numerical Method for Grain Boundaries}{X. X. Qin, Y. J. Gu, L. C. Zhang, and Y. Xiang}

\title{Continuum Model and Numerical Method for Dislocation Structure and Energy of Grain Boundaries\thanks{Submitted to the editors DATE.
\funding{This work was funded by the Hong Kong Research Grants Council General Research Fund 16301720  and 16302818.}}}

\author{Xiaoxue Qin\thanks{Department of Mathematics, Hong Kong University of Science and Technology, Clearwater Bay, Kowloon, Hong Kong
  (\email{maxqin@ust.hk}, \email{maxiang@ust.hk}).}
\and Yejun Gu\thanks{Institute of High Performance Computing, A*STAR, Fusionopolis, 138632, Singapore, and Department of Mechanical Engineering, Whiting School of Engineering, The Johns Hopkins University, Baltimore, MD, 21218, USA
  (\email{yejungu@ihpc.a-star.edu.sg}, \email{yejungu@jhu.edu}).}
\and Luchan Zhang\thanks{College of Mathematics and Statistics, Shenzhen University, Shenzhen, 518060,  China
  (\email{zhanglc@szu.edu.cn}, \email{malczhang@ust.hk}).}
\and Yang Xiang\footnotemark[2]}

\ifpdf
\hypersetup{
  pdftitle={Continuum Model and Numerical Method for Dislocation Structure and Energy of Grain Boundaries},
  pdfauthor={Xiaoxue Qin, Yejun Gu, Luchan Zhang, Yang Xiang }
}
\fi

\begin{document}

\maketitle

% REQUIRED
\begin{abstract}
  We present a continuum model to determine the dislocation structure and energy of low angle grain boundaries in three dimensions.  The equilibrium dislocation structure is obtained by minimizing the grain boundary energy that is associated with the constituent dislocations subject to the constraint of Frank's formula. The orientation-dependent continuous distributions of dislocation lines on grain boundaries are described conveniently using the dislocation density potential functions, whose contour lines on the grain boundaries represent the dislocations. The energy of a grain boundary is the total energy of the constituent dislocations derived from discrete dislocation dynamics model, incorporating both the dislocation line energy and reactions of dislocations.  The constrained energy minimization problem is solved by the augmented Lagrangian method and projection method. Comparisons with atomistic simulation results show that our continuum model is able to give excellent predictions of the energy and dislocation densities of both planar and curved low angle grain boundaries.
\end{abstract}

% REQUIRED

% REQUIRED
%\begin{AMS}
%35Q74, 49M37, 74A50, 74A10
%\end{AMS}

\section{Introduction}
Grain boundaries (the interfaces between adjacent grains) are important components in polycrystalline materials. The energy and dynamic of grain boundaries play essential roles in the behaviors of polycrystalline materials such as grain growth, recrystallization, plastic deformation, etc \cite{Sutton1995}.
 The classical grain boundary dynamics models are based up the motion by mean curvature to reduce the total  energy of the grain boundaries \cite{Herring1951,Mullins1956,Sutton1995}.
The investigation on grain boundaries have been an active research area of materials science and mathematics for many decades, e.g. \cite{Chenlq1994,Kobayashi2000,Kazaryan2000,liuchun2001,Selim2009,ZhangXiang2018,Liu2019gb,Zhang2020} for dynamics simulation methods and \cite{Cahn2004,Zhang2005,Srolovitz2007,Du2009,ZhangXiang2018,Liu2019gb} for static and dynamic properties of grain boundaries, most of which are based on given grain boundary energies.

The energy and dynamics of grain boundaries crucially depend on their underlying microstructure. Low angle grain boundaries can be modeled as arrays of dislocations (line defects) \cite{ReadShockley1950,Sutton1995,HirthLothe1982}. Dislocations are characterized by Burgers vectors. The Burgers vector of a dislocation describes the lattice distortion caused by its presence and is invariant along the dislocation. There are finite number of possible Burgers vectors in a crystal, e.g., six in an fcc (face centered cubic) crystal. The dislocation structure of a grain boundary is associated with local dislocation line directions, inter-dislocation distances, and Burgers vectors of the dislocations.
Simulations of atomistic and dislocation models have revealed many interesting properties of grain boundary dislocation structure and the associated energy and dynamics of grain boundaries \cite{Olmsted2009,Holm2010,lim2009,lim2012,wu2012phase,Winther2013,Dai2013,Dai2014,
Shen2014,yamanaka2017phase,Voigt2017,Voigt2018}.
 Although these microscopic models provide detailed information on the dislocation or atomistic structures of individual grain boundaries, continuum model is desired for energetic and dynamics of grain boundaries at larger length scales.

\begin{figure}[htbp]
\centering
\includegraphics[width=0.5\linewidth]{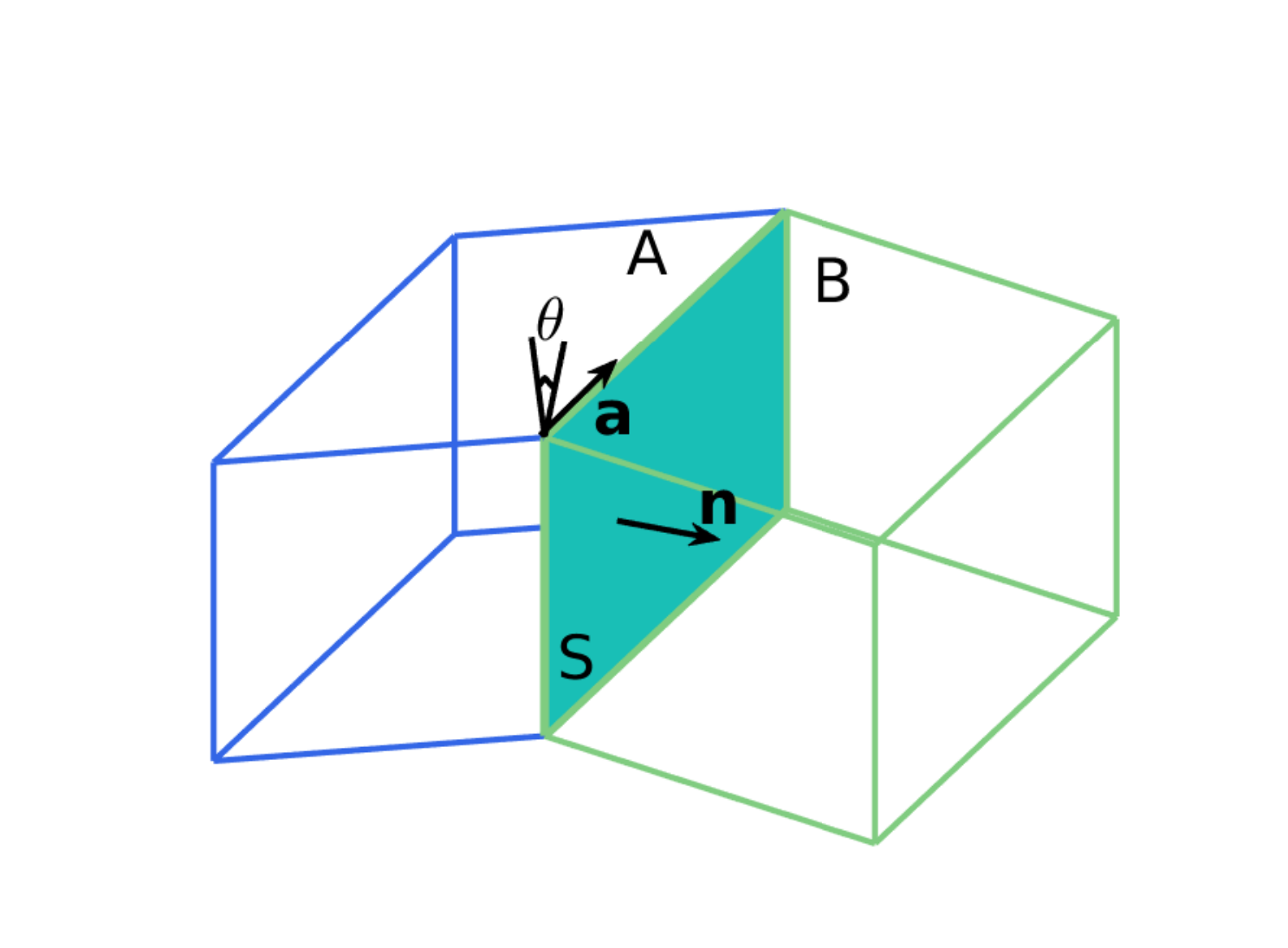}
  \caption{ Illustration of a grain boundary and five degrees of freedom. $A$ and $B$ are two grains and $S$ is the grain boundary. Five degrees of freedom include:  grain boundary normal direction $\mathbf n$, misoreintation angle $\theta$, and rotation axis $\mathbf a$. This example is a tilt boundary.}
\label{fig:5dof}
\end{figure}

On the macroscopic scale, a grain boundary has five  degrees of freedom (DOFs), namely the grain misorientation angle (one DOF), rotation axis (two DOFs) and the boundary plane orientation (two DOFs) \cite{Sutton1995}; see Fig.~\ref{fig:5dof} for an illustration. In the classical theory of Read and Shockley \cite{ReadShockley1950},
the grain boundary energy is $E=E_0\theta(A-\ln\theta)$, where $\theta$ is the misorientation angle and parameters $E_0$ and $A$ depend on the grain boundary orientation and rotation axis.
Olmsted \textit{et al.} and Holm \textit{et al.} \cite{Olmsted2009,Holm2010} calculated the energies of a set of 388 distinct grain boundaries by using atomistic simulations.
Saylor \textit{et al.} \cite{Saylor2003} and Bulatov \textit{et al.} \cite{Bulatov2014} constructed grain boundary energy based on all five DOFs   using energy reconstruction or interpolation. These models  do not directly depend on the dislocation microstructure of grain boundaries except for some special grain boundaries.

Theoretically, the Frank's formula \cite{Frank1950,Bilby1955,HirthLothe1982,Sutton1995} provides a link between the five macroscopic  DOFs and the microscopic dislocation structure.
However, this classical formula is only a necessary condition and in general is not able to uniquely determine the dislocation structure.  In the previously available discrete dislocation dynamics based works on the dislocation structures of low angle grain boundaries \cite{lim2009,lim2012,Winther2013} or heterogeneous interfaces \cite{Quek2011,WangJian2013,Demkowicz2013}, two or three prescribed Burgers vectors informed by experimental observations or atomistic simulations (instead of all six possible Burgers vectors) were adopted so that the Frank's formula was able to give a unique solution of the dislocation structure.
Zhang $et$ $al$.~\cite{zhang2017energy} has developed a continuum model to obtain dislocation structure on planar grain boundaries with all possible Burgers vectors by minimizing the grain boundary energy associated with the dislocation structure subject to the constraint of Frank's formula. The constrained minimization problem is solved by the penalty
method by which it is turned into an unconstrained minimization problem.

In this paper, we generalize the continuum model in Ref.~\cite{zhang2017energy}  to dislocation structure and energy of low angle grain boundaries  in three dimensions, in which both the grain boundaries and their constituent dislocations are curved in general.  Note that in continuum grain boundary dynamics models \cite{Herring1951,Mullins1956,Sutton1995}, a grain boundary has well-defined macroscopic quantities including misorientation angle, rotation axis, and grain boundary energy during the dynamics process. This is under the assumption that the equilibrating process of the microstructure of the grain boundary is much faster than the dynamics process and the grain boundary on the continuum level is considered to be always in equilibrium in terms of its microscopic structure. Therefore, understanding the equilibrium dislocation structure and energy of a fixed low angle grain boundary provides a basis for the study of the dynamics problem.

In our continuum model, the distributions of the constituent dislocations on a general grain boundary are represented by the scalar dislocation density potential functions (DDPFs) \cite{zhu2014continuum} defined on the grain boundary, whose contour lines describe the locations of dislocations on the grain boundary. This simple representation method also guarantees continuity of the dislocation lines on the grain boundaries. The energy of a grain boundary is the total energy of the constituent dislocations derived from discrete dislocation dynamics model \cite{xiang2012continuum,zhu2014continuum}. In this paper, an improved grain boundary energy formula  is proposed to further incorporate dislocation reactions in a dislocation network. we also propose a method to identify the exact dislocation network structures from the  dislocation densities obtained using our continuum model.

As in the continuum model in two dimensions, the dislocation structure on a grain boundary in three dimensions is obtained by solving a constraint energy minimization problem, i.e., minimizing the grain boundary energy  subject to the constraint of the Frank's formula.
The constrained energy minimization problem is solved by the augmented Lagrangian method combined with the projection method. Convergence of this method is discussion. Compared with the penalty method used in Ref.~\cite{zhang2017energy}, the parameter of the penalty function in augmented Lagrangian method in general does not need to go to infinity to achieve convergence, thus the time-consuming and numerical ill-conditioning problems associated with the large parameter in the penalty method are avoided, which is more important for calculations in three dimensions.  A numerical formulation that avoids ill-posedness is proposed for the nonconvex gradient energy in the continuum model, and a numerical method based on projection method is presented to ensure connectivity of the dislocations.

We perform simulations using our continuum model for the  cylindrical and spherical low angle grain boundaries. Comparisons with atomistic simulation results show that our continuum model is able to give excellent predictions of the dislocation structure and energy  of both planar and curved low angle grain boundaries.

The paper is organized as follows. We first review the dislocation model of low angle grain boundaries and the Frank's formula in Sec.~\ref{sec:frank}. In Sec.~\ref{s2}, we present our continuum model. In Sec.~\ref{sec:reaction}, details of incorporation of dislocation reaction in the continuum are presented. A method to identify the exact dislocation network structures from the  dislocation densities obtained using our model is proposed in Sec.~\ref{sec:identification}. In Sec.~\ref{sec:algorithm}, a numerical formulation that avoid ill-posedness is proposed and numerical algorithm based on augmented Lagrangian method and projection method is developed. In Sec.~\ref{sec:NumericalResults}, we perform simulations using our continuum model for the dislocation structure and energy of cylindrical and spherical low angle grain boundaries and compare the results with those of atomistic simulations.

\section{Review of the Frank's formula}\label{sec:frank}

The Frank's formula (or the Frank-Bilby equation) \cite{Frank1950,Bilby1955,HirthLothe1982,Sutton1995} is a condition that governs the dislocation structure of an equilibrium grain boundary. It provides a link between the five macroscopic  DOFs and the microscopic dislocation structure. This condition is equivalent to the cancelation of the long-range elastic field associated with the grain boundary. The classical Frank's formula was obtained based on planar grain boundaries. It has been shown in Ref.~\cite{zhu2014continuum} that for curved grain boundaries, the Frank's formula is equivalent to the cancelation of the continuum long-range elastic fields (i.e. calculated from continuous distributions of the constituent dislocations of the grain boundaries).

 Given a planar grain boundary $S$ with normal direction $\mathbf n$, rotation axis $\mathbf a$ and  misorientation angle $\theta$ (for a low angle grain boundary, $\theta\leq 15^\circ$), the Frank's formula is \cite{Frank1950,Bilby1955,HirthLothe1982,Sutton1995}
 \begin{equation}\label{eqn:frank0}
\theta(\mathbf{V}\times \mathbf{a})= \sum_{j=1}^J \mathbf{b}^{(j)}(\mathbf N^{(j)} \cdot\mathbf{V}),
\end{equation}
where $\mathbf V$ is any vector in the grain boundary, $\mathbf b^{(j)}$,  $j=1,2,\cdots,J$, are the Burgers vectors of the dislocations, and  $\mathbf N^{(j)}$ is a vector that describes the distribution of the dislocations with Burgers vector $\mathbf b^{(j)}$ and is defined as
\begin{equation}\label{eqn:vectorN}
\mathbf N^{(j)}=\frac{1}{D_j}\mathbf n\times\pmb \xi^{(j)},
\end{equation}
with $\pmb \xi^{(j)}$ and $D_j$ respectively being the direction and inter-dislocation distance of the dislocations with Burgers vector $\mathbf b^{(j)}$. Here we follow the definition of vector $\mathbf N^{(j)}$ in Ref.~\cite{Sutton1995}. (The definition of $\mathbf N^{(j)}$ in Ref.~\cite{HirthLothe1982} also includes the misorientation angle $\theta$.)

\begin{figure}[htbp]
\subfigure[]{\includegraphics[width=2.5in]{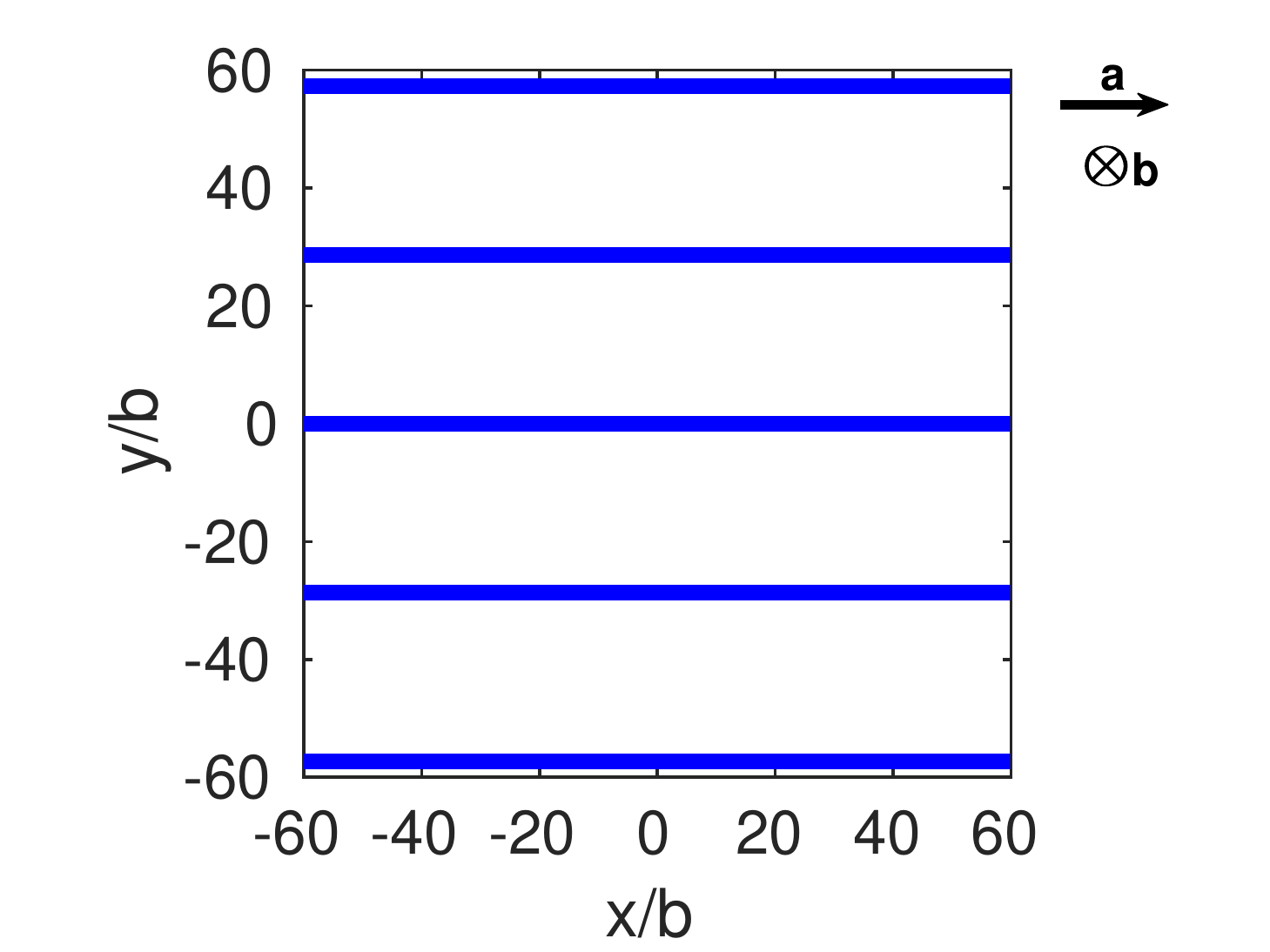}}
\subfigure[]{\includegraphics[width=2.5in]{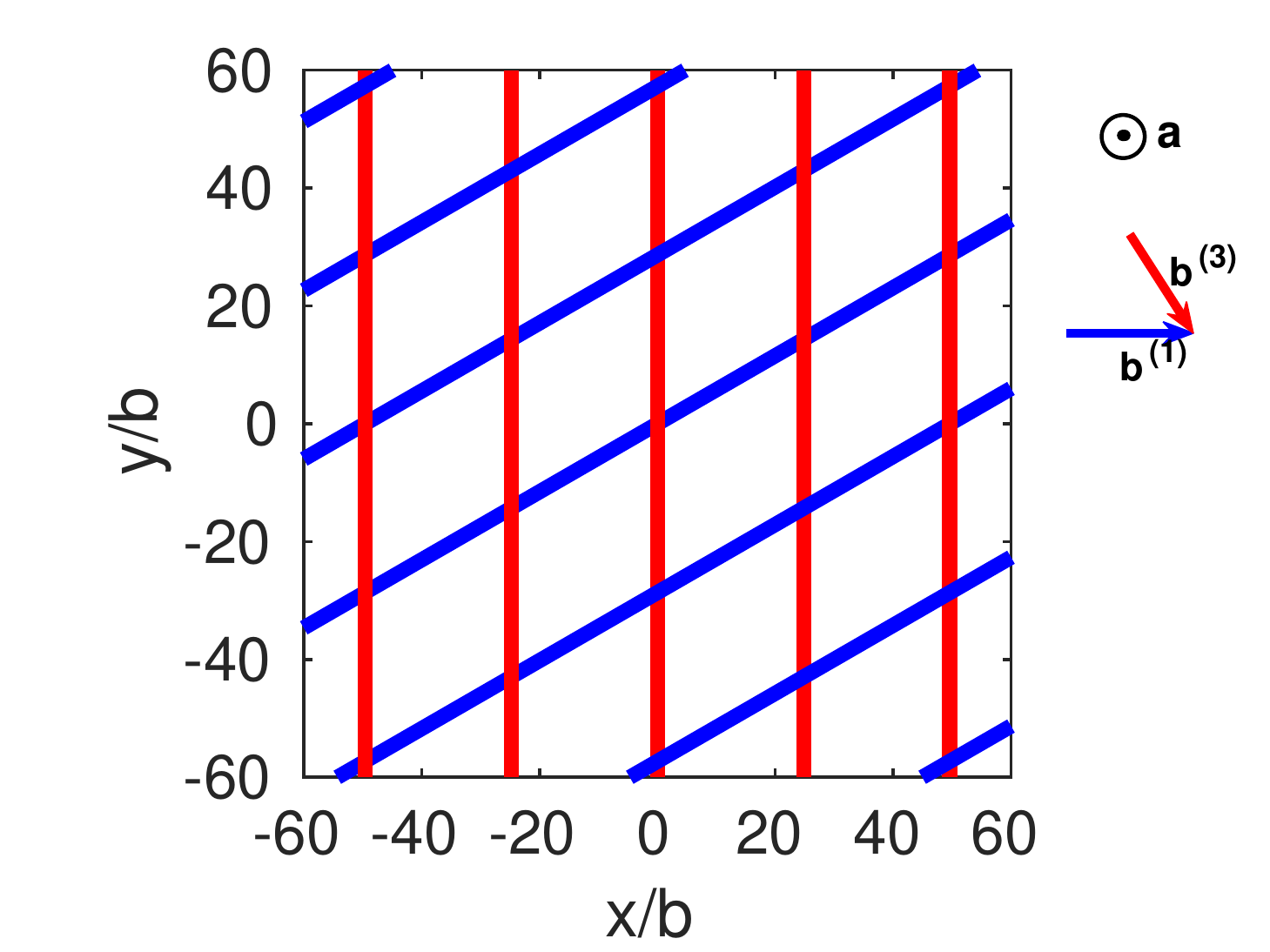}}\\
\subfigure[]{\includegraphics[width=2.5in]{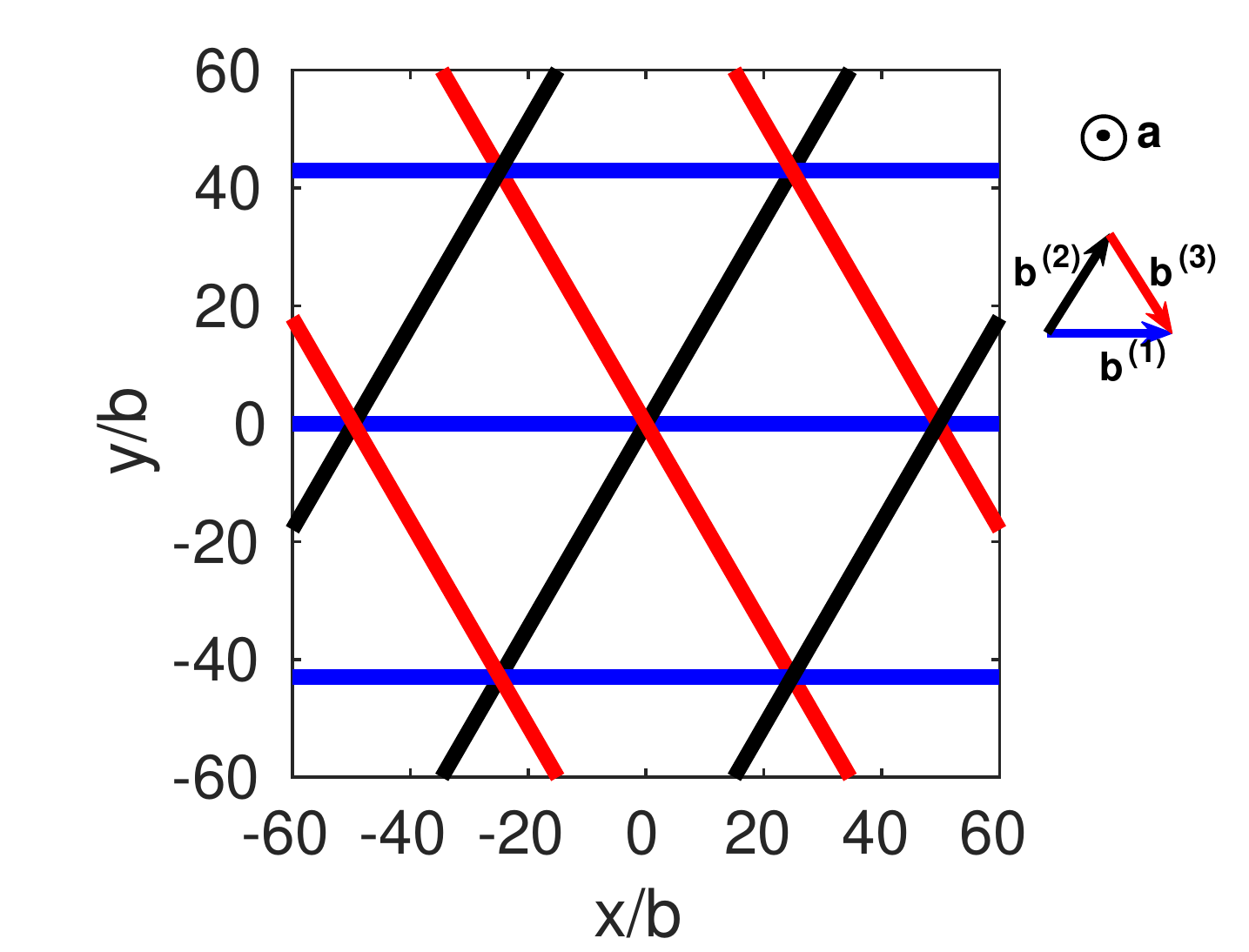}}
\subfigure[]{\includegraphics[width=2.5in]{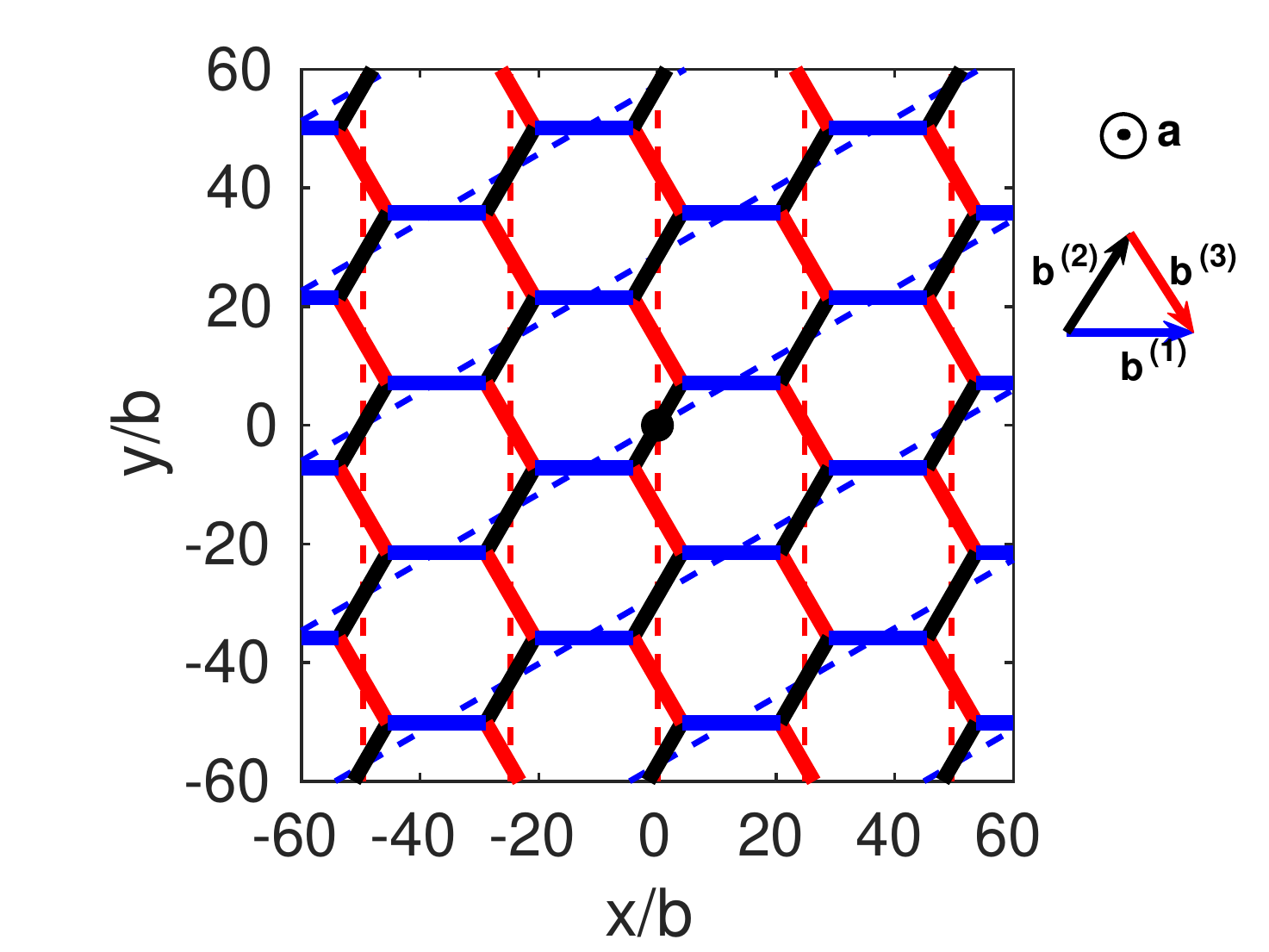}}\\
  \caption{Examples of dislocation structures that satisfy  the Frank's formula  \eqref{eqn:frank0}. The grain boundary is the $xy$ plane with normal direction $\mathbf n=(0,0,1)$. The misorientation angle $\theta=2^\circ$. (a) Dislocation structure on a symmetric tilt boundary,  which is the unique solution of the Frank's formula when all the dislocations have the same Burgers vector $\mathbf b=b(0,0,-1)$.  (b)-(d) Different dislocation structures on a pure twist boundary, with three possible Burgers vectors $\mathbf b^{(1)}=b(1,0,0)$, $\mathbf b^{(2)}=b(\frac{1}{2},\frac{\sqrt{3}}{2},0)$, and $\mathbf b^{(3)}=b(\frac{1}{2},-\frac{\sqrt{3}}{2},0)$ that are in the grain boundary plane. (b) Dislocation structure on this twist boundary that consists of two sets of dislocations with Burgers vectors $\mathbf b^{(1)}$ and $\mathbf b^{(3)}$. (c) Dislocation structure on this twist boundary that consists of three sets of dislocations with Burgers vectors $\mathbf b^{(1)}$, $\mathbf b^{(2)}$ and $\mathbf b^{(3)}$. (d) Dislocation structure on this twist boundary after dislocation reaction from the structure in (b). Two intersecting dislocation segments with Burgers vectors   $\mathbf b^{(1)}$ (in blue) and $\mathbf b^{(3)}$ (in red) react to form a dislocation segment with Burgers vector   $\mathbf b^{(2)}$ (in black), see for example, at the intersection point in the middle of this image. This is the actual dislocation structure on this twist boundary.}
\label{fig:01}
\end{figure}

First consider a case in which all the dislocations have the same Burgers vector $\mathbf b$. Assume that the grain boundary is the $xy$ plane, i.e. $\mathbf n=(0,0,1)$, the Burgers vector is in the $z$ direction, i.e., $\mathbf b=b(0,0,-1)$ with $b$ being the length of the Burgers vector, and the rotation axis is in the $x$ direction, i.e. $\mathbf a=(1,0,0)$. This is a symmetric tilt boundary. In this case, the Frank's formula in Eq.~\eqref{eqn:frank0} gives $bN_1=0$, $bN_2=\theta$ for the $\mathbf N$-vector $\mathbf N=(N_1,N_2,0)$. This gives a unique solution $\mathbf N=(0,\frac{\theta}{b},0)$. Using Eq.~\eqref{eqn:vectorN}, this solution means that the dislocations are in the direction $\pmb \xi=\frac{\mathbf N}{\|\mathbf N\|}\times \mathbf n=(1,0,0)$ and the inter-dislocation distance is $D=\frac{1}{\|\mathbf N\|}=\frac{b}{\theta}$. The dislocation structure of this tilt boundary is shown in Fig.~\ref{fig:01}(a).

Next, we consider a case in which dislocations with three Burgers vectors are present. Assume that the grain boundary is the $xy$ plane, i.e. $\mathbf n=(0,0,1)$,  and the rotation axis is also in the $z$ direction, i.e. $\mathbf a=(0,0,1)$. Dislocations with the following three Burgers vectors are present:  $\mathbf b^{(1)}=b(1,0,0)$, $\mathbf b^{(2)}=b(\frac{1}{2},\frac{\sqrt{3}}{2},0)$, and $\mathbf b^{(3)}=b(\frac{1}{2},-\frac{\sqrt{3}}{2},0)$,
which are all in the grain boundary plane. This is a pure twist boundary in an fcc crystal.
In this case, the Frank's formula in Eq.~\eqref{eqn:frank0} gives the following linear system for the three $\mathbf N$-vectors  $\mathbf N^{(j)}=(N_1^{(j)},N_2^{(j)},0)$, $j=1,2,3$, for the arrangement of the three sets of dislocations:
\begin{align}
\left(
\begin{array}{c}
0\vspace{1ex}\\
-\theta
\end{array}
\right)=N_1^{(1)}b
\left(
\begin{array}{c}
1\vspace{1ex}\\
0
\end{array}
\right)
+N_1^{(2)}b
\left(
\begin{array}{c}
\frac{1}{2}\vspace{1ex}\\
\frac{\sqrt{3}}{2}
\end{array}
\right)
+N_1^{(3)}b
\left(
\begin{array}{c}
\frac{1}{2}\vspace{1ex}\\
-\frac{\sqrt{3}}{2}
\end{array}
\right),\vspace{1ex}\\
\left(
\begin{array}{c}
\theta\vspace{1ex}\\
0
\end{array}
\right)=N_2^{(1)}b
\left(
\begin{array}{c}
1\vspace{1ex}\\
0
\end{array}
\right)
+N_2^{(2)}b
\left(
\begin{array}{c}
\frac{1}{2}\vspace{1ex}\\
\frac{\sqrt{3}}{2}
\end{array}
\right)
+N_2^{(3)}b
\left(
\begin{array}{c}
\frac{1}{2}\vspace{1ex}\\
-\frac{\sqrt{3}}{2}
\end{array}
\right).
\end{align}
This linear system has infinitely many solutions. That is, there are infinitely many dislocation structures that can satisfy the Frank's formula for this twist boundary. Two solutions are shown in Fig.~\ref{fig:01}(b) and (c). The solution in Fig.~\ref{fig:01}(b) has two sets of dislocations with Burgers vectors $\mathbf b^{(1)}$ and $\mathbf b^{(3)}$:  $\mathbf N^{(1)}=(-\frac{\theta}{\sqrt{3}b},\frac{\theta}{b},0)$, $\mathbf N^{(2)}=0$,
$\mathbf N^{(3)}=(\frac{2\theta}{\sqrt{3}b},0,0)$; the solution in Fig.~\ref{fig:01}(c) has three sets of dislocations with Burgers vectors $\mathbf b^{(1)}$, $\mathbf b^{(2)}$, and $\mathbf b^{(3)}$:  $\mathbf N^{(1)}=(0,\frac{2\theta}{3b},0)$,
$\mathbf N^{(2)}=(-\frac{\sqrt{3}\theta}{3b},\frac{\theta}{3b} ,0)$,
$\mathbf N^{(3)}=(\frac{\sqrt{3}\theta}{3b},\frac{\theta}{3b} ,0)$.
The Frank's formula is not able to uniquely determine a dislocation structure in this case.

Note that the dislocations in the structure in Fig.~\ref{fig:01}(b) can react to form a new structure, see  Fig.~\ref{fig:01}(d). (Recall that two intersecting dislocations with Burgers vectors  $\mathbf b^{(1)}$ and $\mathbf b^{(2)}$ may react to form a dislocation with Burgers vector $\mathbf b=\mathbf b^{(1)}+\mathbf b^{(2)}$ if the latter has lower energy.) In fact, this dislocation structure has lower energy and is the actual dislocation structure of the pure twist boundary in an fcc crystal. Since the dislocations are no longer straight lines in this structure, the Frank's formula in Eq.~\eqref{eqn:frank0} is not able to predict this dislocation structure directly. However, this dislocation structure still satisfies the Frank's formula in the averaged sense, i.e., $\theta(\mathbf{V}\times \mathbf{a})= \mathbf B(\mathbf V)$, where $\mathbf B(V)$ is the net Burgers vector of all the dislocations cut by vector $\mathbf V$ and is calculated by averaging the contributions within one period of the dislocation structure.

\begin{figure}[!htb]
  \centering
  \includegraphics[width=3.5in]{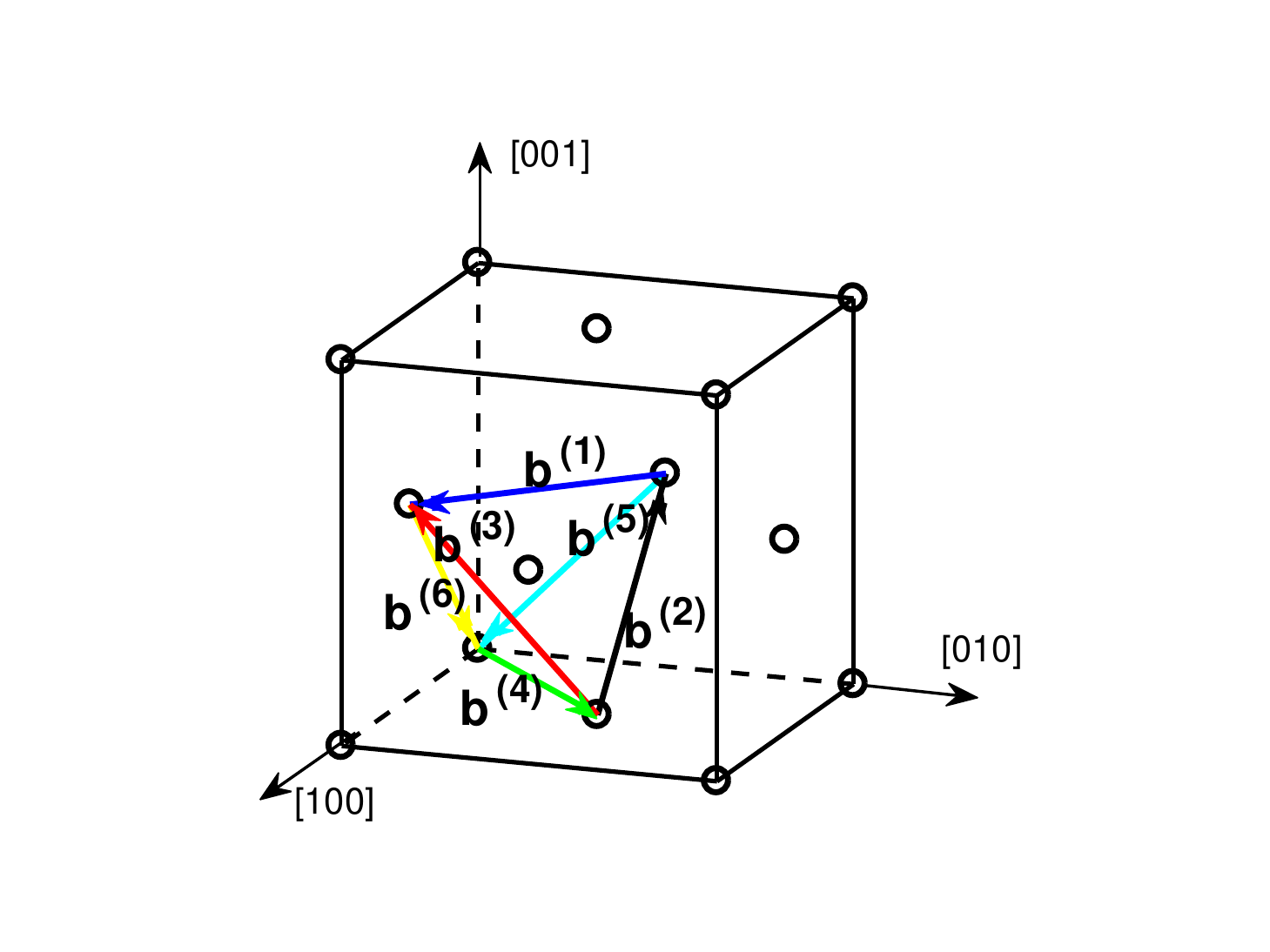}\\
  \caption{There are $6$ Burgers vectors (neglecting the sign) in an fcc crystal: $\mathbf b^{(1)}, \mathbf b^{(2)}, \cdots, \mathbf b^{(6)}$.  They are of the $<110>$ type and have the same length $b=a/\sqrt{2}$, where $a$ is the lattice constant. These Burgers vectors form a Thompson tetrahedron.}
  \label{thompson}
\end{figure}

There are finite number of possible Burgers vectors in a crystal. For example, in an fcc (face-centered cubic) crystal~\cite{HirthLothe1982},  there are $J=6$ Burgers vectors (neglecting the sign), which are of the $<110>$ type and have the same length $b=a/\sqrt{2}$ (where $a$ is the lattice constant of the crystal); and these Burgers vectors form a Thompson tetrahedron, as shown in Fig.~\ref{thompson}.

\section{Continuum model}
\label{s2}

In this section, we present our continuum model for  dislocation structure and energy of low angle grain boundaries in three dimensions. The equilibrium dislocation structure is obtained by minimizing the grain boundary energy that is associated with the constituent dislocations subject to the constraint of Frank's formula.

\begin{figure}[htbp]
\subfigure[]{\includegraphics[width=2.5in]{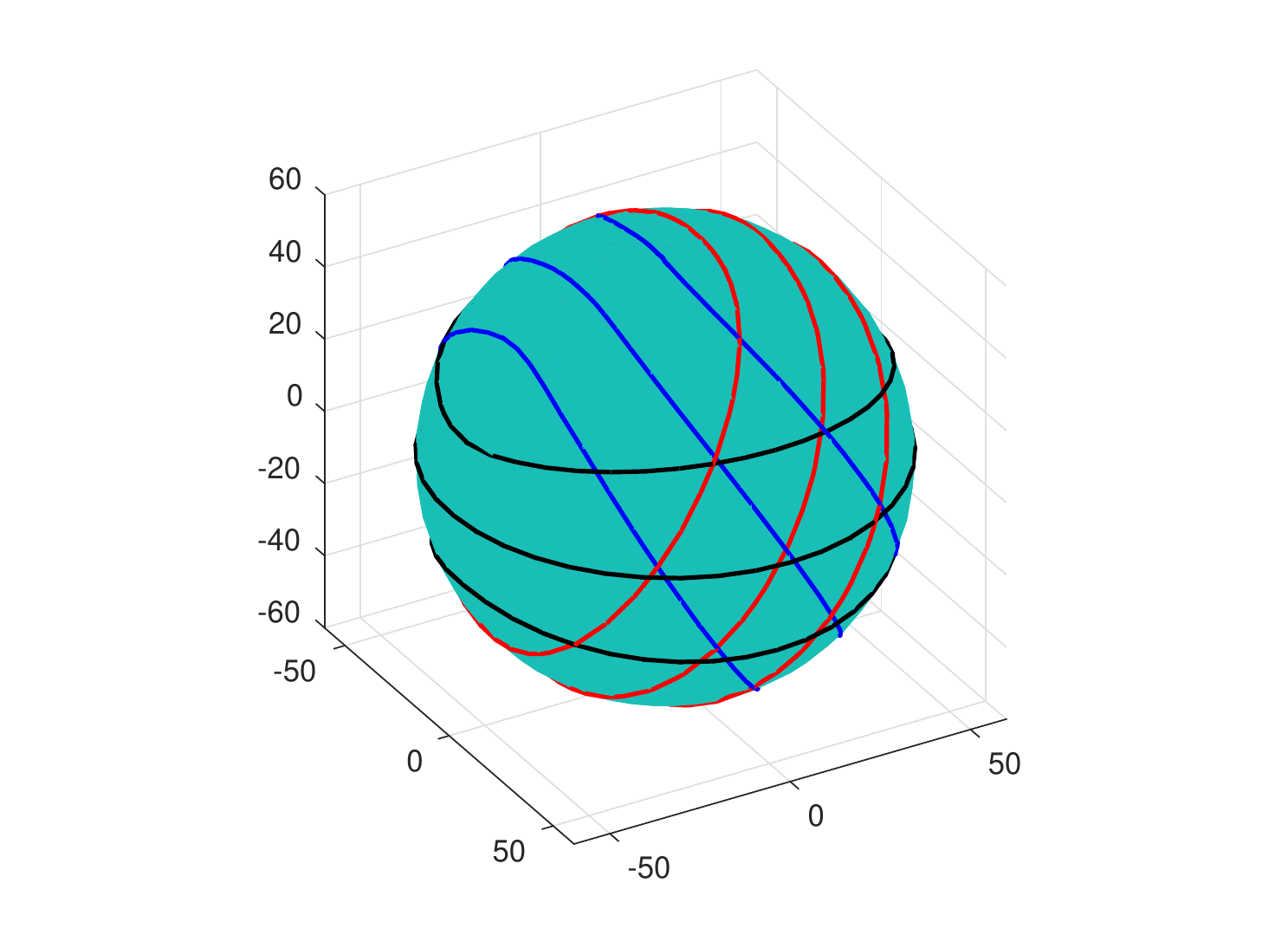}}
\subfigure[]{\includegraphics[width=2.5in]{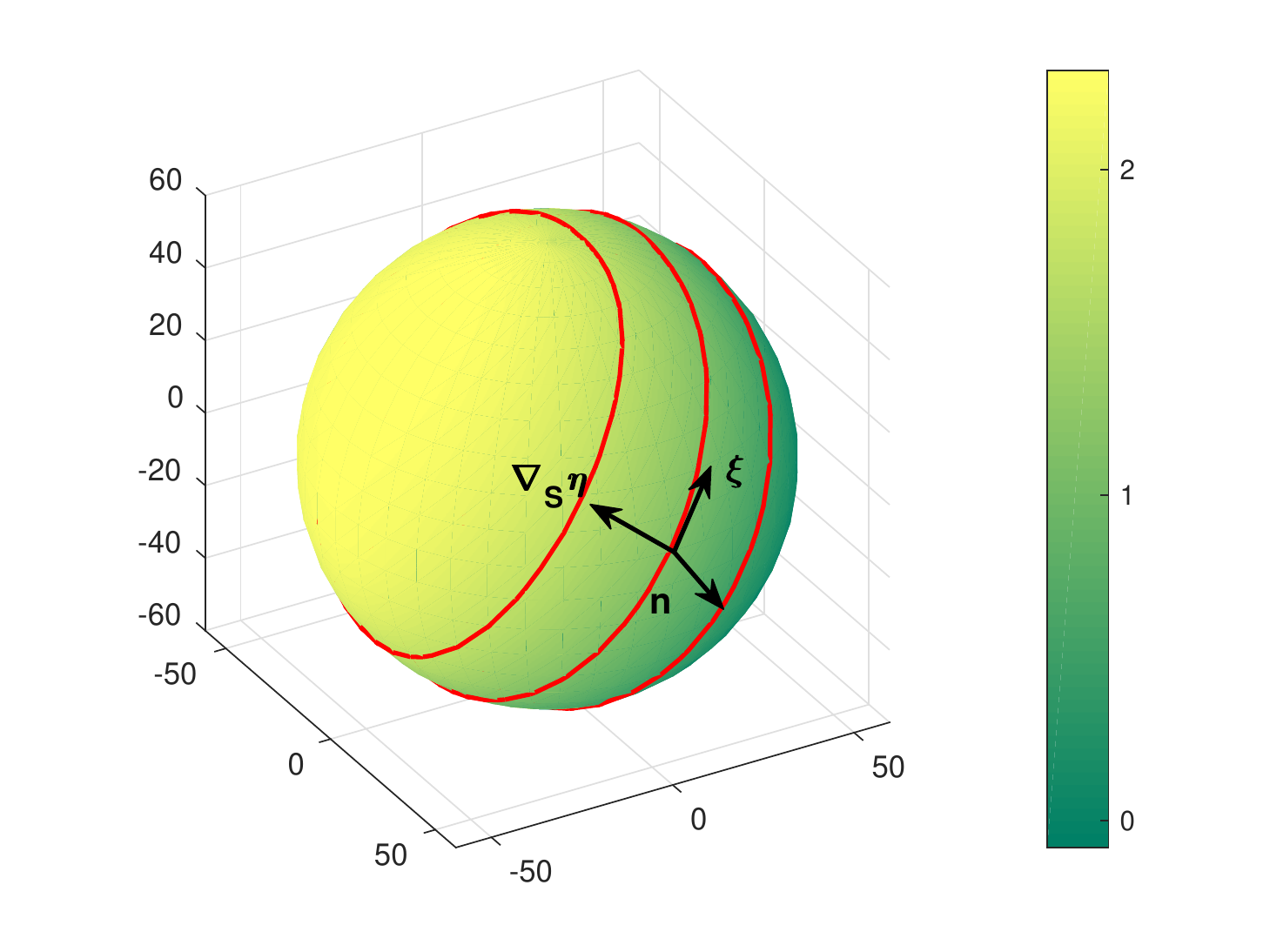}}
  \caption{An example of dislocation structure on a spherical grain boundary and dislocation density potential functions (DDPFs) defined on it. (a) Dislocation structure on the grain boundary. These dislocations have three Burgers vectors $\mathbf b^{(1)}$, $\mathbf b^{(2)}$ and $\mathbf b^{(3)}$ (shown by different colors) represented by contour lines of three DDPFs $\eta_1$, $\eta_2$ and $\eta_3$. (b) DDPF $\eta_1$ for the dislocations with  Burgers vector $\mathbf b^{(1)}$ (color-red dislocations in (a)).}
\label{fig:ddpf}
\end{figure}

We employ the dislocation density potential functions  proposed in Ref.~\cite{zhu2014continuum} to describe the orientation-dependent dislocation densities on the grain boundaries in three dimensions. Assume that the grain boundary  is a surface $S$ in three dimensional space. A
 dislocation density potential function (DDPF) $\eta$ is a scalar function defined on $S$ such that the constituent dislocations of the same Burgers vector $\mathbf b$ are given by the integer-valued contour lines of $\eta$: $\{\eta(x,y) = i, {\rm where}\ i  \ {\rm is \ an \ integer}\}$. The dislocation structure can be described in terms of surface gradient of $\eta$:
 \begin{equation}
 \nabla_s \eta=[\nabla-\mathbf n(\mathbf n \cdot\nabla)]\eta,
  \end{equation}
  where $\mathbf{n}$ is the unit normal vector of the grain boundary $S$. Given the surface gradient $\nabla_s \eta$,  microscopic quantities including the local dislocation line direction  $\pmb \xi$ and the inter-dislocation distance  $D$ can be calculated by
 \begin{equation}\label{eqn:direction-distance}
  \pmb\xi = \frac{\nabla_s\eta }{\|\nabla_s\eta\|}  \times \mathbf{n} \ \ {\rm and} \ \ D = \frac{1}{\|\nabla_s \eta\|},
  \end{equation}
  and the local dislocation density is
  \begin{equation}
  \rho=\|\nabla_s \eta\|.
  \end{equation}
  In fact, $\nabla_s\eta$ is a generalization of the $\mathbf N$-vector for a planar grain boundary as reviewed in the previous section, i.e. $\nabla_s\eta=\nabla\eta=\mathbf N$  for a planar grain boundary.   See Fig.~\ref{fig:ddpf} for an example of dislocation structure on a spherical grain boundary and DDPFs defined on it.

 Assume that on the grain boundary $S$, there are $J$ dislocation arrays represented by $\eta_j$, $j=1,2,\cdots,J$, corresponding to $J$ different Burgers vectors $\mathbf b^{(j)}$,  $j=1,2,\cdots,J$, respectively. Given the  rotation axis $\mathbf a$ and misorientation angle $\theta$, we solve the following constrained minimization problem for the dislocation structure and energy of the grain boundary:

\vspace{0.1in}
\noindent
\underline{\bf Constrained Minimization Problem}
\begin{eqnarray}
\text{minimize}\hspace{0.1in}
&& E = {\displaystyle \int_S \gamma_{\rm gb}  dS},
\ \ {\rm for \ all}\  \eta_j, \  j=1,2,\cdots,J,  \label{eqn:gb_energy}\\
\text{with}\hspace{0.1in} &&\gamma_{\rm gb}={\displaystyle \sum_{j=1}^J  \frac{\mu(b^{(j)})^2}{4\pi(1-\nu)}\!\left[1-\nu\frac{(\nabla_s \eta_j\! \times \!\mathbf{n} \!\cdot\! \mathbf{b}^{(j)})^2}{(b^{(j)})^2 {\|\nabla_s \eta_j\|}^2}\right]\|\nabla_s \eta_j\| \log\! \frac{D_j}{r_g }}, \label{eqn:gb_energy1}\\
\text{subject to}\hspace{0.1in}
&&\mathbf{h}=\theta(\mathbf{V}\times \mathbf{a}) - {\displaystyle \sum_{j=1}^J} \mathbf{b}^{(j)}(\nabla_s\eta_j\cdot\mathbf{V})=\mathbf 0,\label{eqn:frank}\\
&&{\rm for \ any \ vector}\ \mathbf V \ {\rm in \ the \ tangent \ plane \ of}\ S.\nonumber
\end{eqnarray}

In Eqs.~\eqref{eqn:gb_energy} and \eqref{eqn:gb_energy1}, $\gamma_{\rm gb}$  is the grain boundary energy density, where $\mu$ is the shear modulus, $\nu$ is the Poisson's ratio,  $b^{(j)}$ is the length of Burgers vector $\mathbf b^{(j)}$, $r_g$ is a parameter associated with the dislocation core width,  and $D_j$ is an approximation of the inter-dislocation distance of $\mathbf b^{(j)}$-dislocations.
In the continuum model for planar grain boundaries presented in Ref.~\cite{zhang2017energy},  the inter-dislocation distance $D_j$ in $\gamma_{\rm gb}$ in Eq.~\eqref{eqn:gb_energy1} is given by
   \begin{equation}\label{eqn:Dj00}
   D_j=\frac{1}{\|\nabla_s \eta_j\|}.
\end{equation}
This energy formula was derived
   in Refs.~\cite{xiang2012continuum,zhu2014continuum} from discrete dislocation dynamics model.

   Here we present an improved formula of $D_j$ to further incorporate  reactions of dislocations in a dislocation network. The new estimate of inter-dislocation distance is
\begin{flalign}\label{eqn:Nj}
D_j=\left\{
\begin{array}{ll}
 \displaystyle \min\left\{\frac{b^{(j)}}{\theta}, \ \frac{1}{\|\nabla_s \eta_j\|}\right\},&{\rm if \ (R)\ holds},\vspace{1ex}\\
{\displaystyle \frac{1}{\|\nabla_s \eta_j\|}},& {\rm otherwise}.
 \end{array}
 \right.
\end{flalign}
Here (R) collects the   conditions for dislocation reactions:

\underline{\bf (R)}: $\mathbf b^{(j)}=\mathbf b^{(k_{j1})}\pm\mathbf b^{(k_{j2})}$,  $\|\nabla_s \eta_{k_{j1}}\|$, $\|\nabla_s \eta_{k_{j2}}\|>\|\nabla_s \eta_j\|$, and $\pmb \xi^{(k_{j1})}\cdot \pmb \xi^{(k_{j2})}\neq0$,  for  some $k_{j1}$, $k_{j2}$.

\noindent
Here the dislocation line direction $\pmb \xi$ can be calculated by Eq.~\eqref{eqn:direction-distance}.
These conditions mean that a $\mathbf b^{(j)}$-dislocation can be generated by reaction of a $\mathbf b^{(k_{j1})}$-dislocation and a $\mathbf b^{(k_{j2})}$-dislocation when these two types of dislocations are present with smaller inter-dislocation distances ($\|\nabla_s \eta_{k_{j1}}\|$, $\|\nabla_s \eta_{k_{j2}}\|>\|\nabla_s \eta_j\|$) and they are not parallel to each other ($\pmb \xi^{(k_{j1})}\cdot \pmb \xi^{(k_{j2})}\neq0$).
 More details of the derivation and validation of this improved formulation by comparisons with atomistic simulation results will be given in Sec.~\ref{sec:reaction}.

   Note that we also propose a more accurate formula to calculate the inter-dislocation distance $D_j$ from dislocation densities taking into account dislocation reactions in a dislocation network; see Eq.~\eqref{eqn:Dnew}. This alternative formulation is more complicated, although it is in principle able to lead more accurate results when replacing Eq.~\eqref{eqn:Dj00} or \eqref{eqn:Nj}  for $D_j$ in Eq.~\eqref{eqn:gb_energy1}. We develop a method to use this alternative formula of $D_j$ in Eq.~\eqref{eqn:Dnew} for the identification of dislocation network structure from the obtained dislocation densities; see Sec.~\ref{sec:identification}.

The constraint in Eq.~\eqref{eqn:frank} is the classical Frank's formula in Eq.~\eqref{eqn:frank0} generalized to a curved grain boundary using DDPFs $\eta_j$'s. As we discussed above, $\nabla_s\eta$ in this equation is a generalization of the $\mathbf N$-vector for a planar grain boundary.

Recall that for a planar grain boundary, $\nabla_s\eta=\nabla\eta=\mathbf N$. With the approximation of straight dislocations, the grain boundary energy $\gamma_{\rm gb}$ and all the vectors $\mathbf N^{(j)}$'s are constant/constant vectors, and the continuum model in Eqs.~\eqref{eqn:gb_energy}--\eqref{eqn:Nj} for planar grain boundaries can also be written in the alternative form using $\mathbf N^{(j)}$'s as
\begin{eqnarray}
\text{minimize}\hspace{0.1in} &&\gamma_{\rm gb}={\displaystyle \sum_{j=1}^J  \frac{\mu(b^{(j)})^2}{4\pi(1-\nu)}\!\left[1-\nu\frac{(\mathbf N^{(j)}\! \times \!\mathbf{n} \!\cdot\! \mathbf{b}^{(j)})^2}{(b^{(j)})^2 {\|\mathbf N^{(j)}\|}^2}\right]\|\mathbf N^{(j)}\| \log\! \frac{D_j}{r_g}}, \label{eqn:N1} \\
&&{\rm for \ all}\ \mathbf N^{(j)}=(N_1^{(j)},N_2^{(j)}), \  j=1,2,\cdots,J, \nonumber\\
\text{subject to}\hspace{0.1in}
&&\mathbf{h}=\theta(\mathbf{V}\times \mathbf{a}) - {\displaystyle \sum_{j=1}^J} \mathbf{b}^{(j)}(\mathbf N^{(j)}\cdot\mathbf{V})=\mathbf 0,\label{eqn:N2}\\
&&{\rm for \ any \ vector}\ \mathbf V \ {\rm in \ the \ grain \ boundary}.\nonumber
%&&\frac{\partial \eta_{ju}}{\partial v}-\frac{\partial \eta_{jv}}{\partial u}=0.\label{eqn:cc1}
\end{eqnarray}
 Here $D_j=1/\| \mathbf N^{(j)}\|$ following Eq.~\eqref{eqn:Dj00} which was adopted in the study of planar low angle grain boundaries in Ref.~\cite{zhang2017energy}, or when the improved formulation in  Eq.~\eqref{eqn:Nj} is used: $D_j=b^{(j)}/\theta$ if the reaction conditions in (R) hold and $1/\| \mathbf N^{(j)}\|$ otherwise.

 For general, curved grain boundaries, as can be seen from the above formulation, the major advantage of the representation using DDPFs $\{\eta_j\}$ is that simple, scalar functions are employed to describe distributions of orientation dependent dislocation lines on these grain boundaries. This representation also ensures the connectivity of dislocations. It is a generalization of the $\mathbf N$-vector representation of structures of straight dislocations on a planar grain boundary in the classical Frank's formula \cite{Frank1950,Bilby1955,HirthLothe1982,Sutton1995}.

In summary, in our proposed formulation, the equilibrium dislocation structure is obtained by minimizing the grain boundary energy associated with the constituent dislocations given in Eqs.~\eqref{eqn:gb_energy} and \eqref{eqn:gb_energy1}, subject to the constraint of Frank's formula in Eq.~\eqref{eqn:frank}. Note that this formulation is based on dislocation densities, which are described by DDPFs $\{\eta_j\}$. Identification of the equilibrium dislocation network structure from  dislocation densities will be discussed in Sec.~\ref{sec:identification}.

\section{Incorporation of dislocation reaction in the continuum model}\label{sec:reaction}

 In this section, we present details of the derivation of the improved energy formulation in Eqs.~\eqref{eqn:gb_energy1} and Eq.~\eqref{eqn:Nj} that incorporates dislocation reaction.

The grain boundary energy density formula used in the continuum model in Ref.~\cite{zhang2017energy} that was derived in  Refs.~\cite{xiang2012continuum,zhu2014continuum}, is given by Eq.~\eqref{eqn:gb_energy1} and Eq.~\eqref{eqn:Dj00}, i.e.,
\begin{equation}\label{eqn:energy00}
\gamma_{\rm gb}= \sum_{j=1}^J  \frac{\mu(b^{(j)})^2}{4\pi(1-\nu)}\!\left[1-\nu\frac{(\nabla_s \eta_j\! \times \!\mathbf{n} \!\cdot\! \mathbf{b}^{(j)})^2}{(b^{(j)})^2 {\|\nabla_s \eta_j\|}^2}\right]\|\nabla_s \eta_j\| \log\! \frac{1}{r_g \|\nabla_s \eta_j\|}.
\end{equation}
This is the energy density of all the constituent dislocations of the grain boundary. Recall that the energy per unit length of a single dislocation in an array of parallel dislocations with inter-dislocation distance $D$  is  $\frac{\mu b^2}{4\pi(1-\nu)}(1-\nu\cos^2\lambda)\log\frac{D}{r_g}$, where $\lambda$ is the angle between the dislocation and its Burgers vector \cite{ReadShockley1950,HirthLothe1982}. This suggests that we should have inter-dislocation distance inside the logarithm in the continuum energy formula. Since the inter-dislocation distance on the grain boundary is $D_j=1/\|\nabla_s \eta_j\|$ for parallel connected straight dislocations (e.g. those in Fig.~\ref{fig:01}(a)), the energy formula in  Eq.~\eqref{eqn:energy00} recovers the above classical formula.
The $\|\nabla_s \eta_j\|$ factor before the logarithm in the continuum formula in Eq.~\eqref{eqn:energy00} accounts for the dislocation density (per unit area) on the grain boundary.

Note that the purpose of the continuum model is to provide accurate formulations for dislocation density and grain boundary energy that are suitable for simulations on the continuum level. As shown by the extensive comparisons with atomistic simulations, the continuum formula in Eq.~\eqref{eqn:energy00} works well for most of the planar grain boundaries. However, for some planar grain boundaries, there are noticeable differences between the dislocation densities calculated using Eq.~\eqref{eqn:energy00} and atomistic simulations~\cite{zhang2017energy} (See also Fig.~\ref{fig:compare1}(a) below). We find that these differences are caused by the fact that  reactions of intersecting dislocations are not accurately included in the continuum formula in Eq.~\eqref{eqn:energy00} as explained below. %although it is partially incorporated in the average sense in the energy minimization process.

\begin{figure}[htbp]
\centering
\subfigure[]{\includegraphics[width=2.in]{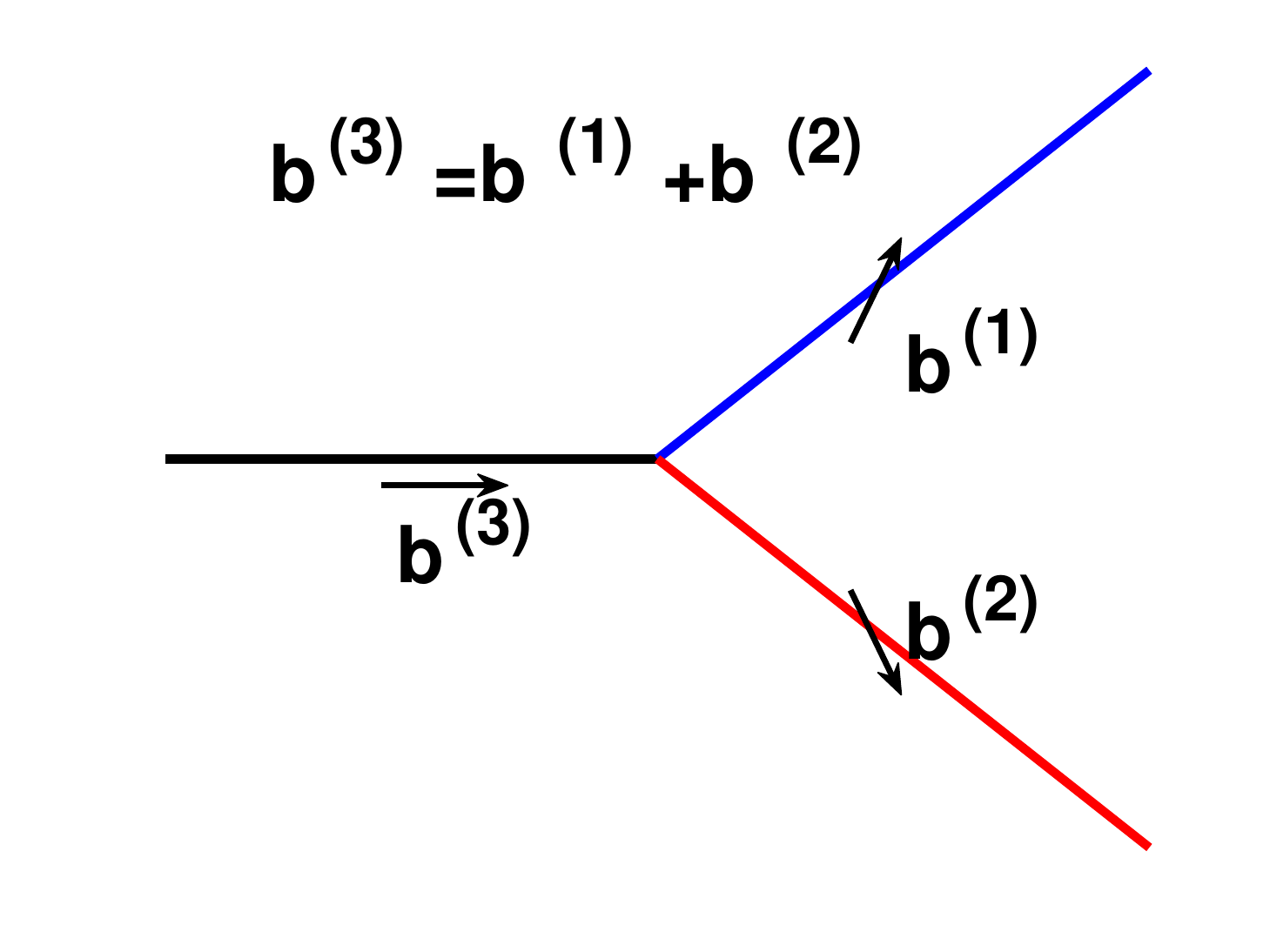}}
\subfigure[]{\includegraphics[width=2.in]{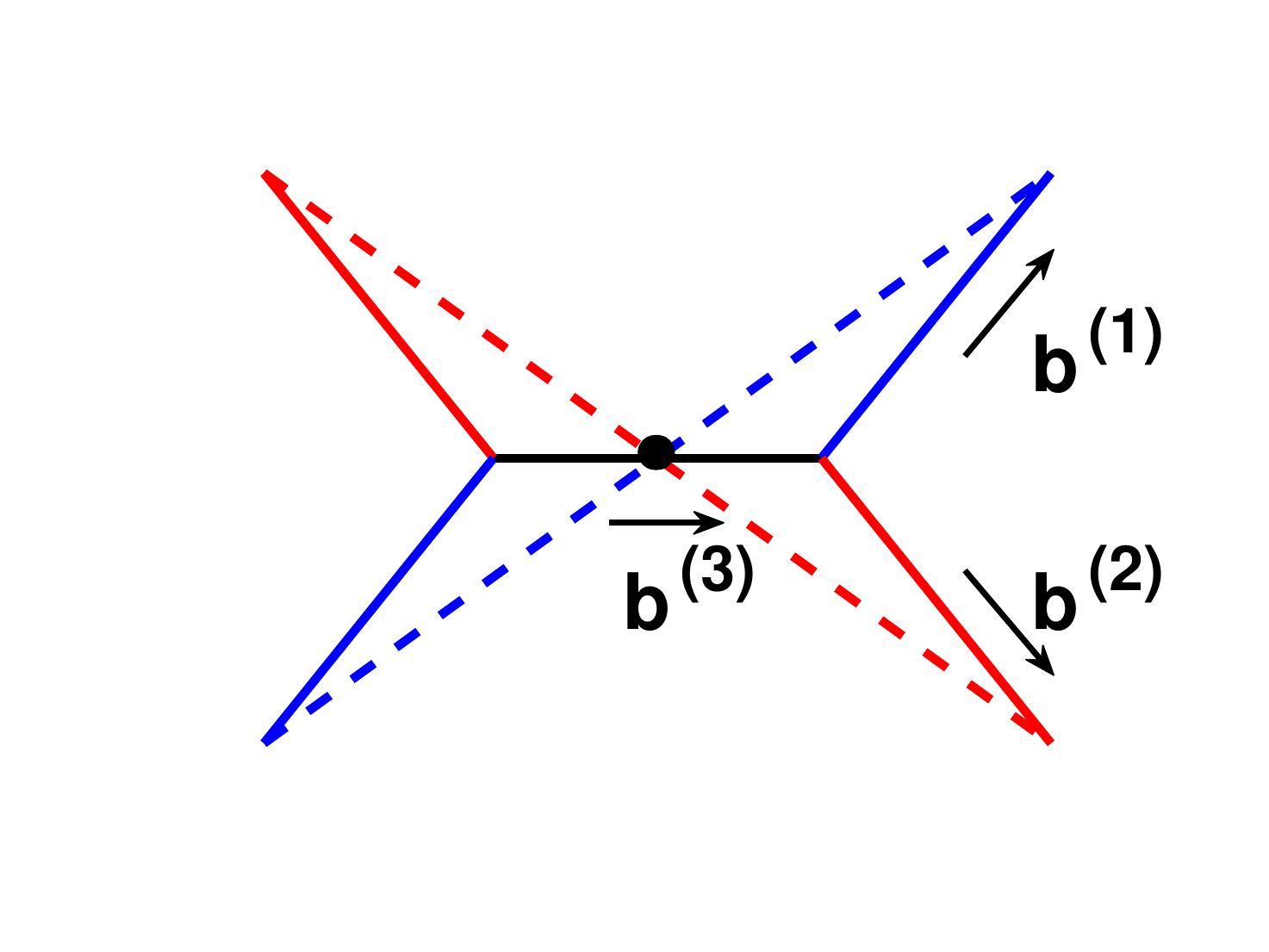}}
  \caption{(a) Dislocation reaction: Two intersecting dislocations with Burgers vectors  $\mathbf b^{(1)}$ and $\mathbf b^{(2)}$ may react to form a dislocation with Burgers vector $\mathbf b^{(3)}=\mathbf b^{(1)}+\mathbf b^{(2)}$. (b) Dislocation reaction in a network:  Two intersecting dislocations with Burgers vectors  $\mathbf b^{(1)}$ and $\mathbf b^{(2)}$ (dashed lines) may react to form a new dislocation segment with Burgers vector $\mathbf b^{(3)}=\mathbf b^{(1)}+\mathbf b^{(2)}$. Solid lines show the dislocation structure after reaction. }
\label{fig:reaction}
\end{figure}

Recall that in the dislocation reaction \cite{HirthLothe1982}, two intersecting dislocations with Burgers vectors  $\mathbf b^{(1)}$ and $\mathbf b^{(2)}$ may react to form a dislocation with Burgers vector $\mathbf b^{(3)}=\mathbf b^{(1)}+\mathbf b^{(2)}$ if the latter has lower energy, see Fig.~\ref{fig:reaction}(a). This reaction can also be written as $\mathbf b^{(3)}=\mathbf b^{(1)}-\mathbf b^{(2)}$ if the $\mathbf b^{(2)}$-dislocation changes into its opposite direction. In a dislocation network, the length of the dislocation segment formed by  reaction is also limited by the increase of the dislocation line energy, leading to a new stable dislocation network, see Fig.~\ref{fig:reaction}(b) and the hexagonal dislocation network on a twist boundary shown in Fig.~\ref{fig:01}(d). In fact, dislocation reaction is  included in an approximate way by the energy minimization process in the continuum formulation in Eq.~\eqref{eqn:gb_energy}, and when the total energy is given by Eq.~\eqref{eqn:energy00}, reactions of parallel straight dislocations are perfectly accounted for.
In the continuum energy in Eq.~\eqref{eqn:energy00}, the inter-dislocation distance in the logarithm is always  $1/\|\nabla_s \eta\|$, which is the inverse of the density of the dislocations. When the density of dislocation segments formed by reaction of intersecting dislocation arrays is small, the inter-dislocation distance calculated using  $1/\|\nabla_s \eta\|$, i.e., inverse of the dislocation density, is large. As a result, the variational force with respect to $\nabla_s \eta$ (to be used in the numerical algorithm presented in Sec.~\ref{sec:algorithm}), which is proportional to $\log \|\nabla_s \eta\|$, diverges as $\|\nabla_s \eta\|$ goes to $0$. Whereas in the actual reaction, the inter-dislocation distance of the segments formed by reaction (which are not necessarily connected) is no longer $1/\|\nabla_s \eta\|$: the inter-dislocation distance in this case is determined by the two arrays of dislocations before the reaction (see, e.g. Fig.~\ref{fig:01}(d)), and is not necessarily large even when the density of dislocation segments formed by reaction is small.  That is, the divergence in variational force as $\|\nabla_s \eta\|$ goes to $0$ is not physical for a hexagonal network formed by dislocation reaction.  In this case, the energy in Eq.~\eqref{eqn:energy00} is not that accurate and may not be able to identify the small amount of segments formed by reaction; see the examples in Fig.~\ref{fig:compare1}(a), (c) and (d) (i.e., densities of the $\mathbf b^{(2)}$ dislocations when $\phi\geq 45^\circ$ in (a) and the dislocation structure of $\phi=59^\circ$ in (c) and (d)).

Here we propose  an improved  grain boundary energy formula as given in Eqs.~\eqref{eqn:gb_energy1} and \eqref{eqn:Nj}  to account for such dislocation reactions in the dislocation network on a low angle grain boundary.  The unphysical divergence in the variational force for small $\|\nabla_s \eta\|$ is fixed by a physically meaningful cutoff value given as follows. Recall that the logarithm factor $\log D_j$ in the grain boundary energy in Eq.~\eqref{eqn:gb_energy1} comes from the logarithm factor in dislocation energy, which is known not sensitive to the quantity $D_j$ inside the logarithm \cite{HirthLothe1982}. Especially, for dislocation reaction in discrete dislocation model, as shown in Fig.~\ref{fig:reaction}, the logarithm factors for the energies of different dislocations are commonly approximated by a constant when determining whether the reaction is energetically favorable or not (Frank’s rule, \cite{HirthLothe1982}).  For this reason, it is expected that in the small $\|\nabla_s \eta\|$ limit of Eq.~\eqref{eqn:Dj00} where the variational force diverges unphysically for a hexagonal network, a constant cutoff by the length scale of the inter-dislocation distances in a hexagonal network will provide a good approximation for the quantity $D_j$ inside the logarithm in the energy formulation.

 Now we look for an approximation of the length scale of the inter-dislocation distances in a hexagonal network. Recall that  $\log\theta$ term is included in the classical grain boundary energy formula $E=E_0\theta(A-\ln\theta)$ \cite{ReadShockley1950,HirthLothe1982}, which was derived rigorously for pure tilt boundaries and was generalized to all grain boundaries with fitting parameters. This can be considered as an approximation of $D_j/r_g\approx 1/\theta$ in the more accurate grain boundary energy formula in Eq.~\eqref{eqn:gb_energy1}, where $r_g$ is a dislocation core parameter of the order of the Burgers vectors. In particular, $D=b/\theta$ for the tilt boundary shown in Fig.~\ref{fig:01}(a).
When dislocation reaction happens, the length scale of inter-dislocation distance of the newly formed disconnected dislocation segments are set by the inter-dislocation distances of the two reacted dislocation arrays; see Fig.~\ref{fig:01}(b) and (d) for an example (in which $\mathbf b^{(2)}$-dislocation segments are generated by reaction of $\mathbf b^{(1)}$- and $\mathbf b^{(3)}$-dislocations with the reaction $\mathbf b^{(2)}=\mathbf b^{(1)}-\mathbf b^{(3)}$),  where the inter-dislocation distances of the reacted $\mathbf b^{(1)}$- and $\mathbf b^{(3)}$-dislocations are both $\sqrt{3}b/(2\theta)$. These suggest that $D_j =O(b^{(j)}/\theta)$, and we choose $b^{(j)}/\theta$ as the length scale to approximate $D_j$ inside the logarithm factor in the small $\|\nabla_s \eta\|$ limit when dislocation reaction happens.
This leads to the formulation in Eq.~\eqref{eqn:Nj} for the inter-dislocation distances.

\begin{figure}[htbp]
\centering
\subfigure[]{\includegraphics[width=2.in]{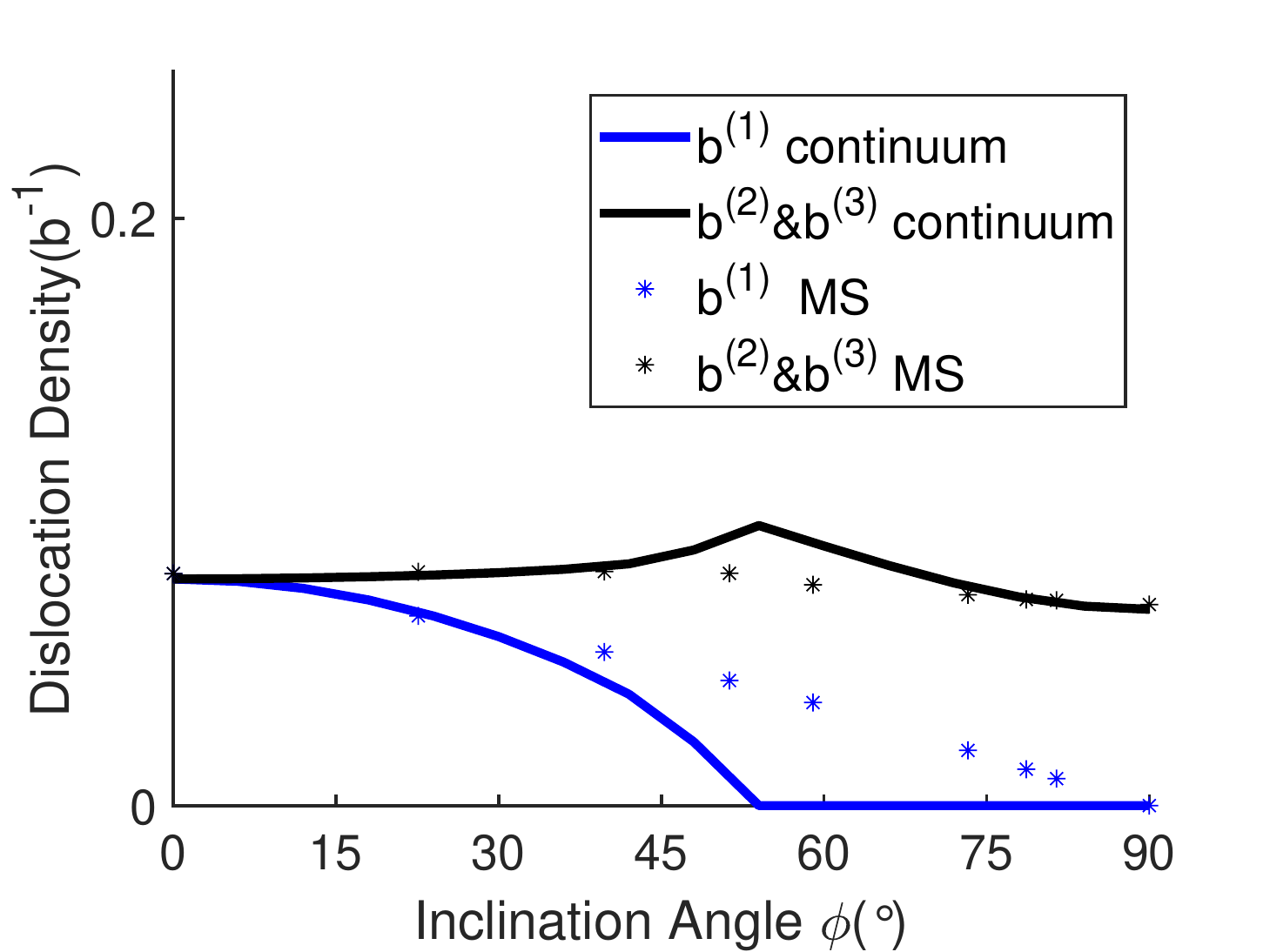}}
\subfigure[]{\includegraphics[width=2.in]{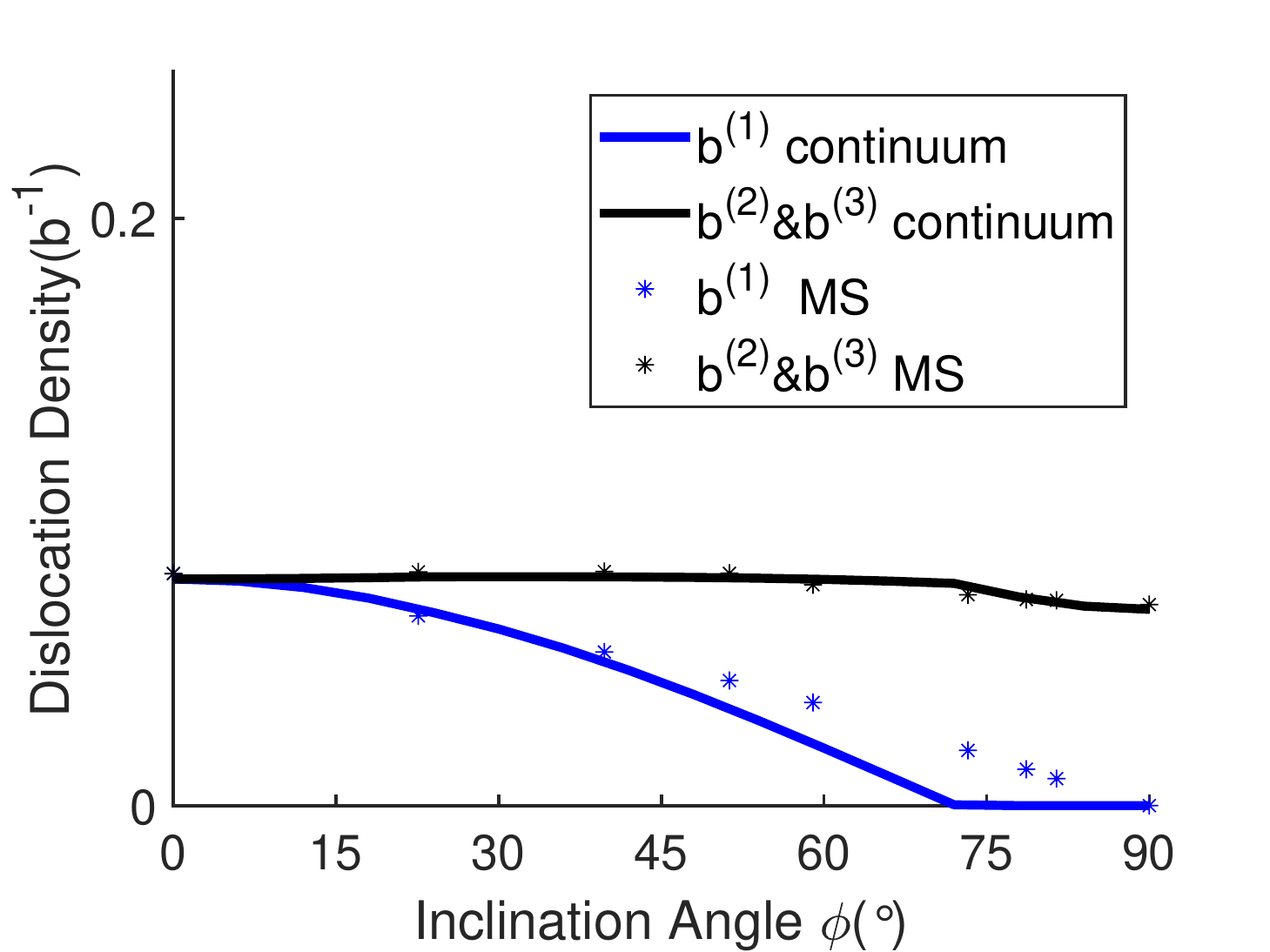}}
%\subfigure{\includegraphics[width=\linewidth]{xtoz_MD_c1.pdf}}
\subfigure[]{\includegraphics[width=\linewidth]{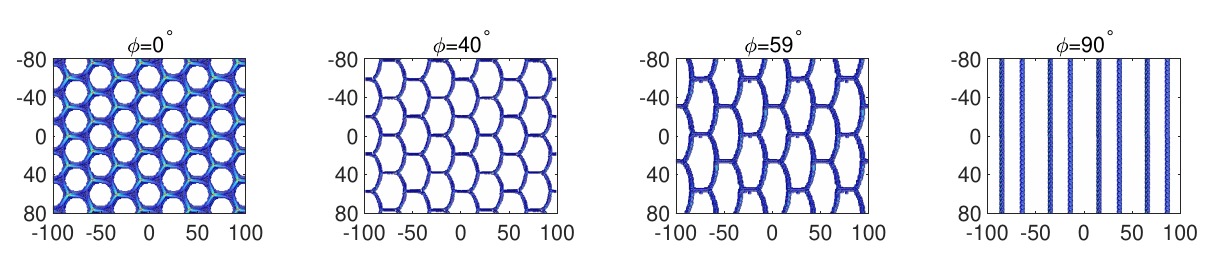}}
\subfigure[]{\includegraphics[width=\linewidth]{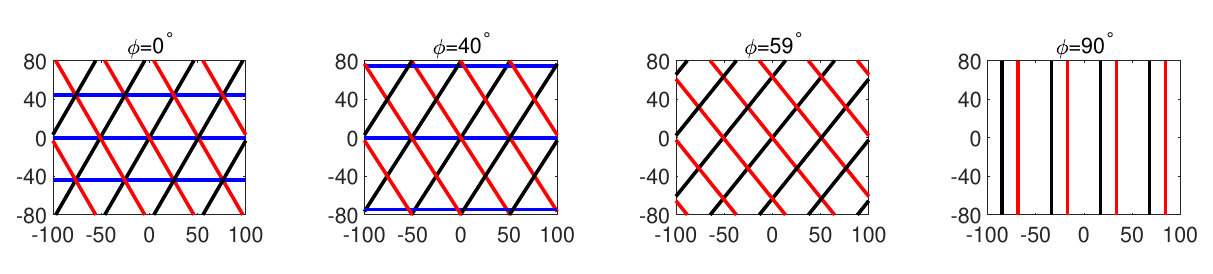}}
\subfigure[]{\includegraphics[width=\linewidth]{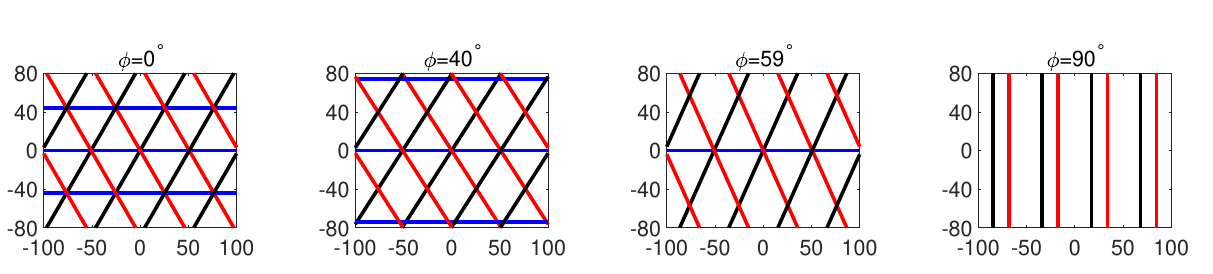}}
\subfigure[]{\includegraphics[width=\linewidth]{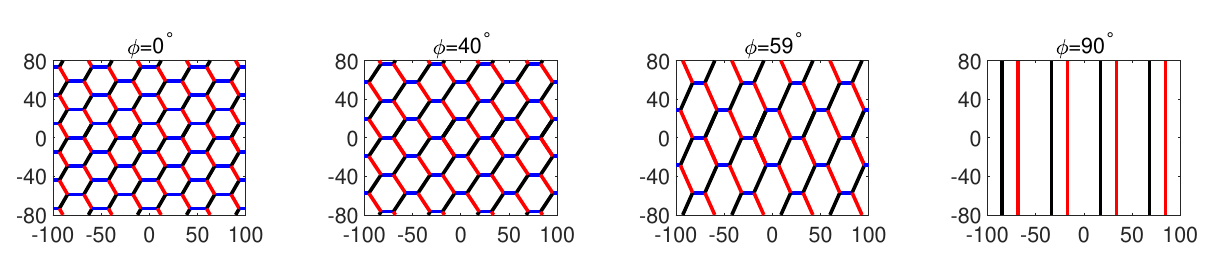}}
\caption{(a) For some planar grain boundaries, there are noticeable differences between the dislocation densities calculated using  Eq.~\eqref{eqn:energy00} (the continuum model in Ref.~\cite{zhang2017energy}) and molecular static (MS) simulations. (b)  Dislocation densities calculated using the new continuum model incorporating dislocation reaction in Eqs.~\eqref{eqn:gb_energy}--\eqref{eqn:frank} and \eqref{eqn:Nj} show significant improvement. These grain boundaries have rotation axis $[111]$ and are parallel to the $[\bar{1}10]$ direction, in fcc Al. The misorientation angle is $\theta= 1.95^\circ$. The horizontal axis is the inclination angle $\phi$, which is the angle between grain boundary normal direction and the rotation axis $[111]$. (c)--(f) Dislocation networks for different inclination angle $\phi$  obtained by (c) MS simulations, (d) the continuum model in Ref.~\cite{zhang2017energy}, (e) the new continuum model, and (f) the new continuum model using the proposed identification method. }
    \label{fig:compare1}
\end{figure}

We examine this new formulation of grain boundary energy in Eqs.~\eqref{eqn:gb_energy1} and \eqref{eqn:Nj} by comparisons with results of molecular static (MS) simulation and the continuum model used previously in Ref.~\cite{zhang2017energy}, for a series of low angle grain boundaries in fcc Al. The results are shown in Fig.~\ref{fig:compare1}. For parameters in the continuum models, see the beginning of Sec.~\ref{sec:NumericalResults}. These grain boundaries have rotation axis $[111]$ and are parallel to the $[\bar{1}10]$ direction. The misorientation angle is $\theta= 1.95^\circ$. The inclination angle $\phi$, which is the angle between grain boundary normal direction and the rotation axis $[111]$, varies from $0^\circ$ to $90^\circ$. Dislocations with Burgers vectors $\mathbf b^{(1)}$, $\mathbf b^{(2)}$, and $\mathbf b^{(3)}$ (see Fig.~\ref{thompson}) appear in the dislocation network on the grain boundary.

Comparisons show significant improvement of the new continuum formulation (Fig.~\ref{fig:compare1}(b)) over the old one (Fig.~\ref{fig:compare1}(a)). Dislocations of Burgers vector $\mathbf b^{(1)}$ with small density (e.g., the small horizontal dislocation segments in the MS results in Fig.~\ref{fig:compare1}(c)) can be  captured by the new continuum formulation (shown in Fig.~\ref{fig:compare1}(e)) for a much larger range of inclination angles compared with the results using the old model (shown in Fig.~\ref{fig:compare1}(d)). Note that the small difference for $\phi>72^\circ$, where the $\mathbf b^{(1)}$-dislocations vanish using the new continuum model whereas the density of these dislocations decreases gradually  in the MS simulations, is not due to the continuum approximation of the Frank's formula, because
such critical inclination angle also exits (about $76^\circ$) when using the Frank's formula in the discrete dislocation model (using the method in Sec.~19-4 of \cite{HirthLothe1982} and Appendix B of \cite{Wang-Xiang}). More accurate continuum results may need incorporation of more atomistic mechanisms beyond dislocation reactions, e.g. dislocation jogs \cite{zhang2017energy}, however, such improvement will lead to more complexity in the continuum model.

An identification method to draw the exact dislocation network based on the dislocation densities obtained by using our continuum model will be presented in the next section. Fig.~\ref{fig:compare1}(f) shows that the resulting dislocation networks are in excellent agreement with the MS results in Fig.~\ref{fig:compare1}(c); see next section for details.

\section{Identification of dislocation structure from dislocation densities}\label{sec:identification}

Note that our continuum model gives approximations of the densities and orientations of the constituent dislocations of low angle grain boundaries.  Recall that the constituent dislocations are the integer-valued contour lines of the DDPFs $\eta_j$'s by their definitions in Sec.~\ref{s2}.
When all the constituent dislocations are continuous straight lines, our continuum model is able to give the grain boundary dislocation structure accurately (see the examples in Ref.~\cite{zhang2017energy}). Whereas if we interpreted a dislocation network with disconnected dislocation segments also by straight lines, the obtained structure will no longer be the exact dislocation structure  (i.e., the dislocation structure from the atomistic model). See comparisons in Fig.~\ref{fig:compare1} (c) and (d) (and that with improved energy density in (e)).

Now we present a method that recovers the exact dislocation structure  as given in the atomistic model based on the densities and orientations of the constituent dislocations obtained in our continuum model. As explained in Sec.~\ref{sec:reaction} and illustrated in Fig.~\ref{fig:reaction},  in the microscopic dislocation structure, due to dislocation reactions, the dislocations may not be continuous straight lines, instead, they form hexagons (not necessarily regular) with disconnected dislocation segments~\cite{HirthLothe1982}, see also examples in Fig.~\ref{fig:compare1}(c).
The identification method is based on calculation of the exact length and orientation of each dislocation segment in the network based on the dislocation densities and orientations in the continuum model.

\begin{figure}[htbp]
	\centering
	\includegraphics[width=3.3in]{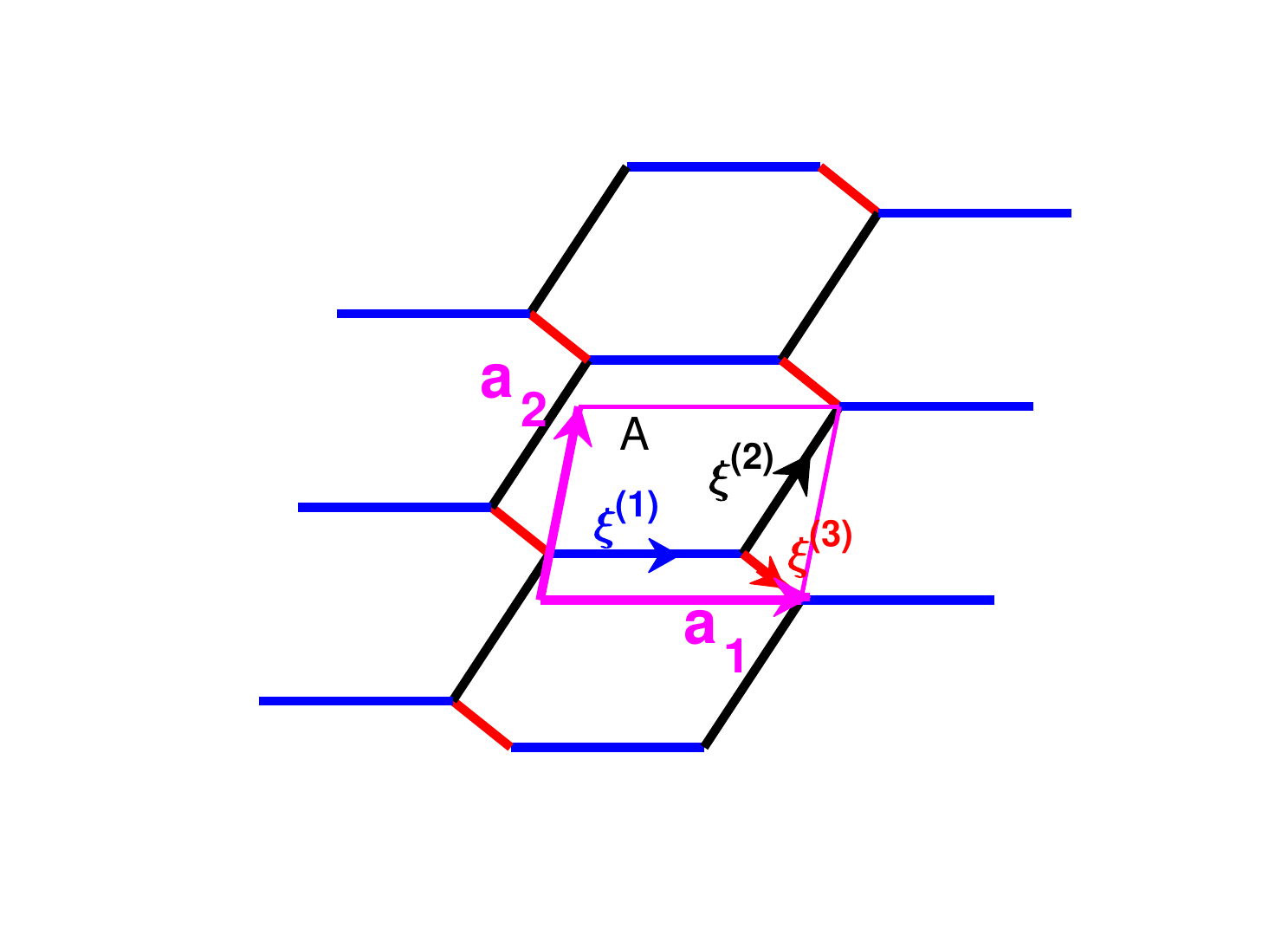}
	\caption{A hexagonal dislocation structure formed by dislocations with Burgers vectors $\mathbf b^{(1)}$, $\mathbf b^{(2)}$ and $\mathbf b^{(3)}$, whose line directions are  $\pmb\xi^{(1)}$, $\pmb\xi^{(2)}$, and $\pmb\xi^{(3)}$, respectively. Vectors $\mathbf a_1$ and $\mathbf a_2$ are the two sides of the periodic parallelogram unit cell.  The area of a unit cell is $A=\| \mathbf a_1\times\mathbf a_2\|$.  }
	\label{fig:hex}
\end{figure}

Consider a hexagonal network formed by dislocation reaction $\mathbf b^{(1)}=\mathbf b^{(2)}+\mathbf b^{(3)}$; see Fig. \ref{fig:hex}.
 As we discussed before, the direction of dislocation generated by $\mathbf b^{(j)}$ is $ \pmb\xi^{(j)} = \frac{\nabla_s\eta_j }{\|\nabla_s\eta_j\|}  \times \mathbf{n}  $, and the dislocation density is $\|\nabla_s\eta_j\| $. This means that the length of a $\mathbf b^{(j)}$-dislocation segment in a periodic parallelogram cell with area $A$ is $\|\nabla_s\eta_j\|A$.
% We can rewrite vectors of the two sides of the periodic parallelogram cell $\mathbf a_1$ and $\mathbf a_2 $ in terms of  $\nabla_s\eta_j$, $j=1,2,3$, and $A$:

From Fig. \ref{fig:hex}, the  $\mathbf b^{(1)}$-,   $\mathbf b^{(2)}$, and  $\mathbf b^{(3)}$-dislocation segments in the parallelogram, written in the vector form, are
\begin{eqnarray}
\mathbf l_1=A\nabla_s\eta_1\times \mathbf n, \ \
\mathbf l_2=A\nabla_s\eta_2\times \mathbf n, \ \
\mathbf l_3=A\nabla_s\eta_3\times \mathbf n,
\end{eqnarray}
respectively, and the area of the  periodic parallelogram cell
\begin{equation}
A= \|\mathbf l_1\times \mathbf l_2\| +\|\mathbf l_2\times \mathbf l_3\|+\|\mathbf l_3\times \mathbf l_1\|.
\end{equation}
Since
\begin{equation}\label{eqn:times}
\mathbf l_i\times \mathbf l_j=A^2(\nabla_s\eta_i\times \mathbf n)\times(\nabla_s\eta_j\times \mathbf n)= A^2( \mathbf n\cdot (\nabla_s\eta_i\times\nabla_s\eta_j ))\mathbf n,
\end{equation}
 it can be solved that
\begin{equation}
A=\frac{1}{\| \nabla_s\eta_1\times \nabla_s\eta_2 \| +\|\nabla_s\eta_2\times \nabla_s\eta_3 \| +\|\nabla_s\eta_3\times \nabla_s\eta_1\| }.
\end{equation}
Thus we have the following identification method.

\vspace{0.1in}
\noindent
\underline{\bf Method to identify dislocation structure from dislocation densities}:

 \vspace{0.05in}
The length of each dislocation segment is
\begin{equation}\label{eqn:segmentlength}
l_j=\frac{\|\nabla_s\eta_j\|}{\| \nabla_s\eta_1\times \nabla_s\eta_2 \| +\|\nabla_s\eta_2\times \nabla_s\eta_3 \| +\|\nabla_s\eta_3\times \nabla_s\eta_1\| }, \ \ j=1,2,3,
\end{equation}
and its direction is
\begin{equation}\label{eqn:segmentdirection}
\pmb\xi^{(j)} = \frac{\nabla_s\eta_j }{\|\nabla_s\eta_j\|}  \times \mathbf{n}.
\end{equation}

Using the formula of length of each dislocation segment given in Eq.~\eqref{eqn:segmentlength} and
its direction in Eq.~\eqref{eqn:segmentdirection},  we can draw the exact  periodic dislocation network structure based on the dislocation densities and orientations in the continuum model represented by DDPFs $\{\eta_j\}$.  Fig.~\ref{fig:reaction}(b) gives an illustration of this improved identification method: Whenever dislocation reaction ($\mathbf b^{(3)}=\mathbf b^{(1)}+\mathbf b^{(2)}$) by a pair of crossing dislocations (dashed segments) is identified, the crossing segments are replaced by the reacted dislocation segments (solid segments) with lengths and directions given by Eqs.~\eqref{eqn:segmentlength} and \eqref{eqn:segmentdirection}.
Using this identification method, we plot dislocation structures for  the results obtained by the new continuum energy formulation shown in Fig.~\ref{fig:compare1}(e), and the generated dislocation networks are shown in Fig.~\ref{fig:compare1}(f). It can be seen that the generated dislocation networks are in excellent agreement with the MS simulation results in Fig.~\ref{fig:compare1}(c).

It is interesting to also have the formulas for the exact distances between parallel dislocation segments based on the dislocation densities and orientations in the continuum model. From Fig. \ref{fig:hex}, we have the distance between $\mathbf b^{(1)}$-dislocation segments can be written as
$D_1=(\|\mathbf l_1\times\mathbf l_2\|+\|\mathbf l_3\times\mathbf l_1 \|)/l_1$, and similar for $D_2$ and $D_3$.
Thus using Eqs.~\eqref{eqn:times} and \eqref{eqn:segmentlength}, we have
\begin{equation}\label{eqn:Dnew}
\begin{aligned}
D_1&= \frac{\|\nabla_s\eta_1\times\nabla_s\eta_2 \| +\|\nabla_s\eta_3\times\nabla_s\eta_1 \| }{\|\nabla_s\eta_1\|\left(\| \nabla_s\eta_1\times \nabla_s\eta_2 \| +\|\nabla_s\eta_2\times \nabla_s\eta_3 \| +\|\nabla_s\eta_3\times \nabla_s\eta_1\|\right)},\\
D_2&=\frac{ \|\nabla_s\eta_1\times\nabla_s\eta_2 \| +\|\nabla_s\eta_2\times\nabla_s\eta_3 \| }{\|\nabla_s\eta_2\|\left(\| \nabla_s\eta_1\times \nabla_s\eta_2 \| +\|\nabla_s\eta_2\times \nabla_s\eta_3 \| +\|\nabla_s\eta_3\times \nabla_s\eta_1\|\right)}, \\
D_3&=\frac{ \|\nabla_s\eta_3\times\nabla_s\eta_1 \| +\|\nabla_s\eta_2\times\nabla_s\eta_3 \| }{\|\nabla_s\eta_3\|\left(\| \nabla_s\eta_1\times \nabla_s\eta_2 \| +\|\nabla_s\eta_2\times \nabla_s\eta_3 \| +\|\nabla_s\eta_3\times \nabla_s\eta_1\|\right)}.
\end{aligned}
\end{equation}
Note that this formula provides a more accurate inter-dislocation distance $D_j$ in the grain boundary energy
$\gamma_{\rm gb}$  in Eq.~\eqref{eqn:gb_energy1}, as an alternative to the formulas of $D_j$ in Eqs.~\eqref{eqn:Dj00} and  \eqref{eqn:Nj}.

For a planar grain boundary,  if the formulation  in Eqs.~\eqref{eqn:N1} and \eqref{eqn:N2} based on the alternative notation $\{\mathbf N^{(j)}\}$  is used, the identification method can be written as
\begin{eqnarray}
&l_j=&\frac{\|\mathbf N^{(j)}\|}{\| \mathbf N^{(1)}\times \mathbf N^{(2)} \| +\|\mathbf N^{(2)} \times \mathbf N^{(3)}\| +\|\mathbf N^{(3)}\times \mathbf N^{(1)}\| }, \ \ j=1,2,3,\\
&\pmb\xi^{(j)} = &\frac{\mathbf N^{(j)} }{\| \mathbf N^{(j)}\|}  \times \mathbf{n},
\end{eqnarray}
and similar for $D_j$ in Eq.~\eqref{eqn:Dnew} with $\nabla_s\eta_j$ replaced by $\mathbf N^{(j)}$.

\section{Numerical Algorithm}
\label{sec:algorithm}

 \subsection{Gradient flow formulation for energy minimum state}

 The continuum minimization problem in Eqs.~\eqref{eqn:gb_energy}, \eqref{eqn:gb_energy1}, \eqref{eqn:frank}, and \eqref{eqn:Dj00} or \eqref{eqn:Nj} is solved by gradient flow of the total energy to equilibrium state.
The major challenge in this formulation is that the energy density $\gamma_{\rm gb}$ is not convex as a function of $\|\nabla_s \eta_j\|$.
This nonconvexity will lead to an ill-posed formulation when using gradient flow of the total energy to find the solution of the minimization problem. In fact, neglecting constants and adjustments due to dislocation direction and reaction, the contribution of $\mathbf b^{(j)}$-dislocations in the energy density $\gamma_{\rm gb}$ is essentially
 $-\|\nabla_s \eta_j\|\log\|\nabla_s \eta_j\|$,
 which is a concave function of $\|\nabla_s \eta_j\|$  and the gradient flow will give a backward-diffusion-like ill-posed equation of $\eta_j$. For example, when $\eta$ depends only on the single variable $x$, the total energy with energy density $-|\eta_x| \log|\eta_x|$ leads to the gradient flow $\eta_t=-\frac{1}{|\eta_x|}\eta_{xx}$, which is ill-posed.

Note that since  in general it is not easy to guess  the solution of this energy minimization problem except for some well-known cases, a gradient flow based solution method is preferred than  Newton's methods because the latter  rely on good initial guesses.

 In order to obtain a gradient flow formulation that avoids this ill-posedness,
we use the components of $\nabla_s \eta_j$ as independent variables, instead of $\eta_j$ itself. By doing so, gradient flow of the total energy will lead to ODE systems instead of backward-diffusion-like equations. For the simplified example considered above, when $\eta$ depends only on the single variable $x$ and the energy density is $-|\eta_x| \log|\eta_x|$, if we use $\zeta=\eta_x$ as the unknown function, i.e., the energy density is $-|\zeta| \log|\zeta|$, the gradient flow of the total energy with respect to $\zeta$ is $\zeta_t=(\log|\zeta|+1)\frac{\zeta}{|\zeta|}$, which is an ODE and has no such ill-posedness.  However, the components of $\nabla_s \eta_j$ are not independent. Thus, we include these relationships of the components of these surface gradients as further constraints in the energy minimization problem. The detailed formulation is given as follows.

 Consider a grain boundary surface $S$ in three dimensions with  parameterization $(u,v)$, i.e. a point on the grain boundary can be written as  $\mathbf r(u,v)=(x(u,v),y(u,v),z(u,v))$.
 The surface gradient of a DDPF $\eta_j$ defined on the grain boundary can be expressed in terms of $\eta_{ju}$ and $\eta_{jv}$ (which are partial derivatives of $\eta_j$ with respect to $u$ and $v$) as
 \begin{equation}\label{suface-gradient-uv}
 \nabla_s \eta_j = \left(\frac{\|\mathbf r_v\|^2}{\|\mathbf r_u\times\mathbf r_v\|^2}\eta_{ju}
 -\frac{\mathbf r_u\cdot\mathbf r_v}{\|\mathbf r_u\times\mathbf r_v\|^2}\eta_{jv}
\right)\mathbf r_u+\left(\frac{\|\mathbf r_u\|^2}{\|\mathbf r_u\times\mathbf r_v\|^2}\eta_{jv}
 -\frac{\mathbf r_u\cdot\mathbf r_v}{\|\mathbf r_u\times\mathbf r_v\|^2}\eta_{ju}
\right)\mathbf r_v,
 \end{equation}
where  $\mathbf r_u$ and $\mathbf r_v$ are partial derivatives of $\mathbf r$ with respect to $u$ and $v$.
When the parameterization $(u,v)$ is orthogonal, i.e., $\mathbf r_u \cdot \mathbf r_v=0$, we have
\begin{equation} \label{suface-gradient-uv0}
 \nabla_s \eta_j = \frac{1}{\|\mathbf r_u\|^2}\eta_{ju}
\mathbf r_u+\frac{1}{\|\mathbf r_v\|^2}\eta_{jv}
\mathbf r_v.
\end{equation}

We use $\eta_{ju}$ and $\eta_{jv}$, $j=1,2,\cdots, J$, as independent variables in the constrained energy minimization problem in Eqs.~\eqref{eqn:gb_energy}--\eqref{eqn:Dj00}/\eqref{eqn:Nj}. When the grain boundary is smooth, we have $\frac{\partial \eta_{ju}}{\partial v}-\frac{\partial \eta_{jv}}{\partial u}=0$. Including these relations  as further constraints,  the energy minimization problem in Eqs.~\eqref{eqn:gb_energy}--\eqref{eqn:Dj00}/\eqref{eqn:Nj} can be written as:

\vspace{0.1in}
\noindent
\underline{\bf Formulation Using Gradient Components}
\begin{eqnarray}
\text{minimize}\hspace{0.1in}
&& E = {\displaystyle \int_S \gamma_{\rm gb}  dS},
\ \ {\rm for \ all}\  \eta_{ju} \ {\rm and}\ \eta_{ju}, \  j=1,2,\cdots,J, \label{eqn:grad1} \\
\text{subject to}\hspace{0.1in}
&&{\textstyle \frac{\partial \eta_{ju}}{\partial v}-\frac{\partial \eta_{jv}}{\partial u}=0}, \ \ j=1,2,\cdots,J,   \label{cc1}\\
&&\mathbf{h}=\theta(\mathbf{V}\times \mathbf{a}) - {\displaystyle \sum_{j=1}^J} \mathbf{b}^{(j)}(\nabla_s\eta_j\cdot\mathbf{V})=\mathbf 0,\label{eqn:grad2}
\end{eqnarray}
where  vector $\mathbf V=\mathbf r_u$ and $\mathbf r_v$,  $\gamma_{\rm gb}$ is given by Eqs.~\eqref{eqn:gb_energy1} and \eqref{eqn:Dj00}/\eqref{eqn:Nj}, and  $\nabla_s \eta_j$ is expressed in terms of $\eta_{ju}$ and $\eta_{jv}$ by Eq.~\eqref{suface-gradient-uv}.

\subsection{Numerical Algorithm}

In the formulation of the constrained energy minimization problem using gradient components of  $\eta_j$ given in Eqs.~\eqref{eqn:grad1}--\eqref{eqn:grad2}, \eqref{eqn:gb_energy1}, \eqref{eqn:Dj00} or \eqref{eqn:Nj}, there are two groups of constraints due to the Frank's formula (Eq.~\eqref{eqn:grad2}) and the relationship between components of $\nabla_s \eta_j$ (Eq.~\eqref{cc1}). Numerically, we use the augmented Lagrangian method and the projection method to handle them, respectively.

For the constraints due to the Frank's formula given Eq.~\eqref{eqn:grad2}, we use the augmented Lagrangian method (e.g., Sec.~4.2 of \cite{Bertsekas1999}) in the numerical implementation.
The augmented Lagrangian function is:
\begin{equation}
L_{\rm A}(\mathbf D\pmb\eta,\pmb\lambda,\alpha)=\int_{S}\left(\gamma_{gb}  +\pmb\lambda \cdot \mathbf{h} +\frac{1}{2}\alpha\|\mathbf{h}\|^2 \right)dS,
\end{equation}
where $\mathbf D\pmb\eta=(\eta_{1u},\eta_{1v},... ,\eta_{Ju},\eta_{Jv}) $, $\pmb \lambda\in \mathbf R^6$ is the Lagrange multiplier vector, and $\alpha$ is a positive scalar.

Numerical implementation using the augmented Lagrangian method is
\begin{flalign}
%&\pmb\eta_t=-\frac{\partial L}{\partial \pmb\eta},\\
&\mathbf D\pmb\eta_{k+1}={\rm argmin}_{\mathbf D\pmb\eta} \ L_{\rm A}(\mathbf D\pmb\eta,\pmb\lambda_k,\alpha_k), \ {\rm for} \ \mathbf D\pmb\eta \ {\rm satisfying \ Eq.~\eqref{cc1}},\label{eqn:subproblem}\\
&\pmb\lambda_{k+1}=\pmb\lambda_k+\alpha_k\mathbf{h},\\
& {\rm update}\  \alpha_{k+1}\geq \alpha_k.
\end{flalign}
We choose constant $\alpha_k$ for all points on the grain boundary $S$.
When the numerical solution using an augmented Lagrangian method converges, it has been shown (e.g., Proposition 4.2.1 of \cite{Bertsekas1999}) that the converged solution is a (local) minimizer of the original constrained minimization problem when $\{\pmb\lambda_k\}$ is bounded and $\alpha_k\rightarrow\infty$ as $k\rightarrow \infty$.  With further the second order sufficient condition of the augmented Lagrangian function, it has been proved that the augmented Lagrangian method converges when $\alpha_k$'s are greater than some threshold value, not necessarily going to $\infty$  (e.g., Proposition 4.2.3 of \cite{Bertsekas1999}). Analysis of the augmented Lagrangian method for this constrained minimization problem will be presented elsewhere.

For the numerical implementation of the constraints in Eq.~\eqref{cc1}, we use the projection method, which is similar to that for solving fluid dynamics problems \cite{Chorin1968}.
 We introduce a new Lagrangian function:
\begin{equation}
L_{\rm P}=L_{\rm A}+\sum_{j=1}^{J}\mu_j\left(\frac{\partial \eta_{ju}}{\partial v}-\frac{\partial \eta_{jv}}{\partial u}\right),
\end{equation}
where $\mu_j$, $j=1,2,\cdots,J$, are Lagrange multipliers associated with the constraints in Eq.~\eqref{cc1}.

The Lagrangian $L_{\rm P}$ is minimized with respect to $\eta_{ju}$, $\eta_{jv}$, $j=1,2,\cdots,J$,  by gradient flow:
\begin{flalign}
&\frac{\partial\eta_{ju} }{\partial t}=-\frac{\delta L_{\rm P}}{\delta \eta_{ju}} =-\frac{\delta L_{\rm A}}{\delta \eta_{ju}}+\frac{\partial\mu_{j}}{\partial v},\\
&\frac{\delta\eta_{jv} }{\delta t}=-\frac{\delta L_{\rm P}}{\delta \eta_{jv}} =-\frac{\delta L_{\rm A}}{\delta \eta_{jv}}-\frac{\partial\mu_{j}}{\partial u}
\end{flalign}
subject to the constraints in Eq.~\eqref{cc1}, where $t$ is some artificial time. During each time step of the evolution, we separate the contributions from $L_{\rm A}$ and $\mu_j$ into two steps:
\begin{flalign}
&\eta_{ju}^* = \eta_{ju}^n-\left.\frac{\delta L_{\rm A}}{\delta \eta_{ju}}\right|_{t_n}\cdot  \delta t, \ \ \ \eta_{jv}^* = \eta_{jv}^n-\left.\frac{\delta L_{\rm A}}{\delta \eta_{jv}}\right|_{t_n}\cdot \delta t, \vspace{1ex} \label{eqn:p1}\\
&  \eta_{ju}^{n+1}=\eta_{ju}^*+\frac{\partial\mu_{j}^{n+1}}{\partial v} \delta t, \ \ \ \eta_{jv}^{n+1}=\eta_{jv}^*-\frac{\partial\mu_{j}^{n+1}}{\partial u} \delta t. \label{dd}
\end{flalign}
In order to satisfy the constraint $\frac{\partial \eta_{ju}^{n+1}}{\partial v}-\frac{\partial \eta_{jv}^{n+1}}{\partial u}=0 $, using Eq.~\eqref{dd}, we have the formula for updating $\mu_j$:
\begin{equation}
  \bigtriangleup \mu_{j}^{n+1}=\frac{1}{\delta t}\left(\frac{\partial \eta_{jv}^{*}}{\partial u}- \frac{\partial \eta_{ju}^{*}}{\partial v}\right), \label{eqn:p2}
\end{equation}
where $\bigtriangleup$ is the Laplace operator.

In summary, the constrained minimization problem in Eq.~\eqref{eqn:subproblem} for each step in the augmented Lagrangian algorithm can be solved by projection method following the steps in Eqs.~\eqref{eqn:p1}, \eqref{eqn:p2}, and \eqref{dd}.

The complete numerical algorithm for solving the constrained energy minimization problem  in Eqs.~\eqref{eqn:grad1}--\eqref{eqn:grad2} is summarized below.

\vspace{0.1in}
\noindent
\underline{\bf Numerical Algorithm}
\vspace{0.05in}

1. Solve the constrained minimization problem $\mathbf D\pmb\eta_{k+1}={\rm argmin}_{\mathbf D\pmb\eta} \ L_{\rm A}(\mathbf D\pmb\eta,\pmb\lambda_k,\alpha_k)$ subject to the constraints in Eq.~\eqref{cc1} by the following iteration of projection method to equilibrium:
\begin{flalign*}
&\eta_{ju}^* = \eta_{ju}^n-\left.\frac{\delta L_{\rm A}}{\delta \eta_{ju}}\right|_{t_n}\cdot \delta t, \ \ \eta_{jv}^* = \eta_{jv}^n-\left.\frac{\delta L_{\rm A}}{\delta \eta_{jv}}\right|_{t_n}\cdot \delta t, \\
&\bigtriangleup \mu_{j}^{n+1}=\frac{1}{\delta t}\left(\frac{\partial \eta_{jv}^{*}}{\partial u}- \frac{\partial \eta_{ju}^{*}}{\partial v}\right),\\
&  \eta_{ju}^{n+1}=\eta_{ju}^*+\frac{\partial\mu_{j}^{n+1}}{\partial v}\delta t, \ \ \eta_{jv}^{n+1}=\eta_{jv}^*-\frac{\partial\mu_{j}^{n+1}}{\partial u}\delta t.
\end{flalign*}

2. Update $\pmb\lambda_{k+1}=\pmb\lambda_k+\alpha_k\mathbf{h}$.

3. Update  $\alpha_{k+1}\geq \alpha_k$.

4. Repeat the above steps until convergence.

\section{Numerical results}
\label{sec:NumericalResults}

In this section, we present some numerical simulation results using our continuum model and compare them with those of molecular static (MS) simulations.

We focus on low angle grain boundaries in fcc Al. The six possible Burgers vectors are shown in Fig.~\ref{thompson}, and their  length is $b=0.286$nm. The Poisson ratio is $\nu=0.347$. Following Ref.~\cite{zhang2017energy}, we choose $r_g = 3.5e^{-\sin\phi}b$ in the formula of $\gamma_{\rm gb}$ in Eq.~\eqref{eqn:gb_energy1}, where $\phi$ is the angle between rotation axis $\mathbf{a}$ and the out normal vector $\mathbf{n}$ of the grain boundary. The improved formula in Eq.~\eqref{eqn:Nj} is used. In numerical simulations, $\|\nabla_s \eta_j\|$ in Eqs.~\eqref{eqn:Nj} and is regularized as $\sqrt{\|\nabla_s \eta_j\|^2+\varepsilon}$, and we chose
the regularization parameter $\varepsilon$  to be $8 \times 10^{-7}b^{-2}$.

We use the central difference schemes to calculate the partial derivatives in Eqs.~\eqref{dd} and \eqref{eqn:p2}. The Laplace equation in Eq.~\eqref{eqn:p2} is solved by Gauss-Jordan elimination in which the inverse of the coefficient matrix after discretization is calculated only once during the energy minimization process.
%
%In this numerical scheme, we use central difference scheme to deal with the derivative operator, and the boundary condition is periodic.

We use the EAM potential for Al developed by Mishin et al.~\cite{MishinEAM} and the
LAMMPS code \cite{LAMMPS} in the MS simulations.
 Periodic boundary conditions are applied in all three dimensions in the MS simulations. The simulation volume consists of two grains. The inner grain is rotated with a misorientation angle about a given rotation axis. Accordingly, the interface between the inner and outer grains forms a grain boundary with the desired misorientation angle and rotation axis. The constituent dislocations of the grain boundary are identified and visualized by using the software AtomEye~\cite{AtomEye} (which is based on the atom energy). The simulation volume size is chosen to be large enough (about three times of the largest dimension of the cylindrical or spherical grain boundary) to eliminate the size effect. As a result, the number of atoms in the simulation volume varies from $4$ million to $11$ million.
The MS simulations were running on $8$ processors, and the wall-clock time varied from $10$ minutes to one hour for the presented examples. Note that in the continuum simulations, a grain boundary was discretized into a mesh of size $40\times20$ in the domain of parametrization $(u,v)$, and the calculation of each presented example took about 30 minutes on a single i7-6500u processor. The continuum simulations took much less memory and computational time compared to the MS simulations.

\subsection{Boundary of a finite cylindrical grain}
 We first consider the grain boundary when a finite cylindrical grain is embedded in an infinite matrix. The grain boundary $S$ is the surface of the finite cylindrical grain. We choose the directions $[\bar{1}10]$, $[\bar{1}\bar{1}2]$ and $[111]$ to be the $x$, $y$ and $z$ directions, respectively.
The six burgers vectors in this coordinate system are $\mathbf{b}^{(1)}=(1,0,0)b$,
 $\mathbf{b}^{(2)}=(\frac{1}{2},\frac{\sqrt{3}}{2},0)b$,
$\mathbf{b}^{(3)}=(\frac{1}{2},-\frac{\sqrt{3}}{2},0)b$,
$\mathbf{b}^{(4)}=(0,\frac{\sqrt{3}}{3},-\frac{\sqrt{6}}{3})b$,
$\mathbf{b}^{(5)}=(\frac{1}{2},\frac{\sqrt{3}}{6},
\frac{\sqrt{6}}{3})b$, and $\mathbf{b}^{(6)}=(-\frac{1}{2},\frac{\sqrt{3}}{6},\frac{\sqrt{6}}{3})b$.
The axis of the cylindrical grain is the $z$ direction (i.e. along the $[111]$ direction). The radius of the cylindrical grain is $R=30b$ and its height is $H=50b$. The rotation axis of $S$ is $\mathbf{a}=(0 0 1)$ (i.e. in the $[111]$ direction).  The misorientation angle of $S$ is $\theta=3^\circ$.

\begin{figure}[htbp]
\centering
\hspace{0.6in} \subfigure[]{\includegraphics[width=1.5in]{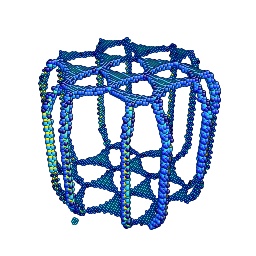}} \hspace{0.1in}
\subfigure[]{\includegraphics[width=2.0in]{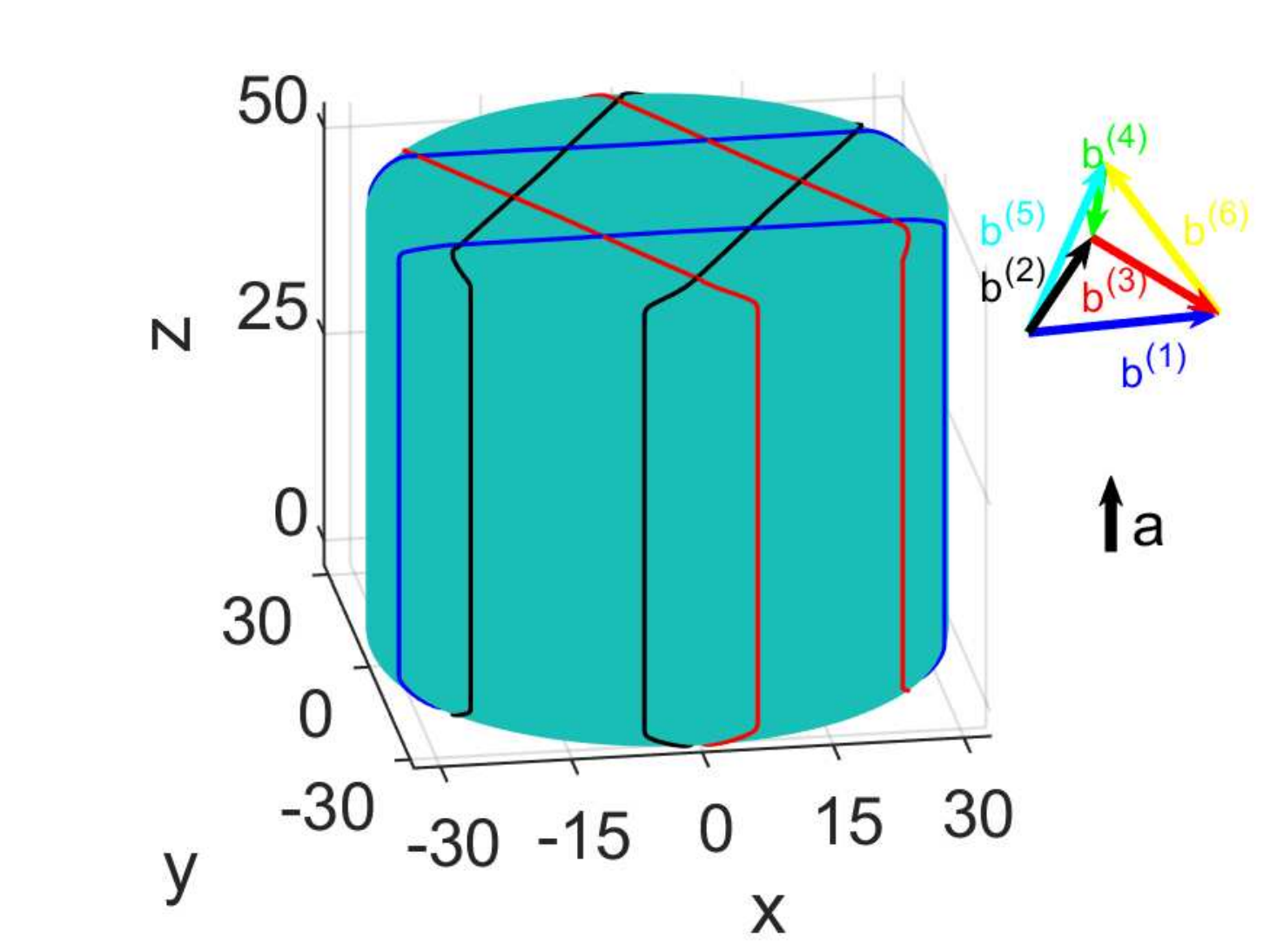}}
\subfigure[]{\includegraphics[width=2.2in]{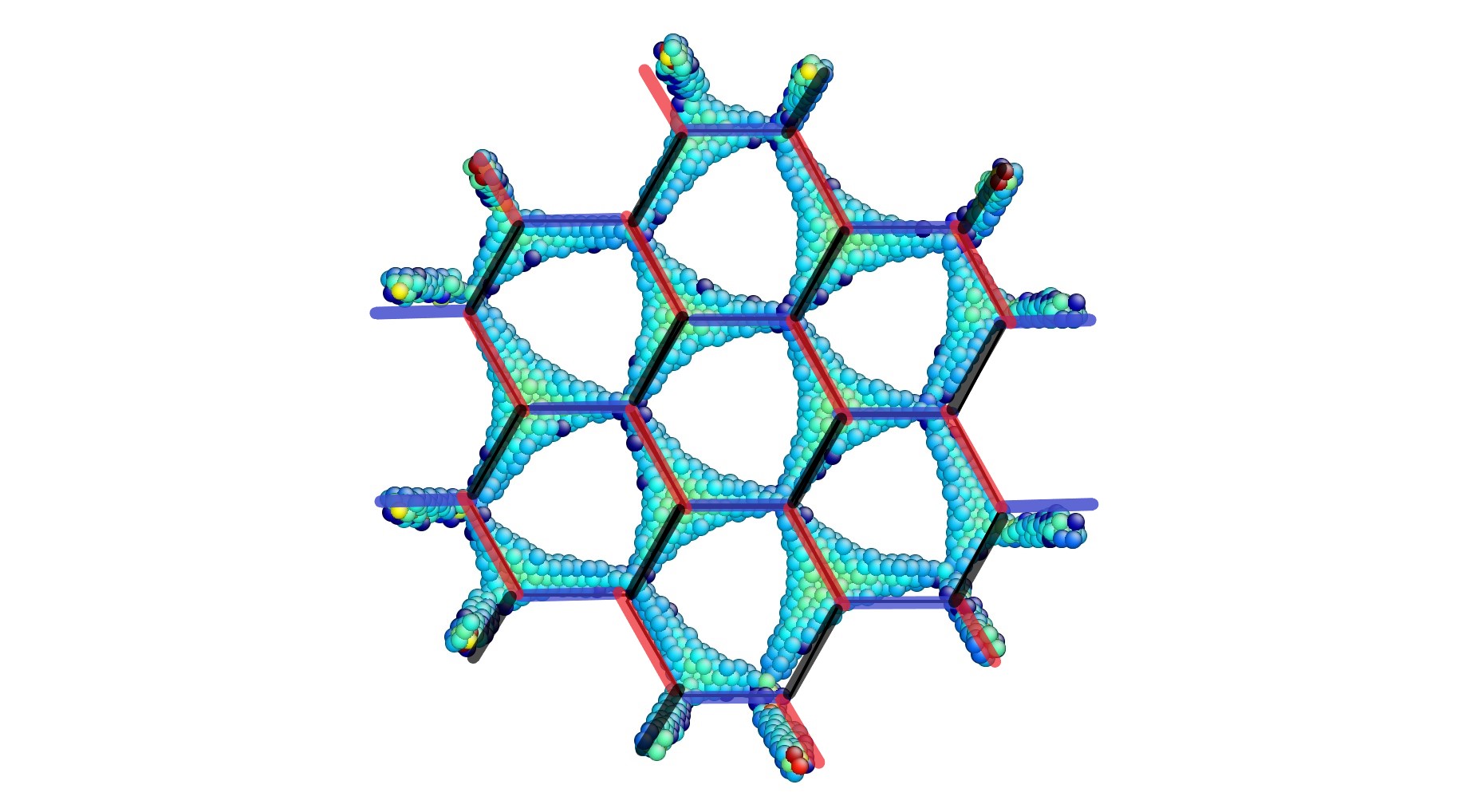}}
\subfigure[]{\includegraphics[width=1.45in]{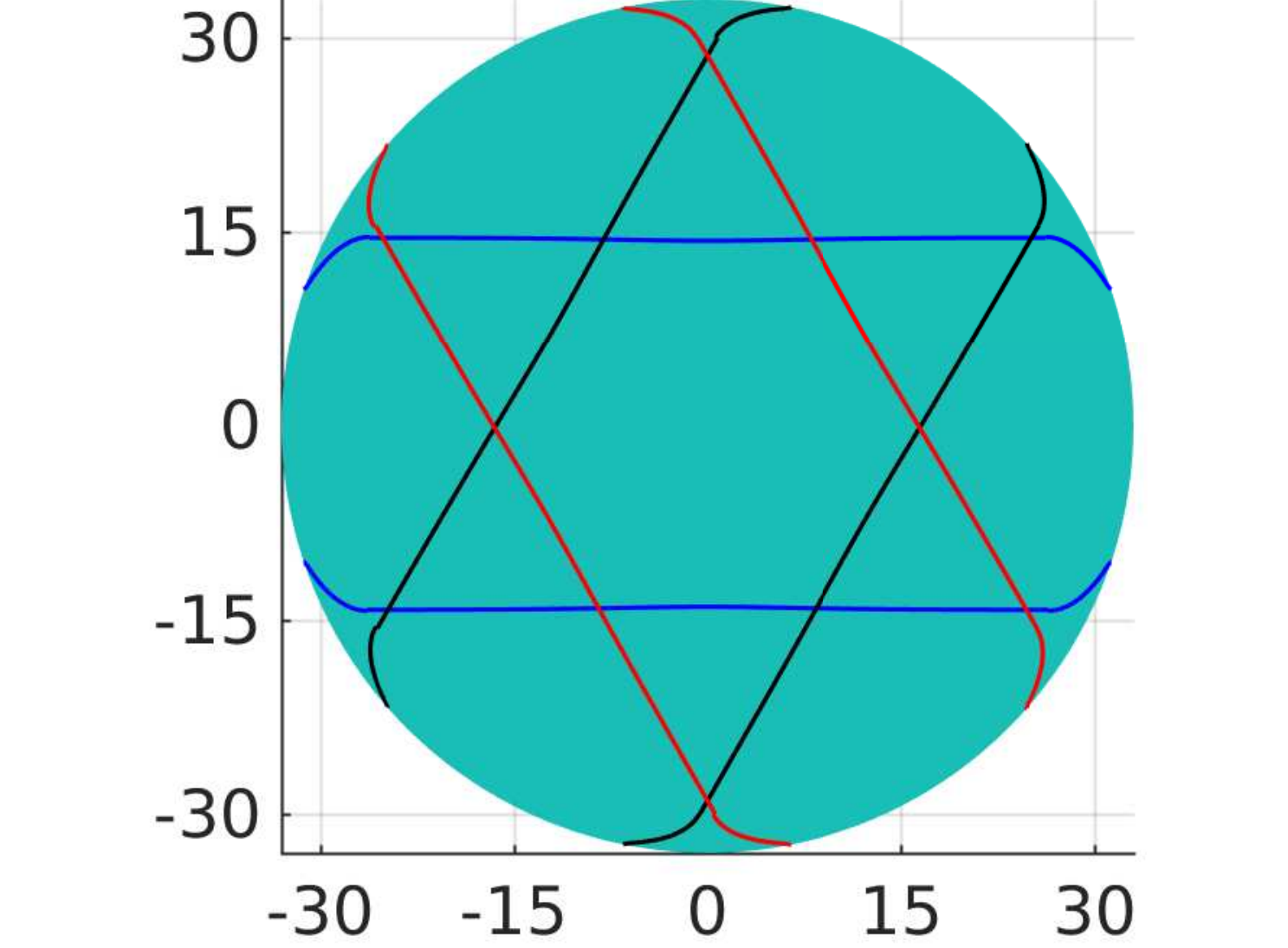}}
\subfigure[]{\includegraphics[width=2.2in]{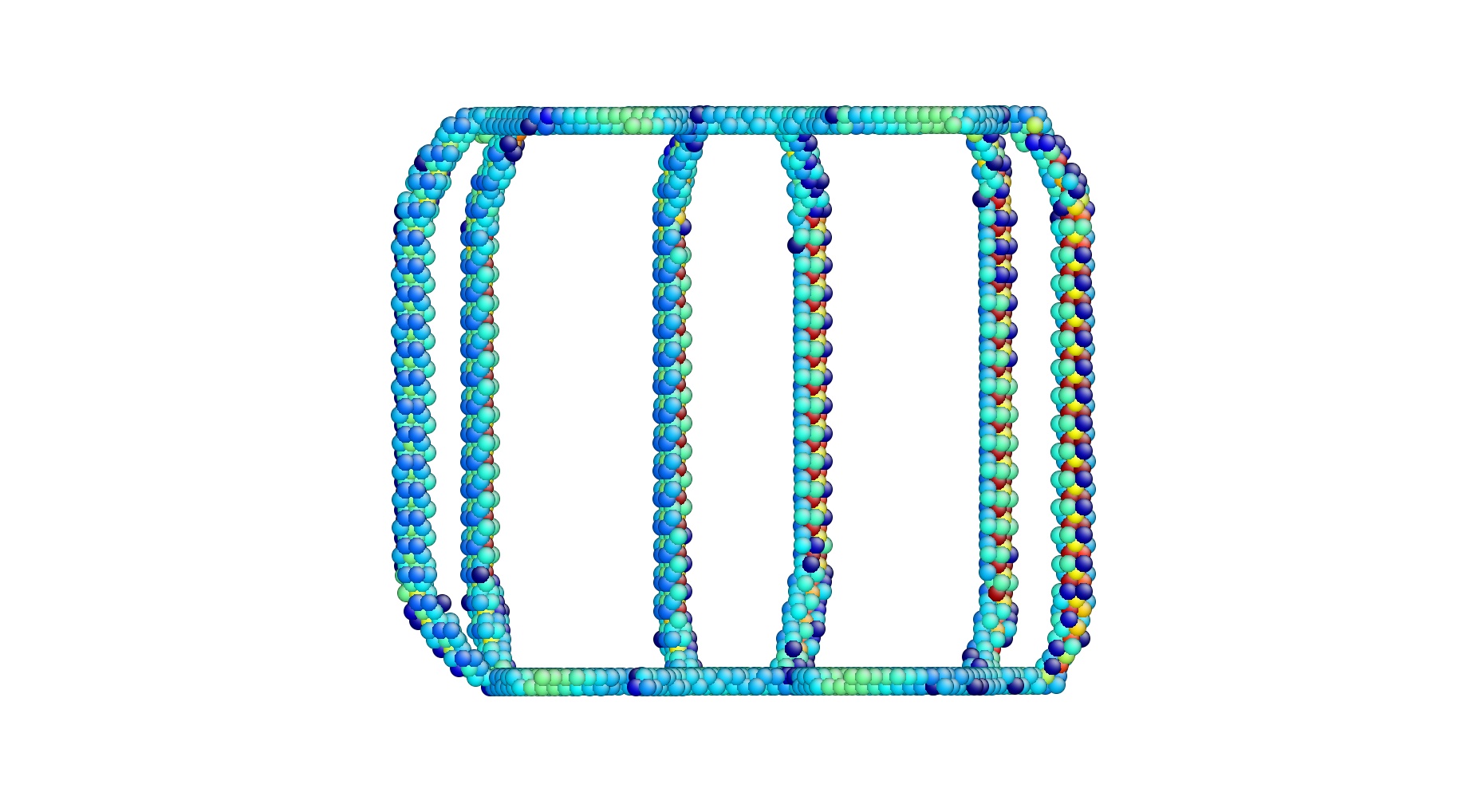}}
\subfigure[]{\includegraphics[width=1.45in]{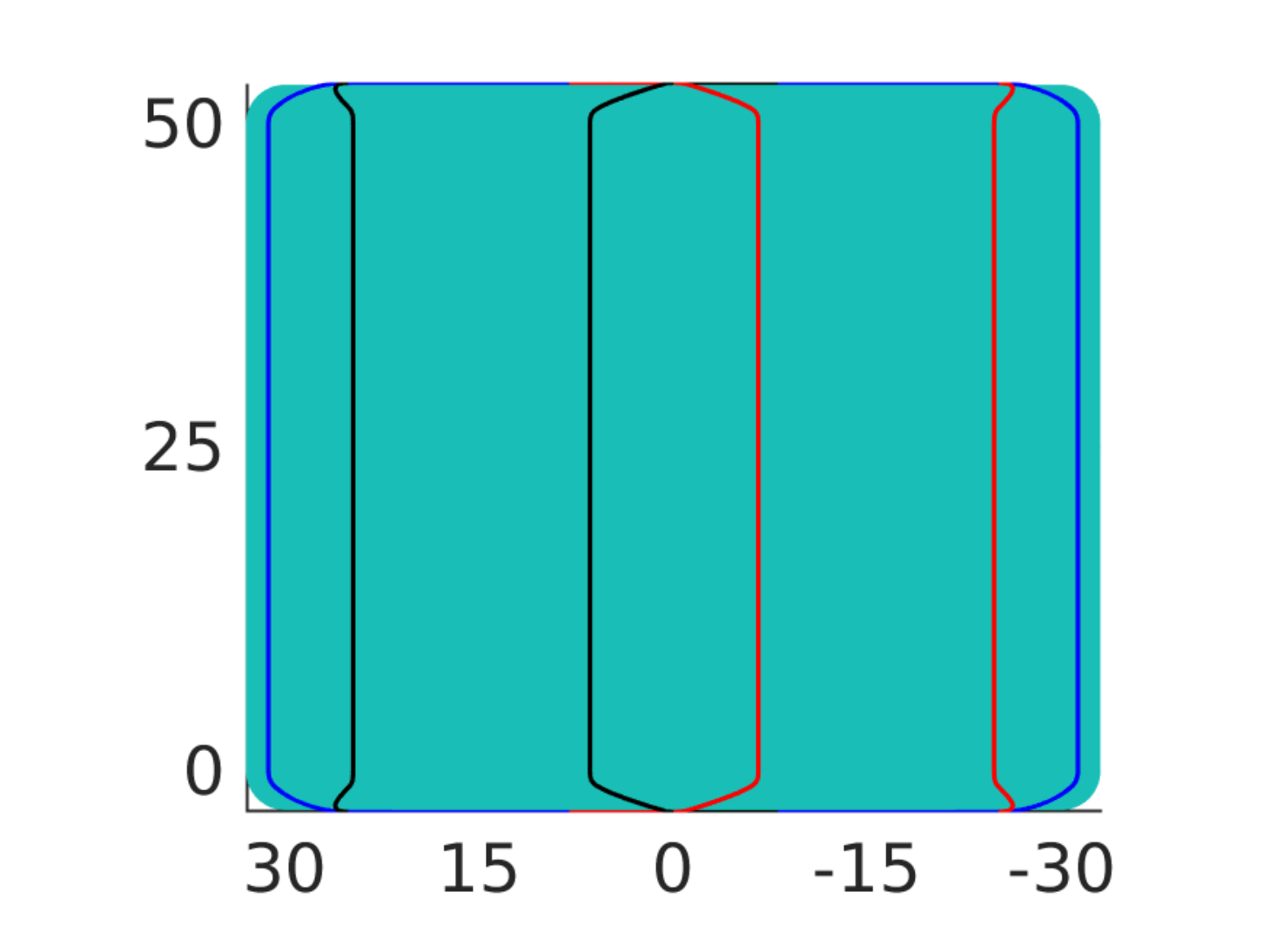}}
\subfigure[]{\includegraphics[width=2.2in]{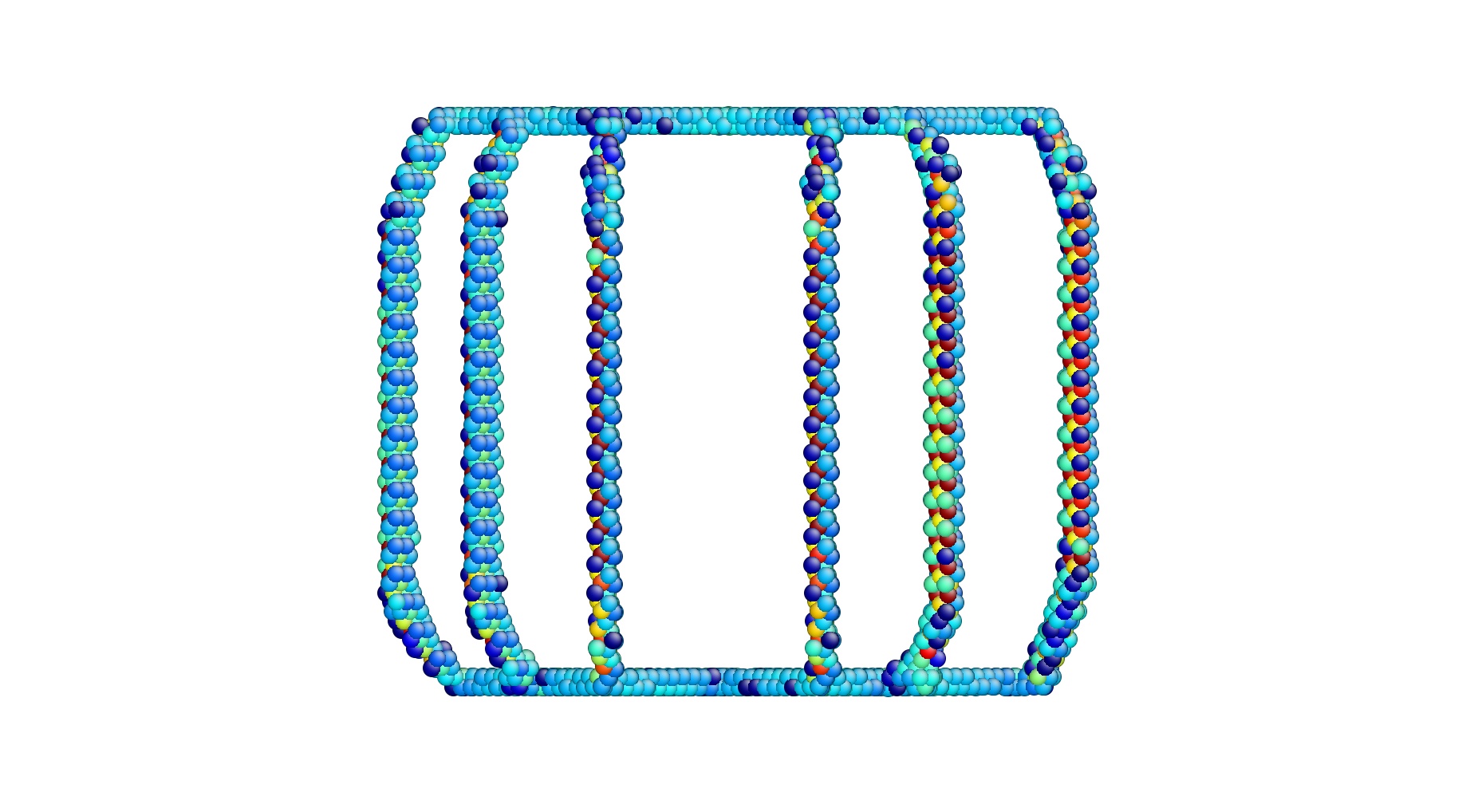}}
\subfigure[]{\includegraphics[width=1.45in]{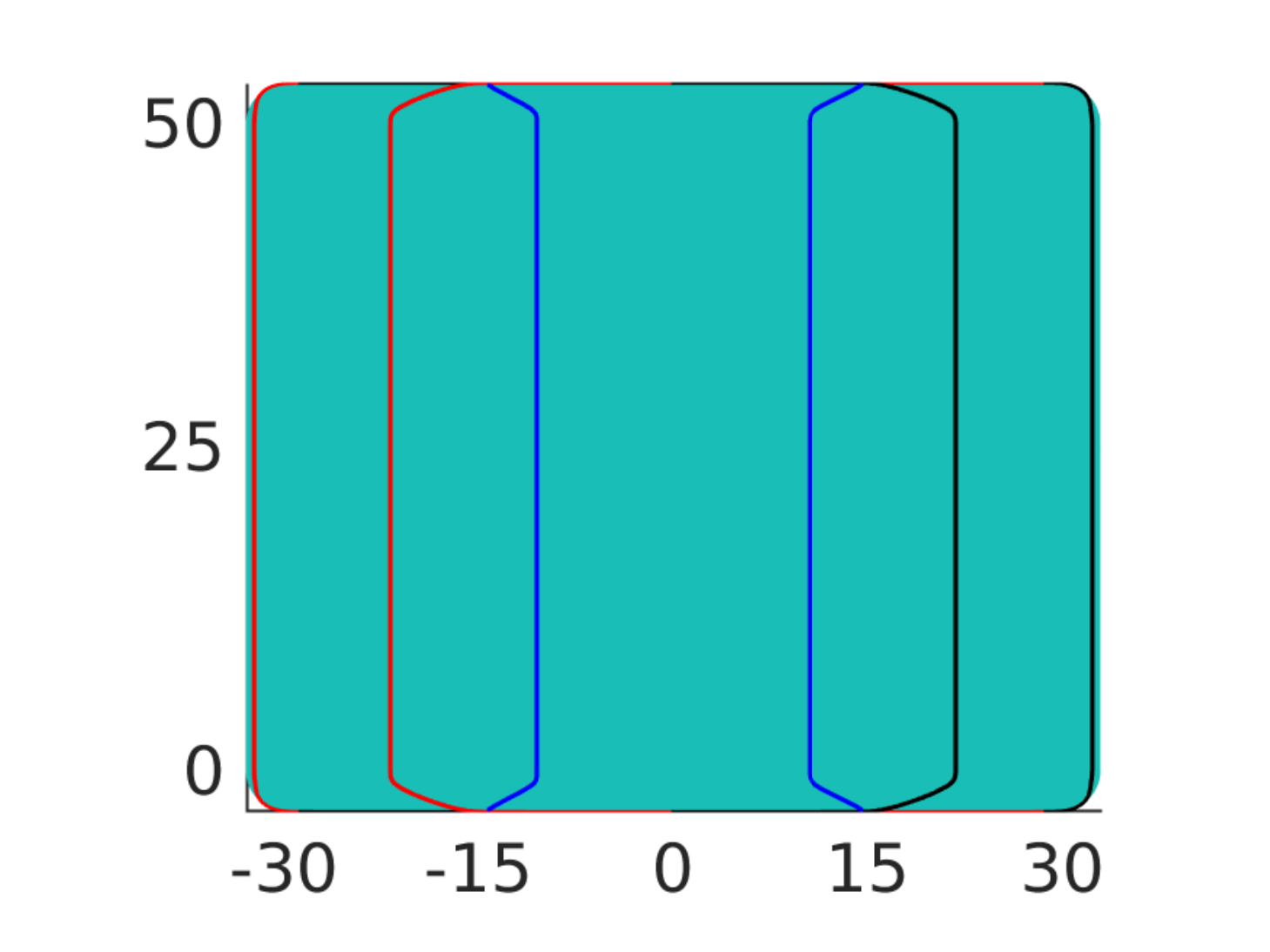}}
\caption{Dislocation structure of the boundary of a cylindrical grain with rotation axis $\mathbf a$ along its axis (in the $+z$ meaning the $[111]$ direction) and misorientation angle  $\theta=3^\circ$. The right panel shows the numerical result obtained using our continuum model, and the left panel shows the result of MS simulation.  The length unit in the images of continuum model result is $b$. (a) and (b) Three-dimensional view. The dislocation structure on this grain boundary consists of dislocations with Burgers vectors $\mathbf{b}^{(1)}$,  $\mathbf{b}^{(2)}$, and $\mathbf{b}^{(3)}$, all of which are in the $xy$ plane (the $(111)$ plane). Dislocations with these three Burgers vectors are shown by blue, black, and red lines, respectively, in the images of result of the continuum model. (c) and (d) Top view. There are partial separations in the MS simulation result in (c), in which the colored lines indicate the locations of dislocation segments with different Burgers vectors without partial separation. (e) and (f) View from the $+y$ direction (the  $[\bar{1}\bar{1}2]$ direction). (g) and (h) View from the $+x$ direction (the  $[\bar{1}10]$ direction). Note that the results of the continuum model have translation invariance, and here we plot those with good agreement with MS results under this translation invariance in these images.}
\label{fig:001cy}
\end{figure}

For the cylindrical surface of the inner grain, we the parametrization $(u,v)=(R\alpha,z)$, where $\alpha$ is the polar angle in the $xy$ plane. The cylindrical surface is $\mathbf r(u,v)=(R\cos\frac{u}{R}, R\sin\frac{u}{R}, v)$. With this parametrization, we have $\mathbf r_u=(-\sin\frac{u}{R}, \cos\frac{u}{R}, 0)$ and $\mathbf r_v=(0,0,1)$. This parametrization is orthogonal, and  $\|\mathbf r_u\|=\|\mathbf r_v\|=1$.
The outer normal vector $\mathbf{n}$ on the cylindrical surface is $\mathbf{n}=(\cos\frac{u}{R}, \sin\frac{u}{R}, 0)$.
For the top and bottom disks of the cylindrical grain, we use the polar coordinates $(u,v)=(r,\alpha)$, $0\leq r \leq R$. Thus we have $\mathbf r(u,v)=(u\cos v, u\sin v,C)$, where $C=H$ on the top and $C=0$ on the bottom,
 $\mathbf r_u=(\cos v, \sin v, 0)$, $\mathbf r_v=(-u\sin v,u\cos v,0)$, and the outer normal vector of the grain boundary is $\mathbf{n} = (0,0,\pm 1)$. With these parametrizations, the surface gradients of $\eta_j$ on these surfaces  are given by
Eq.~\eqref{suface-gradient-uv0}. The sharp corners of this cylinder are regularized by connecting the top/bottom surface with the side surface smoothly by a quarter circle with radius $3b$ for each polar angle $\alpha$.

The dislocation structure on the surface of this cylindrical grain obtained using our method and comparison with the MS result are shown in Fig.~\ref{fig:001cy}. These dislocations have Burgers vectors $\mathbf{b}^{(1)}$,  $\mathbf{b}^{(2)}$, and $\mathbf{b}^{(3)}$. It is pure tilt on the side, cylindrical grain boundary, and the obtained dislocations on this boundary are straight lines, which agrees excellently with the MS simulation result. This result is also consistent with the results of planar low angle grain boundaries calculated in Ref.~\cite{zhang2017energy} (Fig.~11 there) if we consider the dislocation structure on this cylindrical boundary pointwise.

 The planar top and bottom boundaries are pure twist, and the dislocation structure obtained using our method is shown in Fig.~\ref{fig:001cy}(d) and the MS simulation result in Fig.~\ref{fig:001cy}(c). Considering the densities of dislocations on the top boundary, the total length of dislocations with Burgers vector $\mathbf b^{(3)}$ in the result of the continuum model, i.e., the two red lines in Fig.~\ref{fig:001cy}(d), is about $376\AA$ (approximated by straight lines), and this agrees excellently with the total length of $\mathbf b^{(3)}$-dislocations in the MS result in  Fig.~\ref{fig:001cy}(c) marked by red segments, which is $350\AA$. The orientations of the dislocations in the continuum model result in Fig.~\ref{fig:001cy}(d) are also the same as those of the corresponding dislocations in the MS result in Fig.~\ref{fig:001cy}(c).

\begin{figure}[htbp]
\centering
\subfigure[]{\includegraphics[width=1.6in]{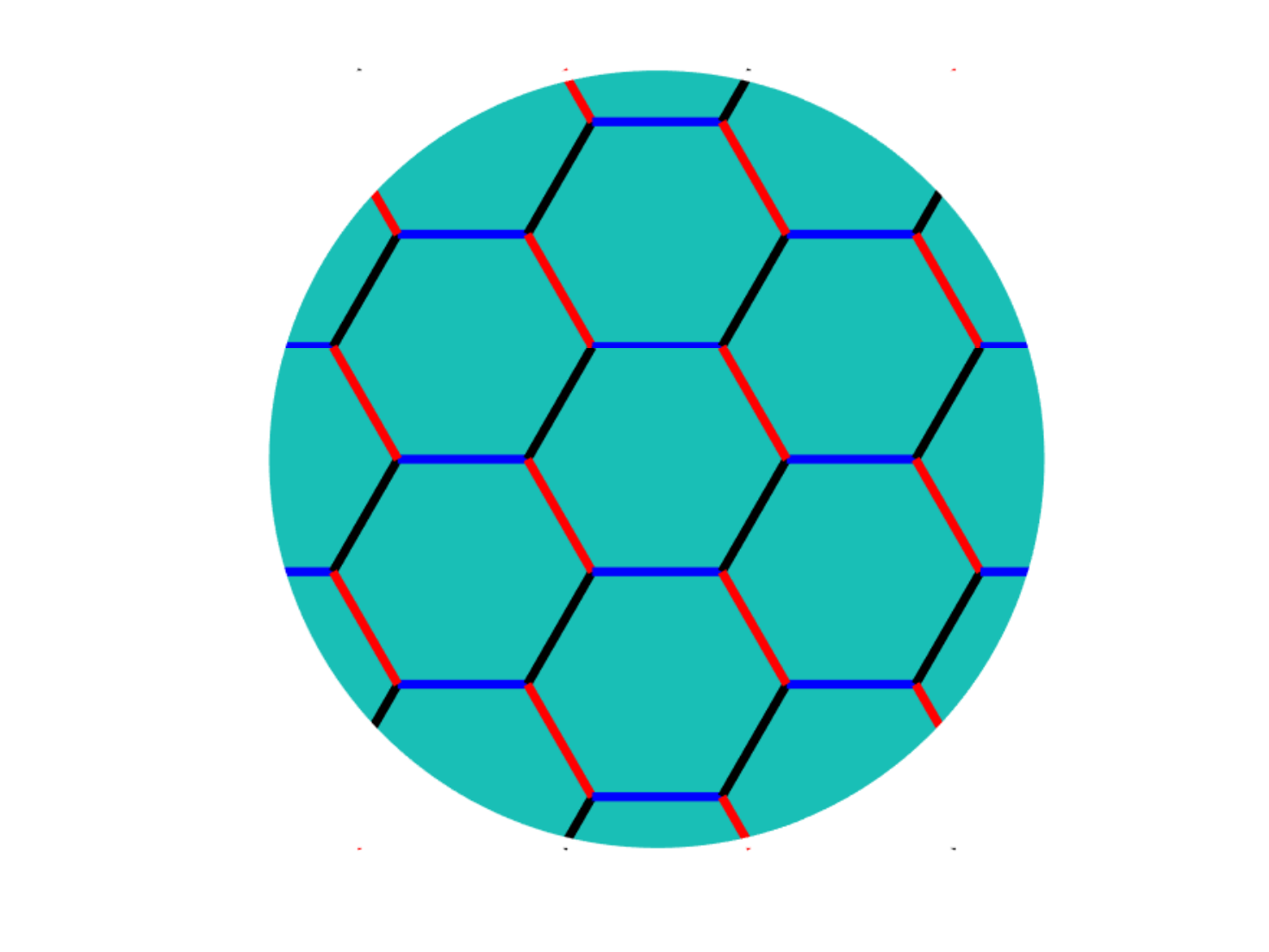}}
\subfigure[]{\includegraphics[width=1.75in]{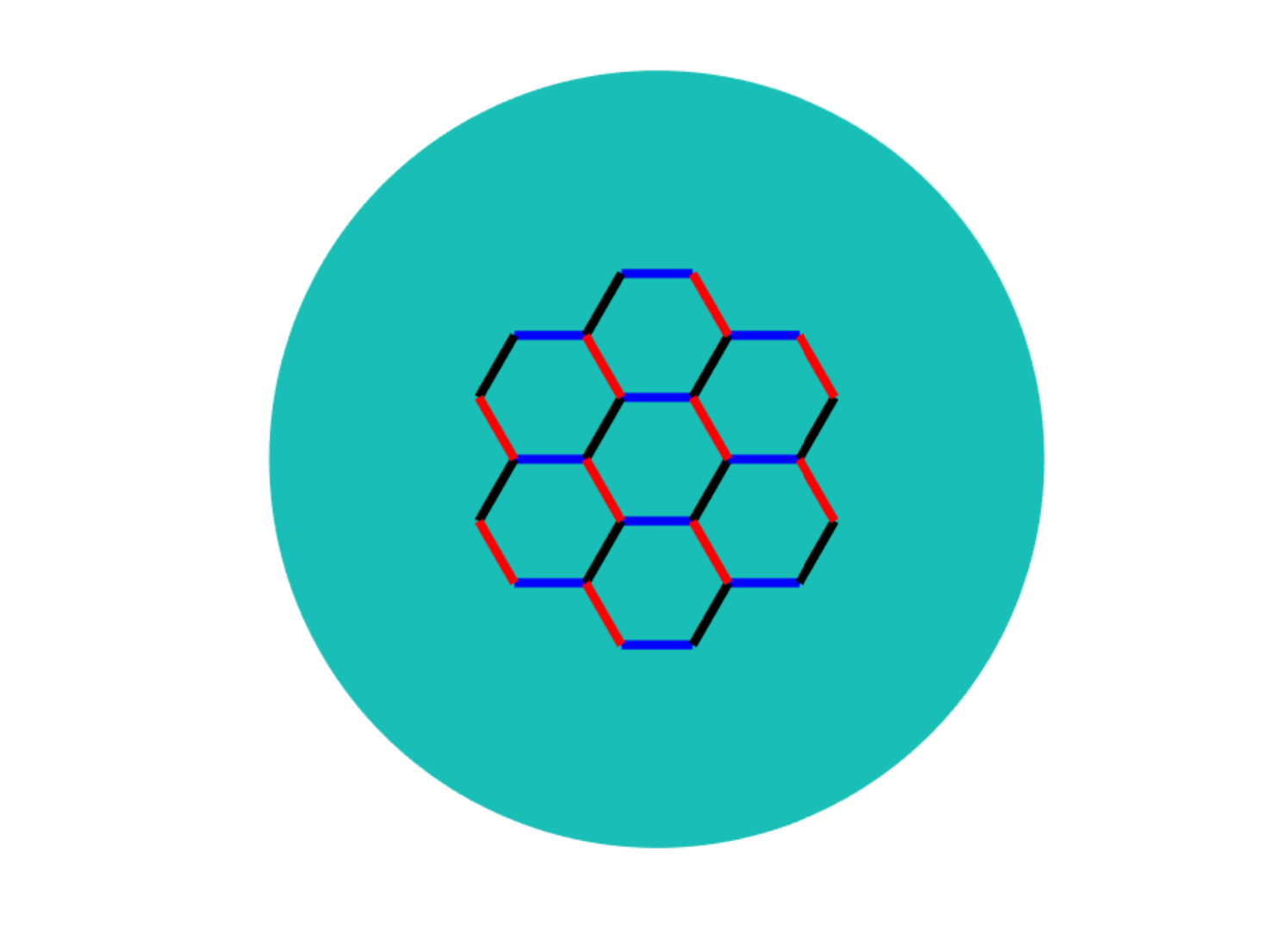}}
\caption{ The exact dislocation network structures on grain boundaries generated using the identification method presented in Sec.~\ref{sec:identification} from the results of the continuum model. (a) Top boundary of the cylindrical grain in Fig.~\ref{fig:001cy}. (b) Planar dislocation network at the pure twist point $x=R$ on the spherical grain boundary in Fig.~\ref{fig:sphere}.}
\label{fig:networks}
\end{figure}

 From the result of the continuum model, we generate the exact dislocation network structure on the top boundary using the identification method presented in Sec.~\ref{sec:identification}, and the result is shown in Fig.~\ref{fig:networks}(a), which excellently recovers the hexagonal dislocation network in the MS result in Fig.~\ref{fig:001cy}(c).

% Also for planar. agrees with that calculated in Ref.~\cite{zhang2017energy} (Fig.~3 there) using a method for planar grain boundaries. As has been discussed in Ref.~\cite{zhang2017energy}, the results of our continuum models provide excellent approximations to the dislocation densities and energy density of this twist boundary compared with those of MS simulations (as shown by Figs.~3 and 4 and Table 2 in  Ref.~\cite{zhang2017energy}).

 % which is the main purpose of the continuum models instead of resolving individual dislocation segments in the dislocation structure. For example, the total length of dislocations with an individual Burgers vector in the triangular network obtained using the continuum models is the same as the total length of the dislocation segments with the same Burgers vector in the exact dislocation structure of hexagonal network obtained by the discrete dislocation model and MS simulations (as compared in Fig.~3(d) in  Ref.~\cite{zhang2017energy}).

 % This is because our continuum models are based on the Frank's formula, which alone is not able to describe the hexagonal network. Dislocation reaction is needed in order to describe the hexagonal network.

 Note that there are partial separations in the MS simulation results in Fig.~\ref{fig:001cy}(c), which is neglected in the continuum model. The transition of dislocation structures from twist to tilt in the smooth transition region between the top/bottom boundary and the side boundary of this cylindrical grain  is consistent with the results in Fig.~\ref{fig:compare1} (and those in Sec.~4.2 in Ref.~\cite{zhang2017energy}).

\begin{figure}[htbp]
\centering
\subfigure[]{\includegraphics[width=2.5in]{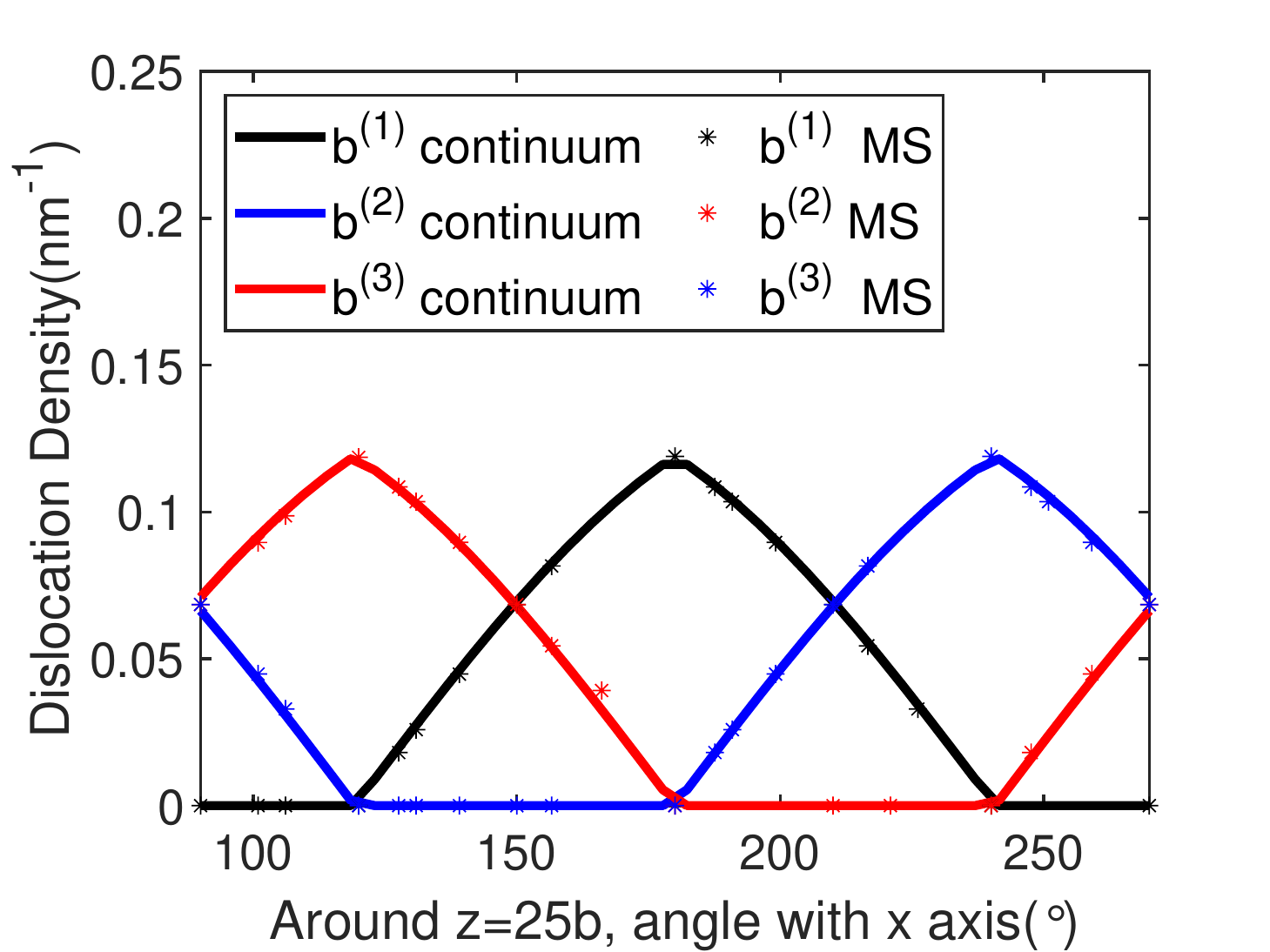}}
\subfigure[]{\includegraphics[width=2.5in]{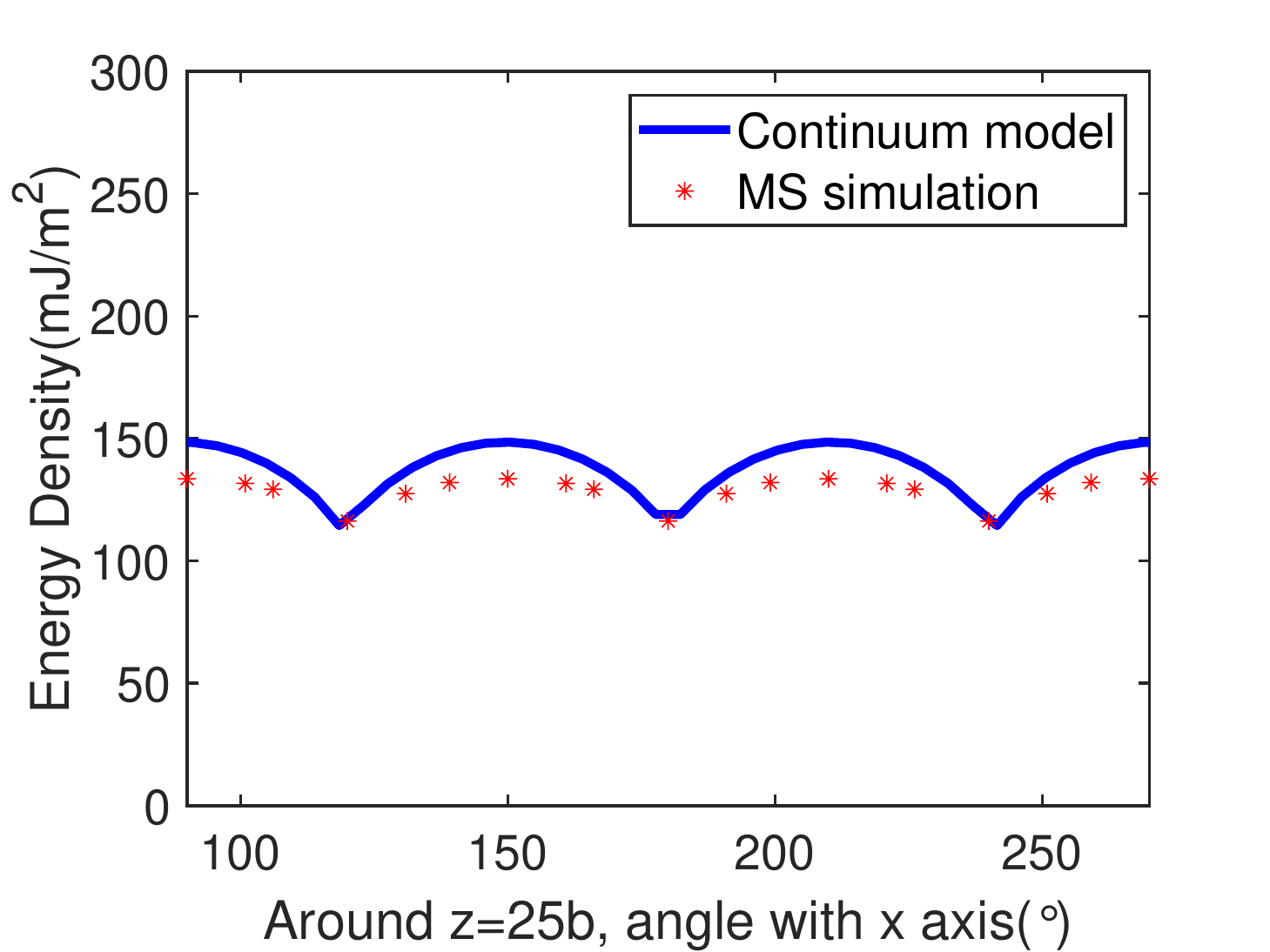}}
\subfigure[]{\includegraphics[width=2.5in]{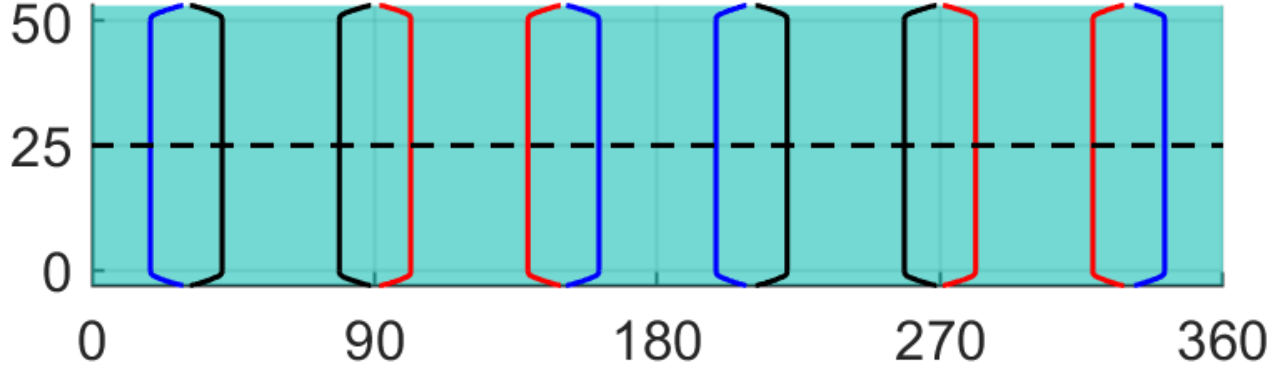}}
\caption{Comparisons of (a) dislocation densities and (b) energy density at every point along the line with height $H/2$ on the cylindrical boundary (the dashed line in (c)) of the finite cylindrical grain I calculated using our continuum model, with the MS results for planar low angle grain boundaries calculated in Ref.~\cite{zhang2017energy}
The misorientation angle is $\theta = 1.95^\circ$.  (c) Dislocation structure on the cylindrical boundary. The unit of the horizontal axis is the polar angle and that of the vertical axis is $b$. }
\label{fig:z0cy001}
\end{figure}

For quantitative comparisons, we compare the
densities of dislocations and energy density along the line with height $H/2$ on this cylindrical boundary, obtained using our method and  MS simulation. The results are shown in Fig.~\ref{fig:z0cy001}.
Note that accurate dislocation  and energy densities using MS simulation can only be obtained for planar low angle grain boundaries. We compare the pointwise results on the cylindrical boundary obtained using our method with the
 MS results for planar low angle grain boundaries calculated in Ref.~\cite{zhang2017energy} (Fig.~10 there), where the misorientation angle $\theta = 1.95^\circ$.
Excellent agreement between the results using these two methods can be seen.  Note the dislocation structure of this grain boundary with $\theta = 1.95^\circ$ is similar to that shown in Fig.~\ref{fig:001cy} with  $\theta = 3^\circ$.

\subsection{Spherical grain boundary}

\begin{figure}[htbp]
\centering
\hspace{0.2in}\subfigure[]{\includegraphics[width=2.6in]{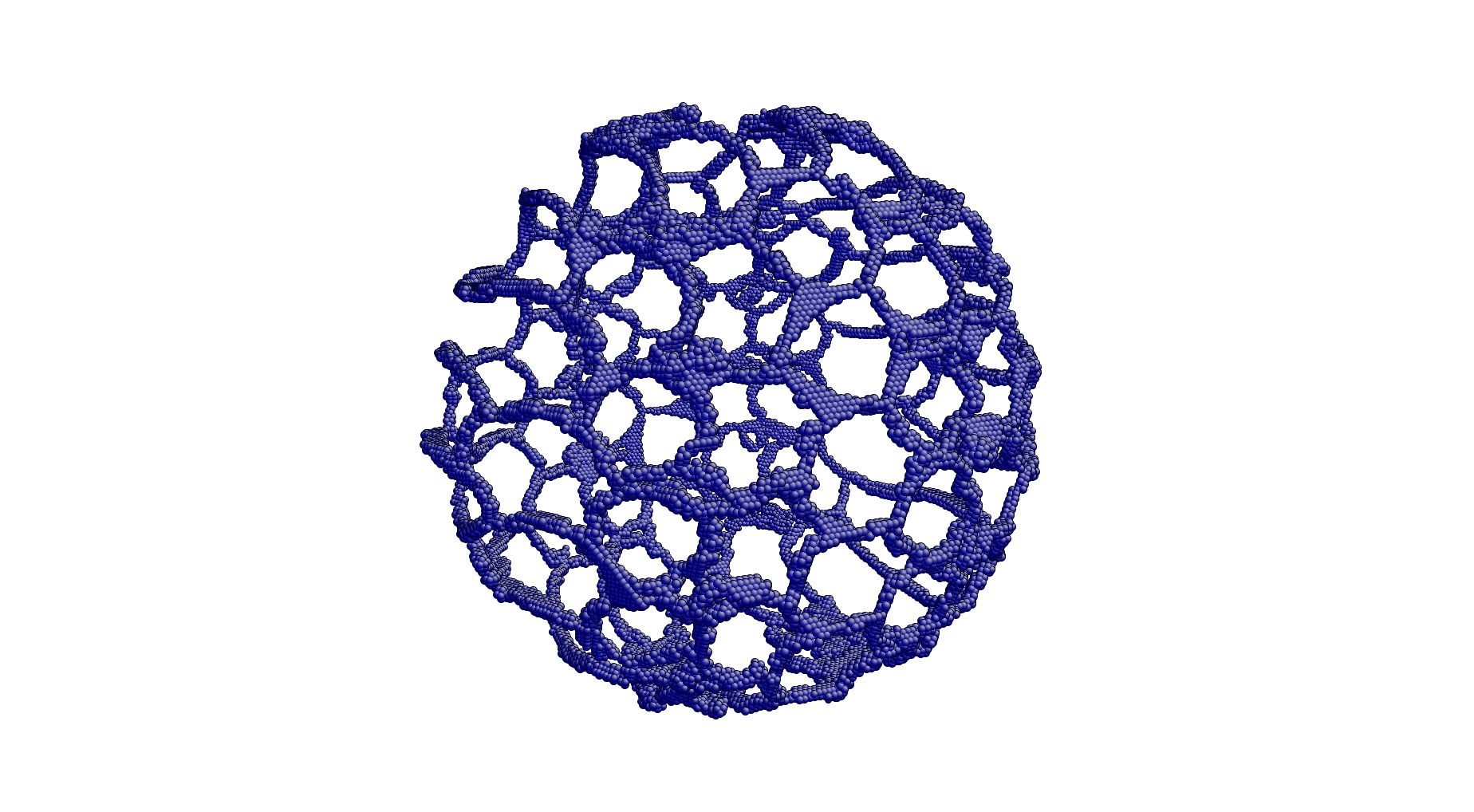}}\hspace{-0.1in}
\subfigure[]{\includegraphics[height=1.6in]{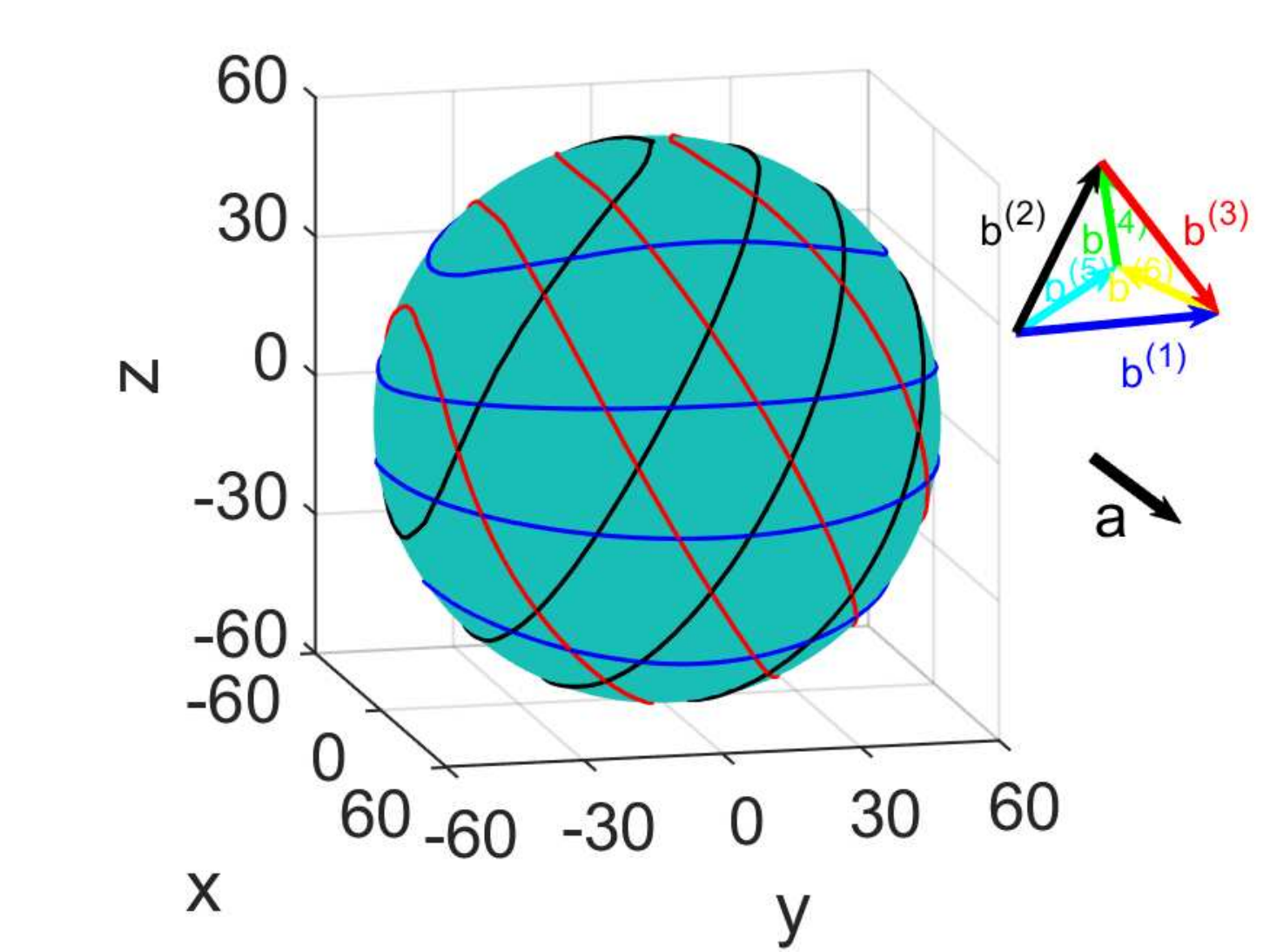}}
\subfigure[]{\includegraphics[width=2.5in]{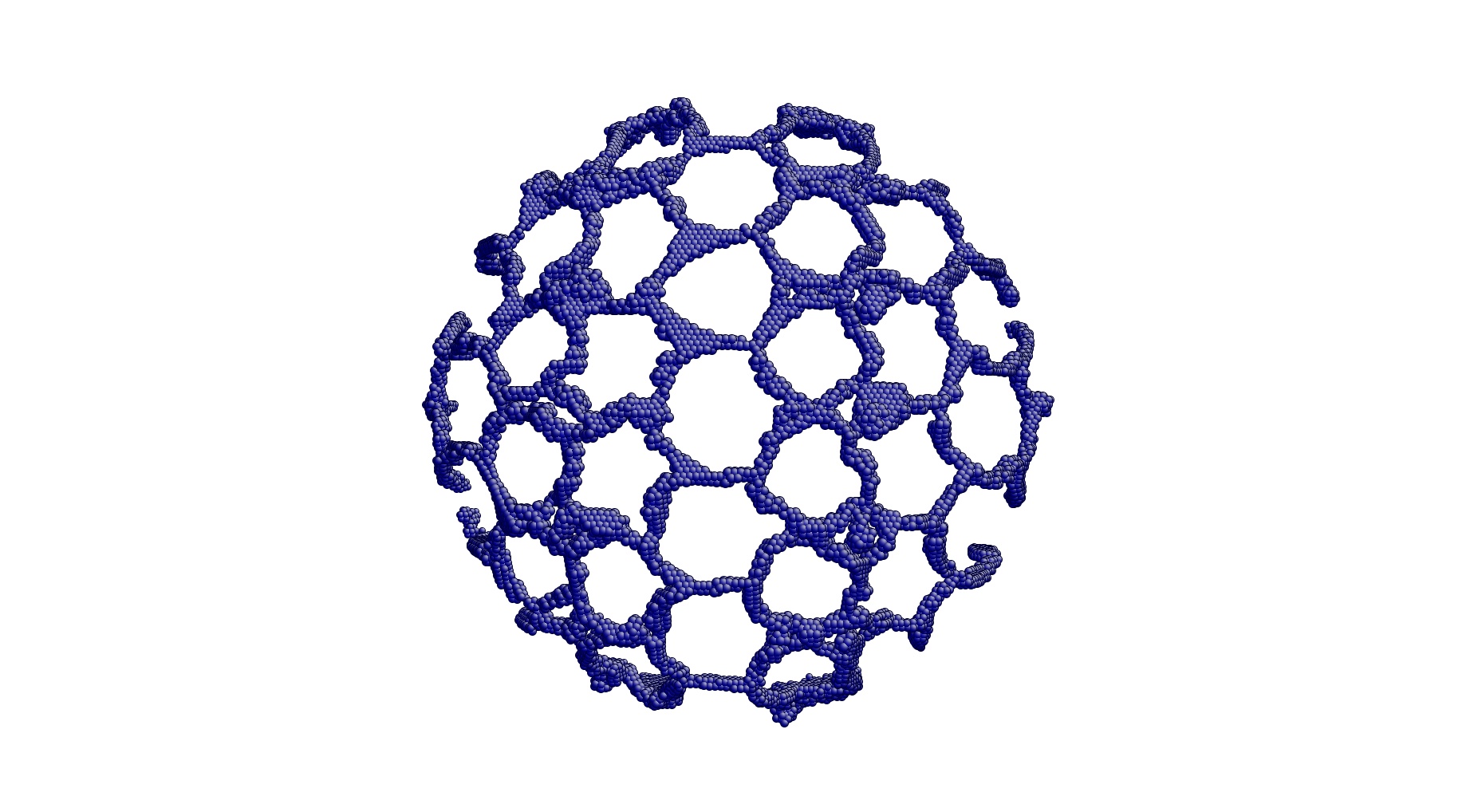}}
\subfigure[]{\includegraphics[height=1.3in]{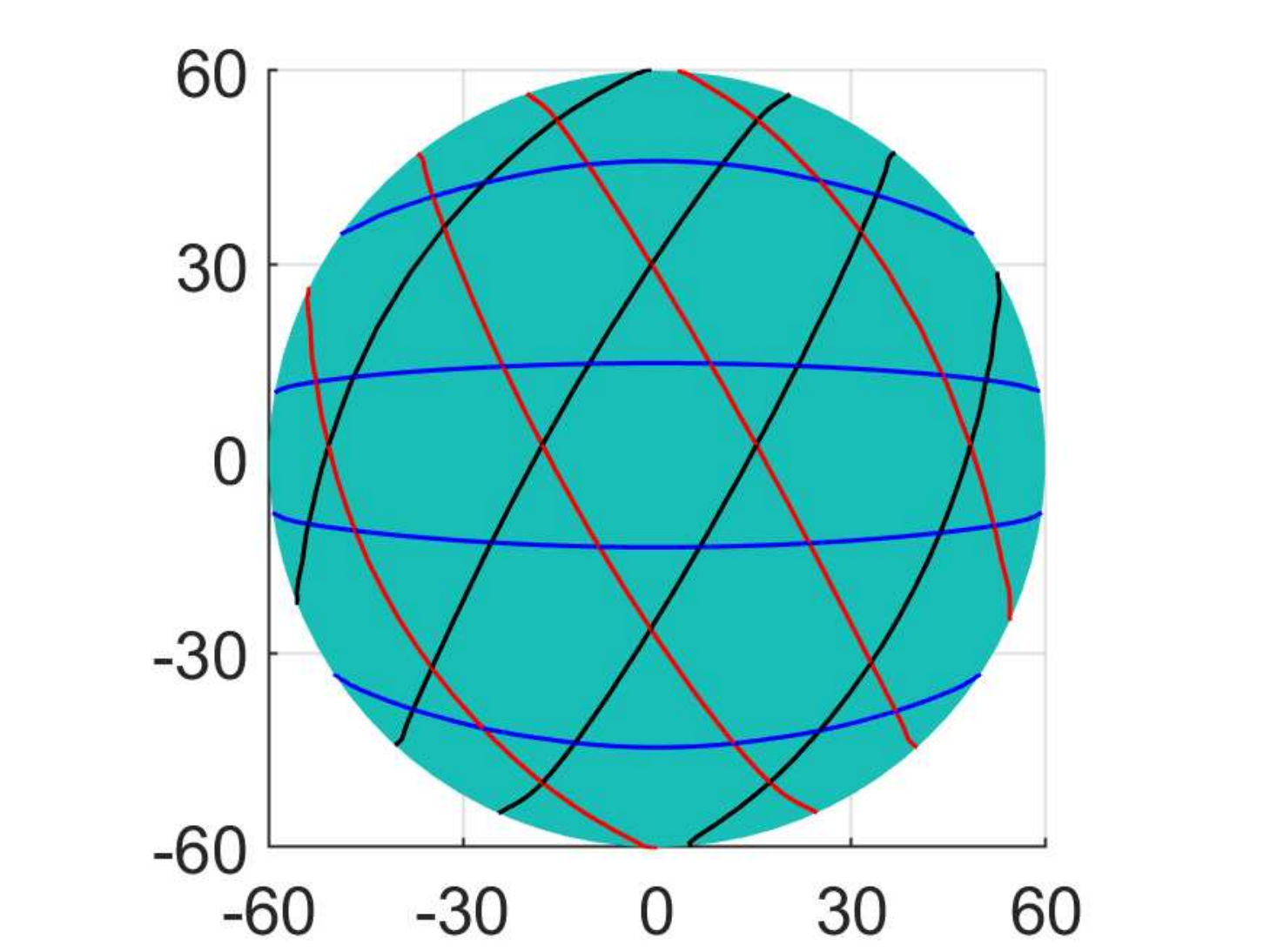}}
\subfigure[]{\includegraphics[width=2.5in]{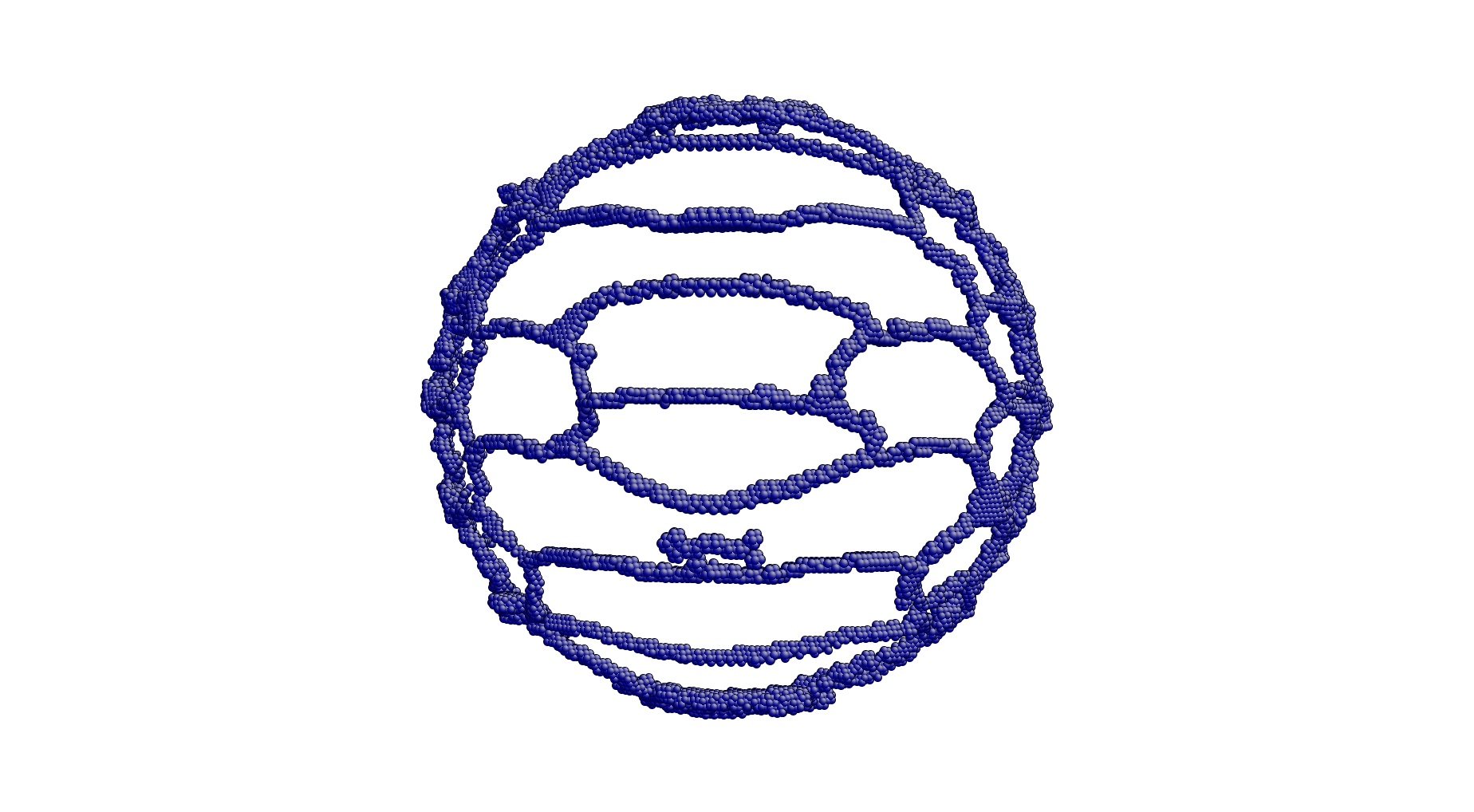}}
\subfigure[]{\includegraphics[height=1.3in]{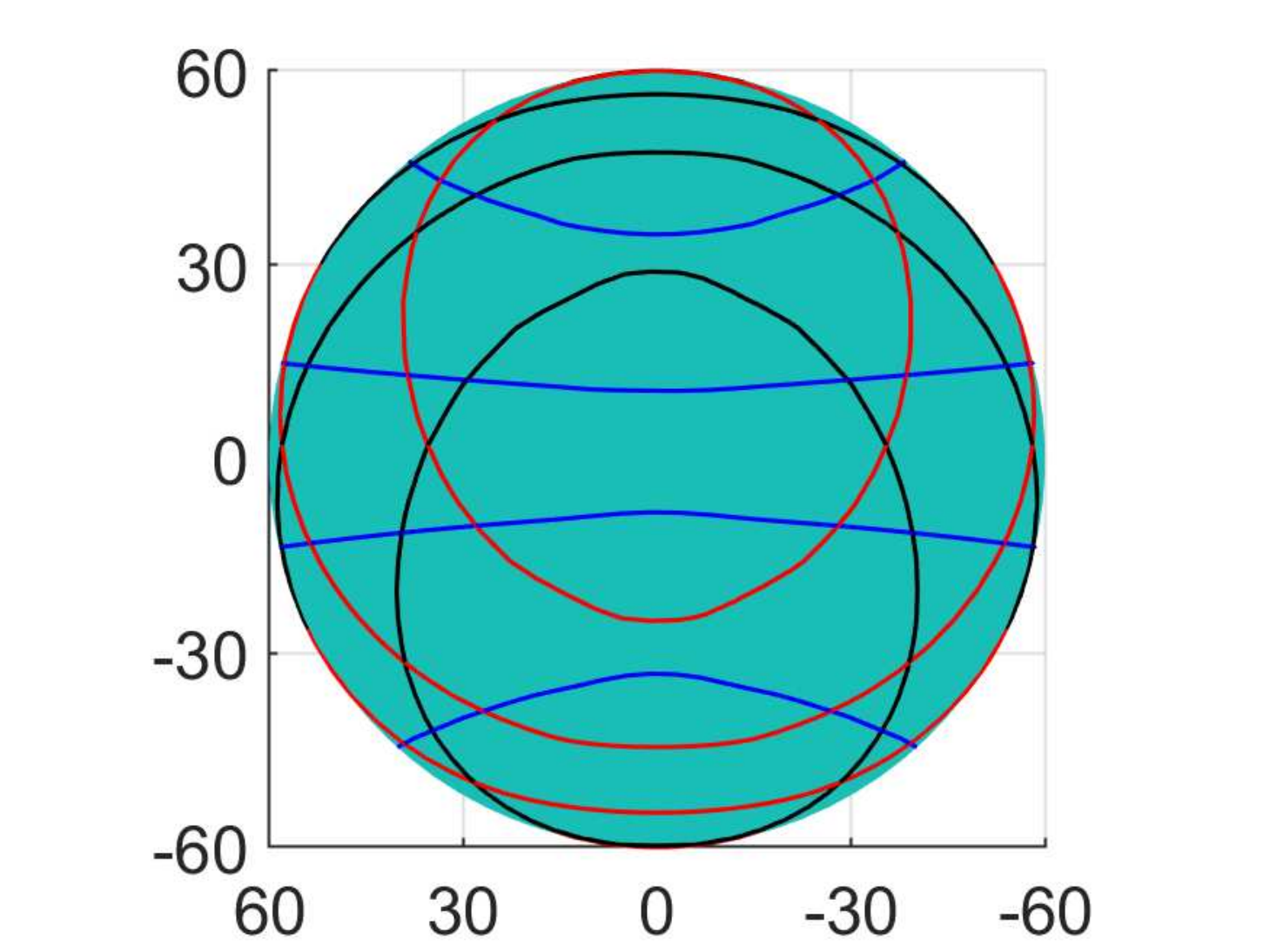}}
\subfigure[]{\includegraphics[width=2.5in]{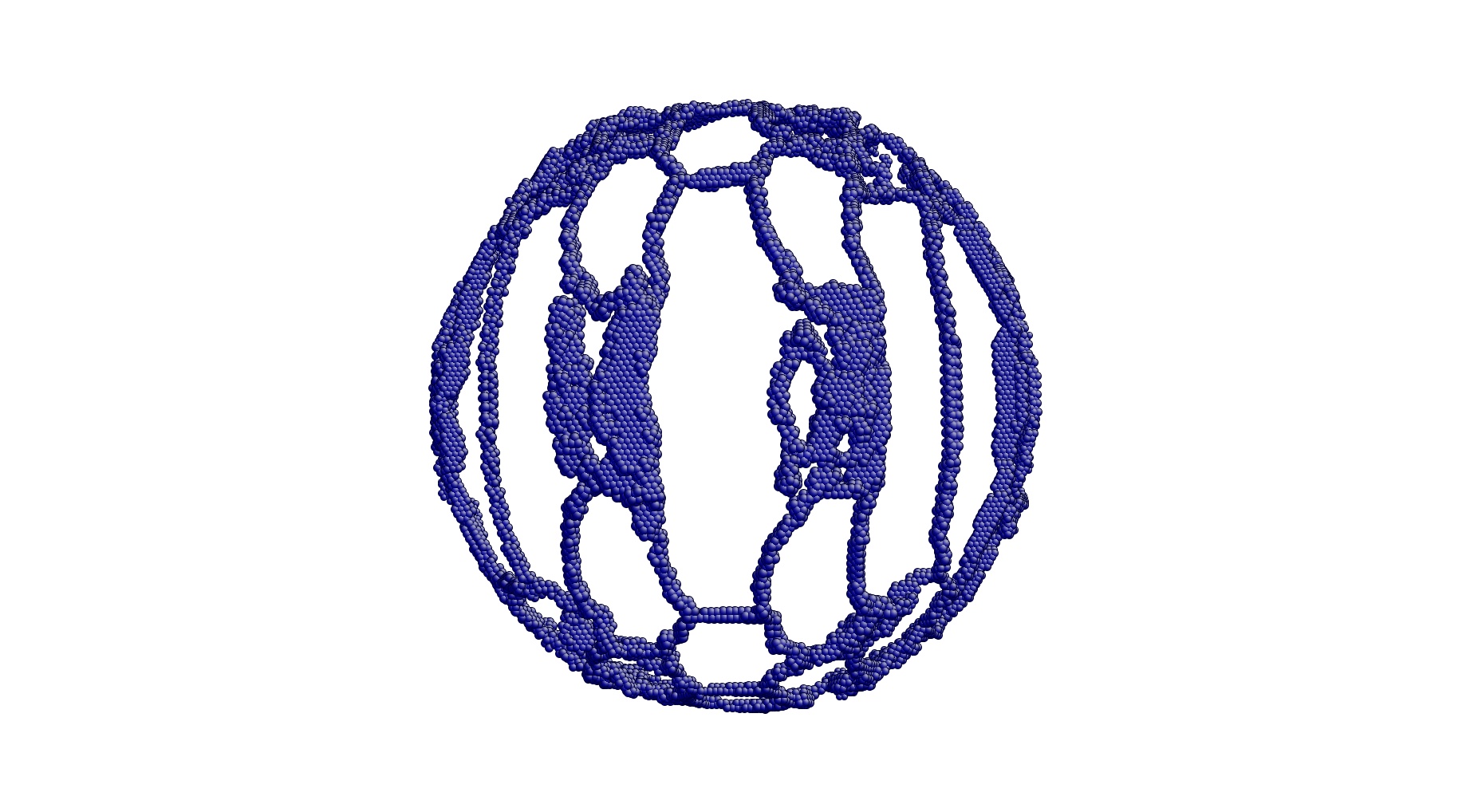}}
\subfigure[]{\includegraphics[height=1.3in]{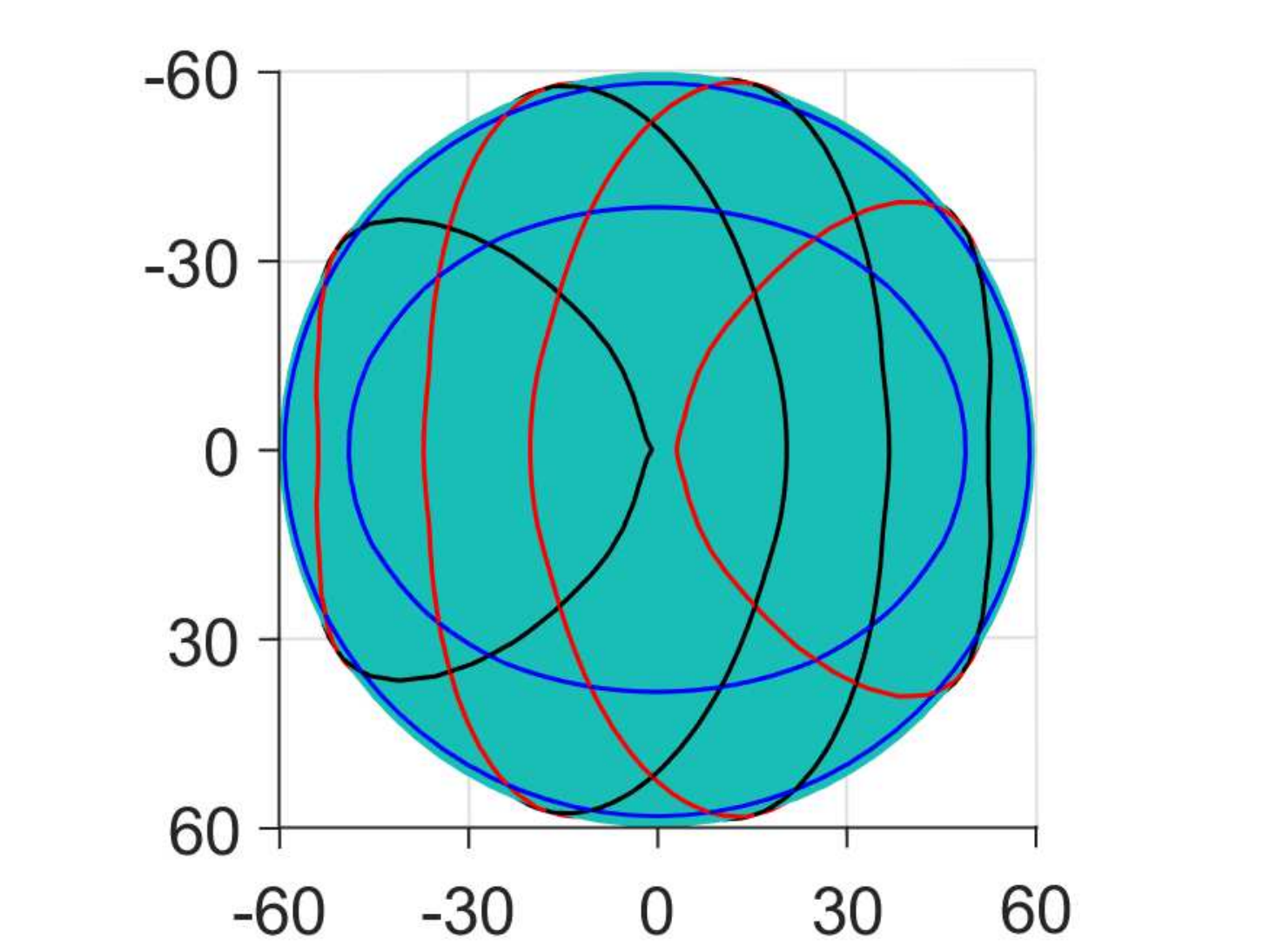}}
\caption{Dislocation structure of the boundary of a spherical grain with rotation axis $\mathbf a$ in the $x$ axis direction (meaning the $[111]$ direction) and misorientation angle  $\theta=3^\circ$.  The right panel shows the numerical result obtained using our continuum model, and the left panel shows the result of MS simulation.  The length unit in the images of continuum model result is $b$. (a) and (b) Three-dimensional view. The dislocation structure on this grain boundary consists of dislocations with Burgers vectors $\mathbf{b}^{(1)}$,  $\mathbf{b}^{(2)}$, and $\mathbf{b}^{(3)}$. Dislocations with these three Burgers vectors are shown by blue, black, and red lines, respectively, in the images of result of the continuum model. (c) and (d) View from the $+x$ direction (the  $[111]$ direction).  (e) and (f) View from the $+y$ direction (the $[\bar{1}10]$  direction). (g) and (h) View from the $+z$ direction (the  $[\bar{1}\bar{1}2]$ direction).  }
\label{fig:sphere}
\end{figure}

We consider a spherical grain boundary in this subsection.
 We choose the directions $[111]$, $[\bar{1}10]$, $[\bar{1}\bar{1}2]$ and  to be the $x$, $y$ and $z$ directions, respectively.
The six burgers vectors in this coordinate system are $\mathbf{b}^{(1)}=(0,1,0)b$, $\mathbf{b}^{(2)}=(0,\frac{1}{2},\frac{\sqrt{3}}{2})b$,
$\mathbf{b}^{(3)}=(0,\frac{1}{2},-\frac{\sqrt{3}}{2})b$,
$\mathbf{b}^{(4)}=(-\frac{\sqrt{6}}{3},0,\frac{\sqrt{3}}{3})b$,
$\mathbf{b}^{(5)}=(\frac{\sqrt{6}}{3},\frac{1}{2}$,
$\frac{\sqrt{3}}{6})b,\mathbf{b}^{(6)}=(\frac{\sqrt{6}}{3},-\frac{1}{2},\frac{\sqrt{3}}{6})b$. The  rotation axis $\mathbf{a}=(1,0,0)$, i.e., in the $[111]$ direction, and the rotation angle $\theta = 3^\circ $.

The spherical grain boundary is parametrized using spherical coordinates $(R,\beta,\alpha)$, with $R=60b$, $0\leq \beta\leq \pi$, and $0\leq \alpha < 2\pi$.
The unit outer normal vector is
$\mathbf{n} = (\sin\beta \cos\alpha,\sin\beta \sin\alpha,\cos\beta)$,
and the surface gradient of $\eta_j$ is
\begin{equation}
\bigtriangledown_s\eta_j = \frac{1}{R}\left(\cos\beta\cos\alpha\frac{\partial\eta_j}{\partial\beta}
-\frac{\sin\alpha}{\sin\beta}\frac{\partial\eta_j}{\partial\alpha},
\cos\beta\sin\alpha\frac{\partial\eta_j}{\partial\beta}
+\frac{\cos\alpha}{\sin\beta}\frac{\partial\eta_j}{\partial\alpha},
-\sin\beta\frac{\partial\eta_j}{\partial\beta}\right).
\end{equation}
We choose the vector $\mathbf{V}$ in Frank's formula (Eq.~\eqref{eqn:grad2}) to be the two unit tangent vectors $\mathbf{t_1}=(\cos\beta\cos\alpha,\cos\beta\sin\alpha,-\sin\beta)$ and $\mathbf{t_2}=(-\sin\alpha,\cos\alpha,0)$.

The dislocation structure on this spherical grain boundary obtained using our method and comparison with the MS result are shown in Fig.~\ref{fig:sphere}. These dislocations also have Burgers vectors $\mathbf{b}^{(1)}$,  $\mathbf{b}^{(2)}$, and $\mathbf{b}^{(3)}$.
The grain boundary is pure twist at $x=\pm R$, and pure tilt at $y=\pm R$ and $z=\pm R$.
It can be seen that the dislocation structures at these extreme points are consistent with the results of planar low angle grain boundaries calculated in Ref.~\cite{zhang2017energy}.
These results obtained using our continuum model agree excellently with the MS simulation results.
At the pure twist point of $x=R$, using the identification method presented in Sec.~\ref{sec:identification},  we generate the exact dislocation network structure of the planar pure twist boundary with the dislocation densities and orientations at this point obtained using the continuum model, and the result is shown in Fig.~\ref{fig:networks}(b). It can be seen that the planar network structure in Fig.~\ref{fig:networks}(b) indeed recovers the dislocation network structure  near the point  $x=R$ on the spherical grain boundary as shown in the MS simulation result in Fig.~\ref{fig:sphere}(c) (where $x=R$ is the center point). Notice that the hexagons on the spherical grain boundary in the MS simulation result in Fig.~\ref{fig:sphere}(c) are slightly greater than those in the planar network structure in Fig.~\ref{fig:networks}(b). This is believed to be due to the discreteness of the number of hexagons in the MS result.

\begin{figure}[htbp]
\centering
\subfigure[]{\includegraphics[width=2.2in]{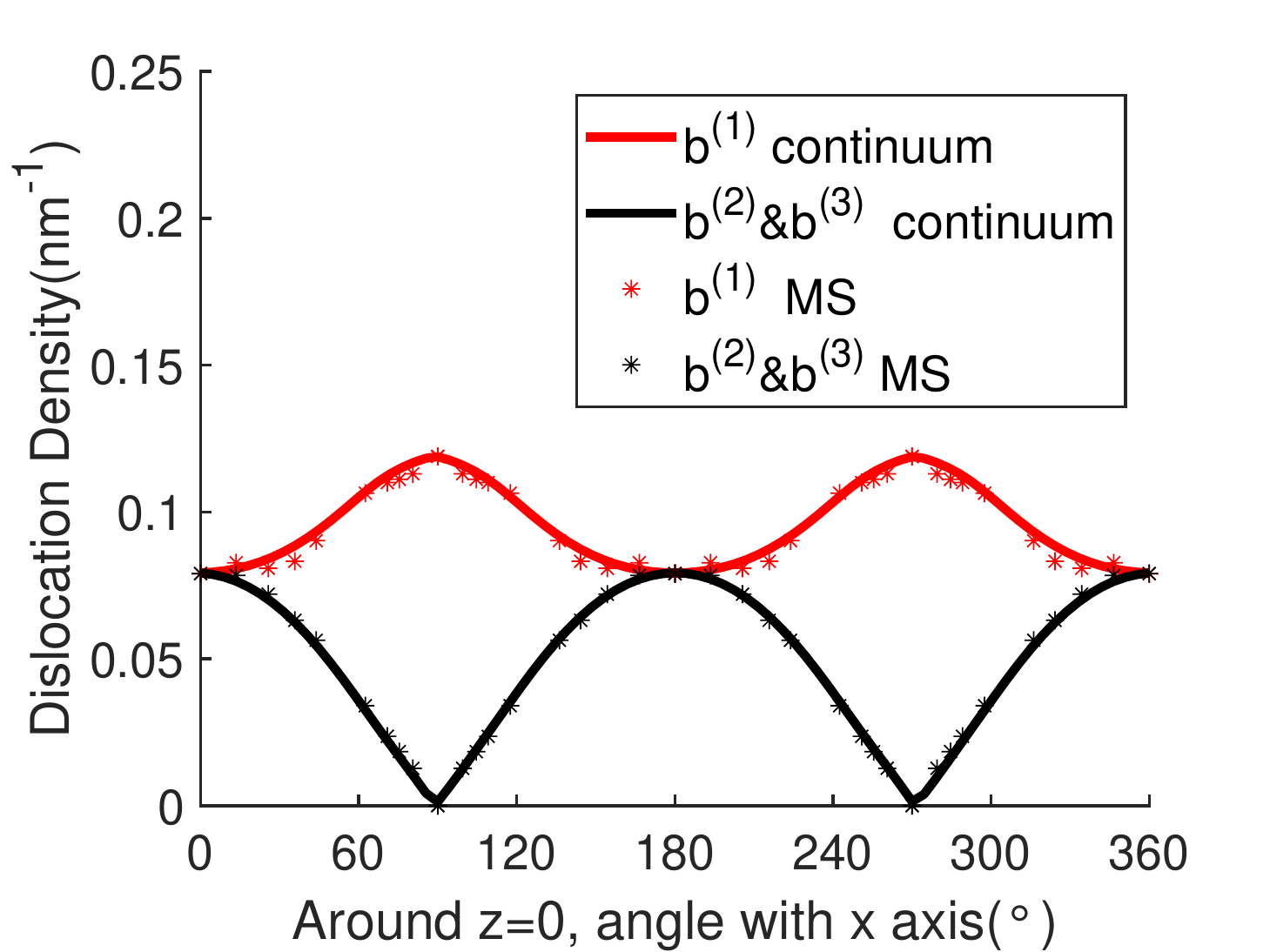}}
\subfigure[]{\includegraphics[width=2.2in]{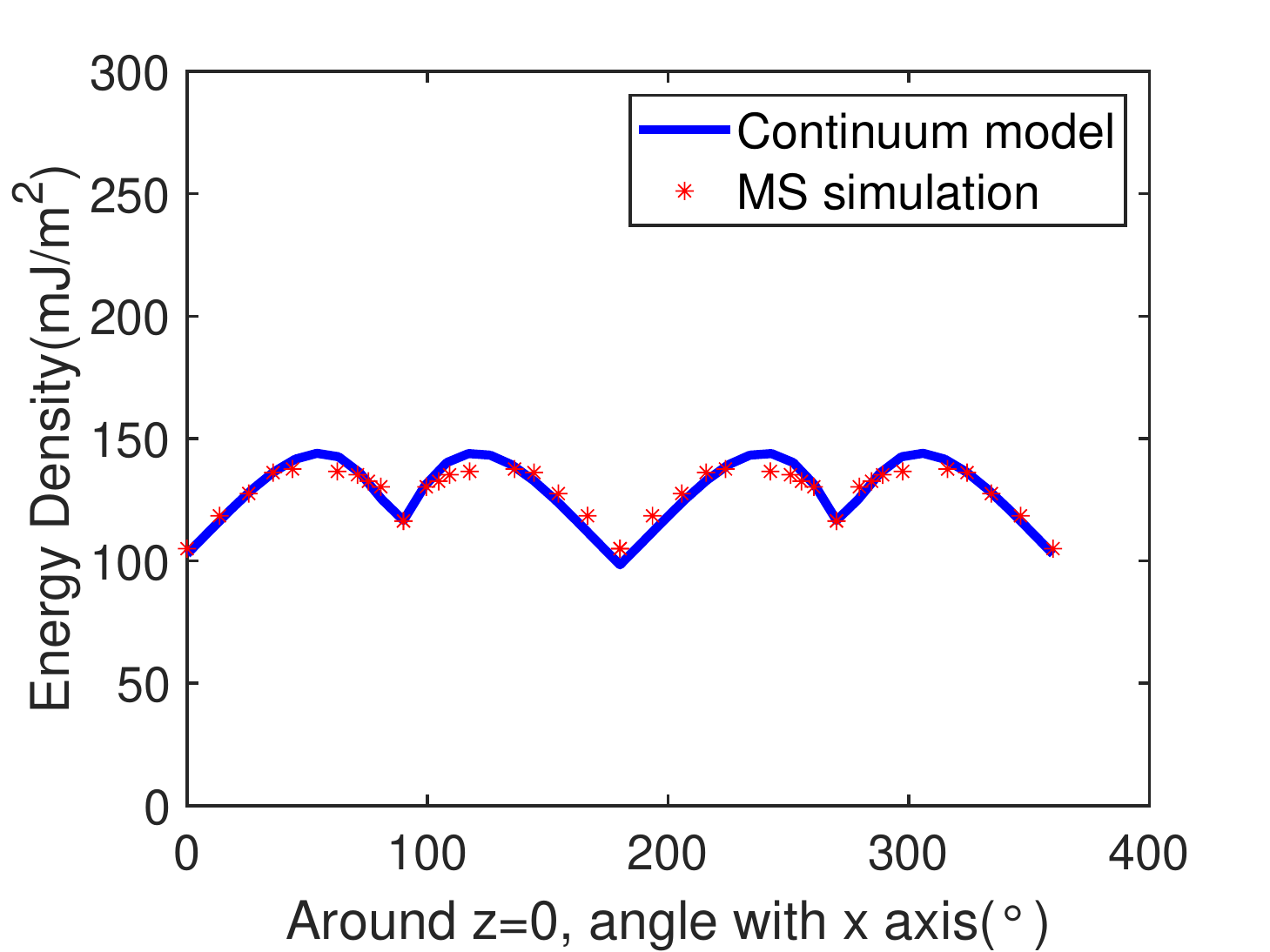}}
\subfigure[]{\includegraphics[width=2.2in]{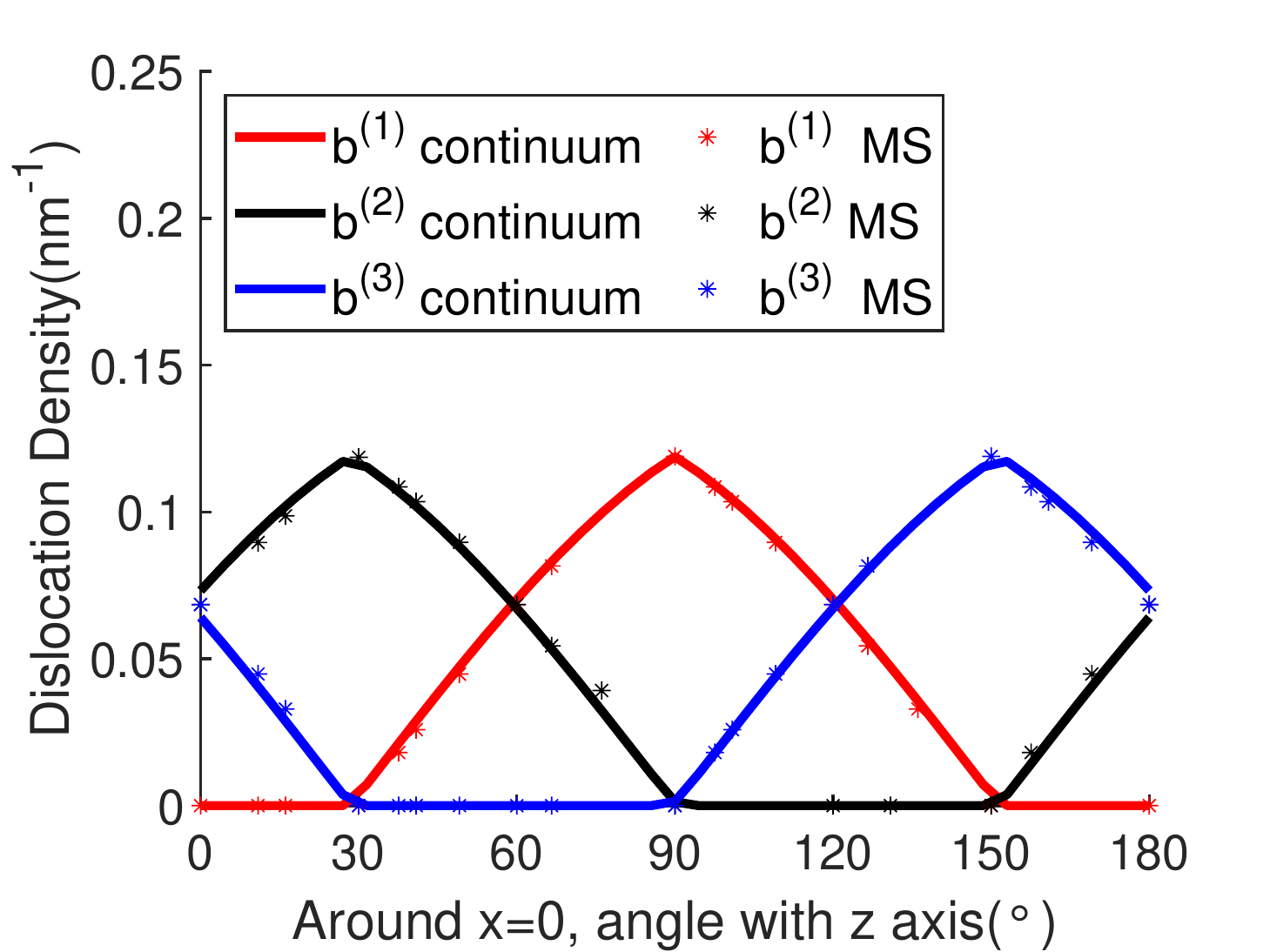}}
\subfigure[]{\includegraphics[width=2.2in]{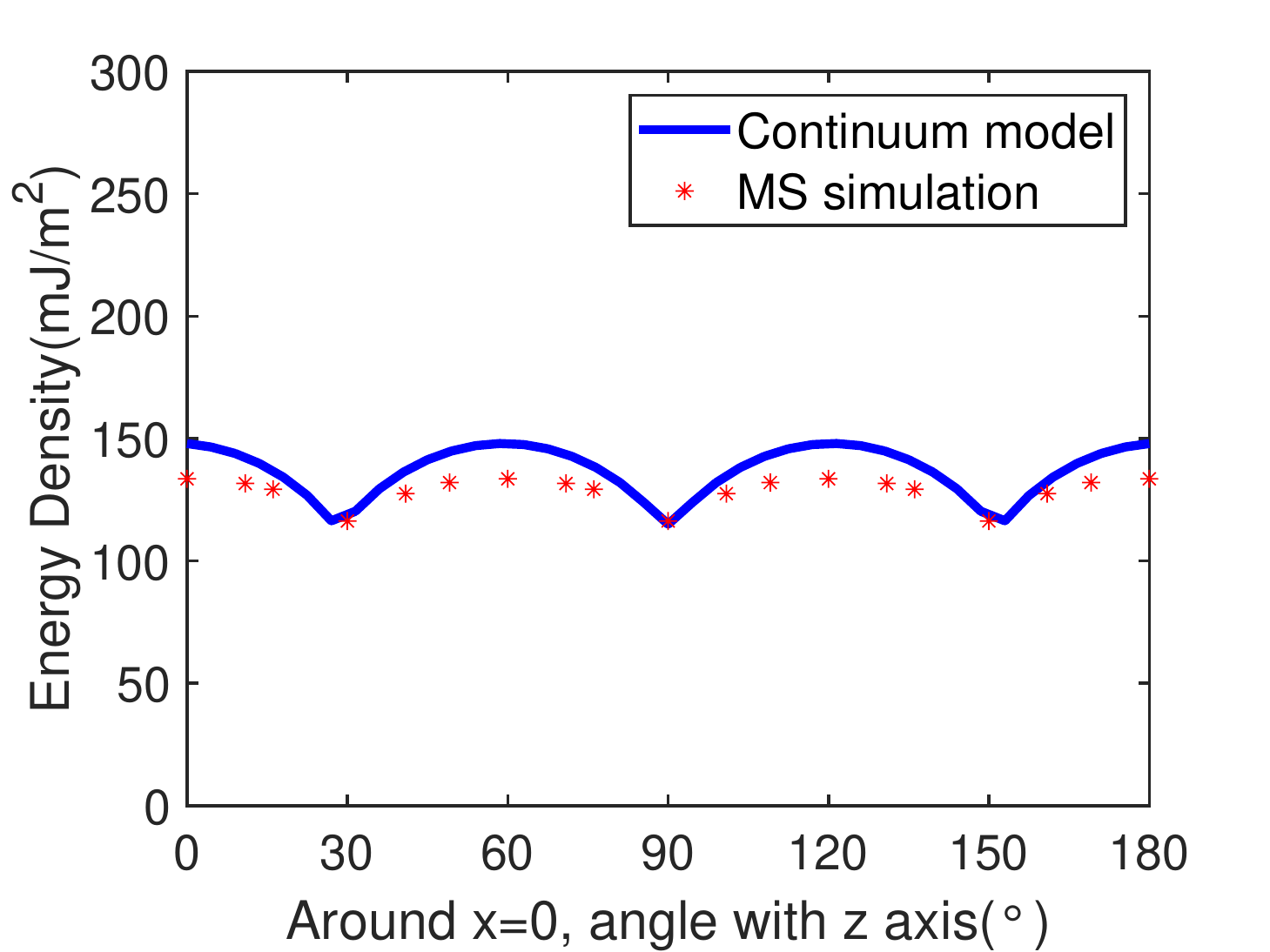}}
\subfigure[]{\includegraphics[width=2.2in]{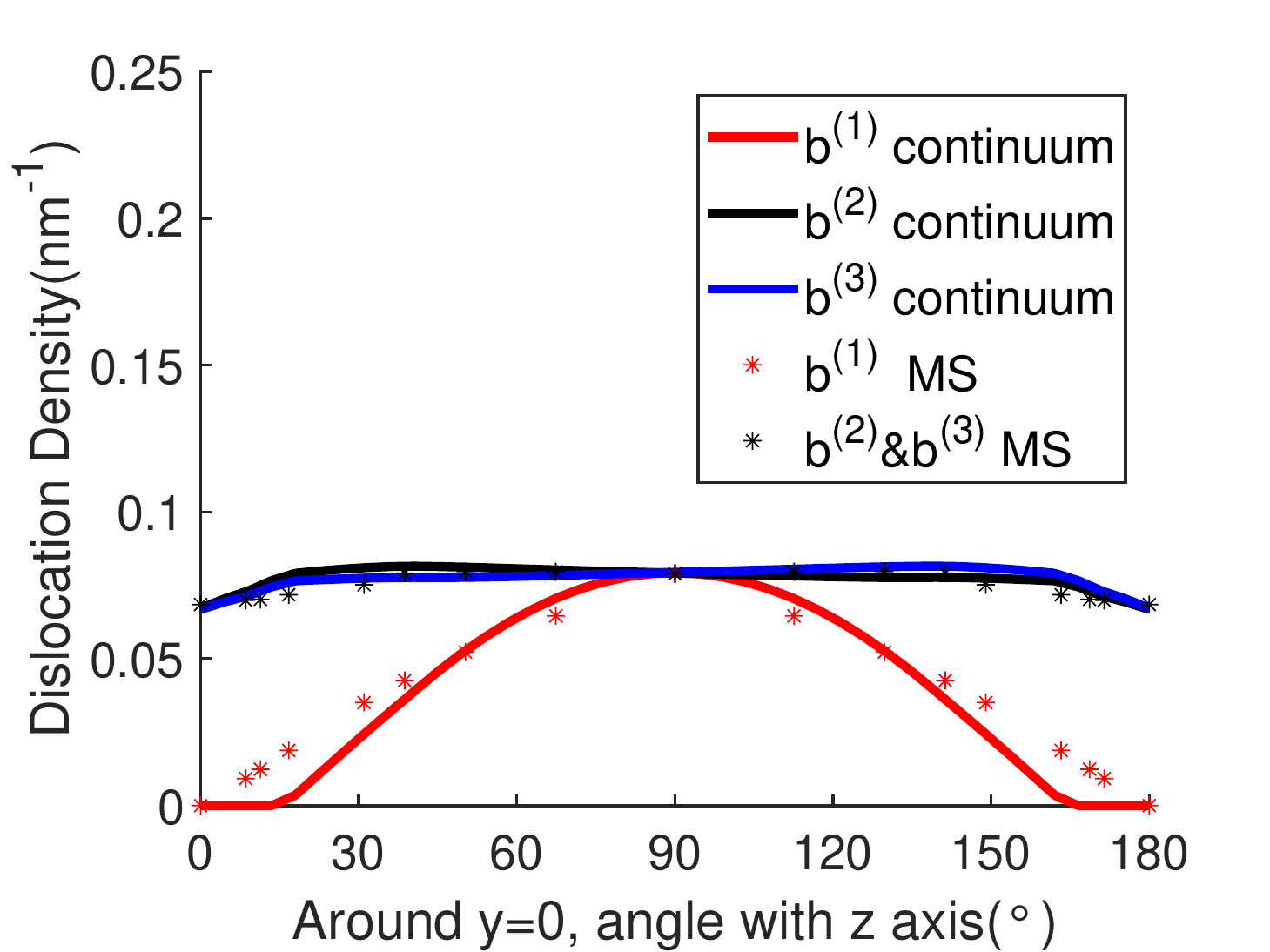}}
\subfigure[]{\includegraphics[width=2.2in]{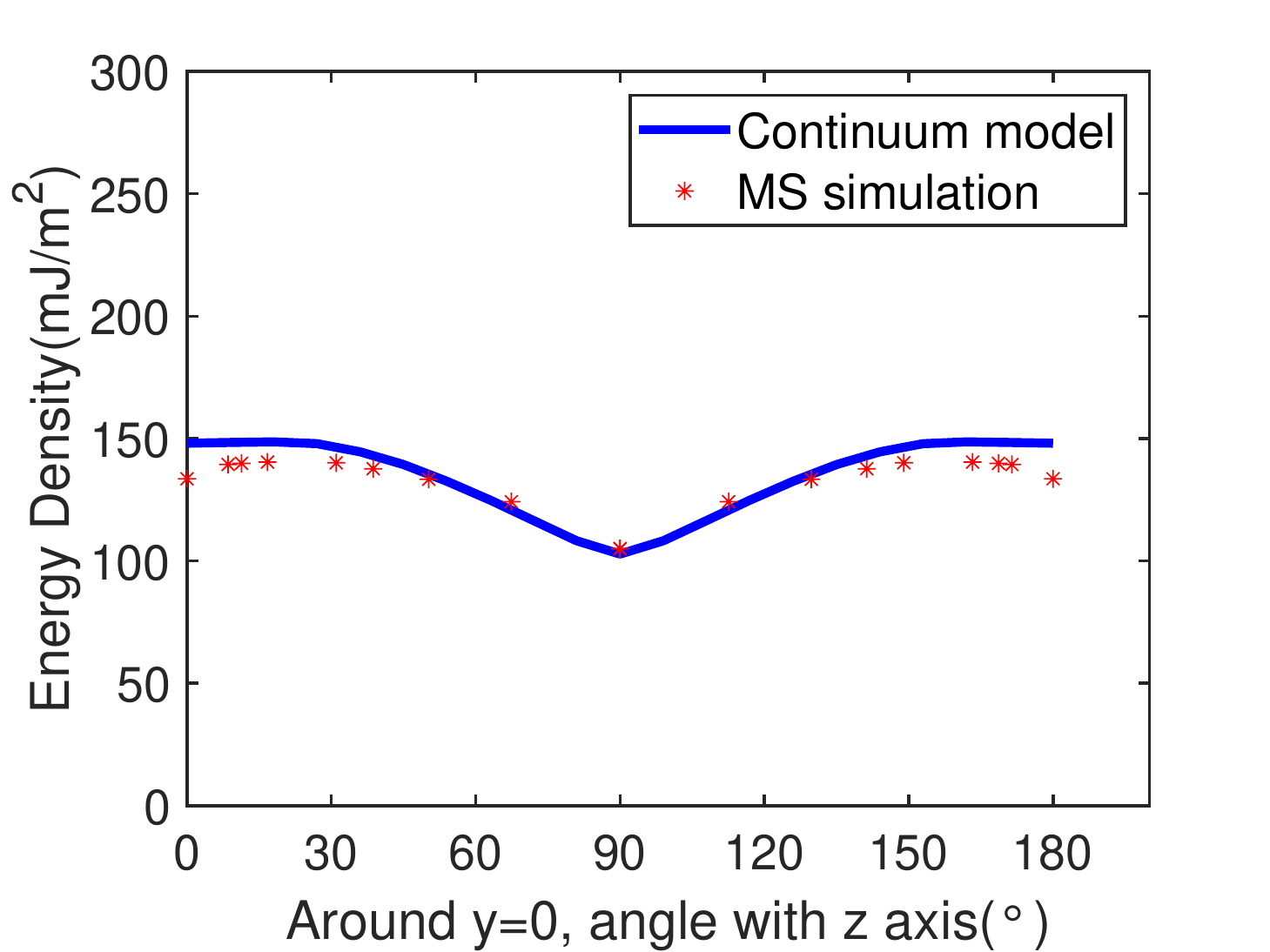}}
\caption{Comparisons of the dislocation density and energy between the results of our continuum model and those MS simulations. The misorientation angle is $\theta = 1.95^\circ$. (a) and (b) are the dislocation density and energy density along $z=0$ on the spherical grain boundary.  (c) and (d) are the dislocation density and energy density along $x=0$.  (e) and (f) are the dislocation density and energy density along $y=0$. }
\label{fig:zxy0s}
\end{figure}

For quantitative comparisons, we compare the
densities of dislocations and energy density along the three lines $x=0$, $y=0$ and $z=0$ on this spherical grain boundary, obtained using our method and  MS simulation. The results for $\theta = 1.95^\circ$ are shown in Fig.~\ref{fig:zxy0s}.
As we have done in cylinder,  we compare the pointwise results on the spherical boundary obtained using our method with the MS results for planar low angle grain boundaries calculated in Ref.~\cite{zhang2017energy} (Figs.~6, 8, 10 there). Again excellent agreement between the results using these two methods can be seen.

%\newpage
\section{Conclusions}
\label{s5}

We have presented a continuum model to determine the dislocation structure and energy of low angle grain boundaries in three dimensions.  The equilibrium dislocation structure is obtained by minimizing the grain boundary energy that is associated with the constituent dislocations subject to the constraint of Frank's formula. The orientation-dependent continuous distributions of dislocation lines on grain boundaries are described conveniently using the dislocation density potential functions, whose contour lines on the grain boundaries represent the dislocations. The energy of a grain boundary is the total energy of the constituent dislocations derived from discrete dislocation dynamics model, incorporating both the dislocation line energy and reactions of dislocations.  We have also proposed a method to identify the exact dislocation network structures from the  dislocation densities obtained using the continuum model.

The constrained energy minimization problem is solved by the augmented Lagrangian method.   A numerical formulation that avoids ill-posedness is proposed for the nonconvex gradient energy in the continuum model, and a numerical method based on projection method is presented to ensure connectivity of the dislocations. Comparisons with atomistic simulation results show that our continuum model is able to give excellent predictions of the dislocation structure and  energy of both planar and curved low angle grain boundaries.

The presented continuum model for the dislocation structure and energy of static low angle grain boundaries in three dimensions provides a basis for the dynamics model, which will be presented elsewhere~\cite{Qin2020}. Future work may also include generalization of the continuum model to the dislocation structures of heterogeneous interfaces \cite{Quek2011,WangJian2013,Demkowicz2013}.
 In this paper, we focus on the equilibrium dislocation structure and energy of low angle grain boundaries. Such an equilibrium state is stable under the strong long-range elastic interaction generated by the constituent dislocations of the grain boundary (described by the Frank's formula which is equivalent to the cancellation of the long-range elastic field)
 \cite{Frank1950,Bilby1955,HirthLothe1982,Sutton1995,zhu2014continuum,Xiang-Yan2018}. Influences of the external stress fields and other defects such as external dislocations are in general considered in the dynamic processes of grain boundaries, which  will also be considered in the future work.

%\section*{Acknowledgments}

\bibliographystyle{siamplain}
\bibliography{ref}
\end{document}